\documentclass[onecolumn,amsmath,amssymb,superscriptaddress,nofootinbib,11pt,a4paper]{revtex4}

\pdfoutput=1
\usepackage[T1]{fontenc}
\usepackage{CJK}
\usepackage{xcolor}
\usepackage{amsfonts}
\usepackage{epstopdf}
\usepackage{wrapfig} 
\usepackage{subfigure}
\usepackage{graphicx}  
\usepackage{dcolumn}   
\usepackage{bm}
\usepackage{float}
\usepackage[T1]{fontenc}
\usepackage{CJK}
\usepackage{xcolor}
\usepackage{amsfonts}
\usepackage{epstopdf}
\usepackage{wrapfig} 
\usepackage{subfigure}
\usepackage{graphicx}  
\usepackage{dcolumn}   
\usepackage{bm}
\usepackage{float}
\usepackage[utf8]{inputenc}
\usepackage[T1]{fontenc}
\newcommand{\be}{\begin{equation}}
\newcommand{\ee}{\end{equation}}
 \newcommand{\bea}{\begin{eqnarray}}
 \newcommand{\ena}{\end{eqnarray}}

\begin{document}

\title{Observational features of the  rotating Bardeen black hole surrounded by perfect fluid dark matter }
\author{Ke-Jian He}\email{kjhe94@.163com}
\affiliation{Department of Mechanics, Chongqing Jiaotong University, Chongqing 400000, People's Republic of China}
\author{Guo-Ping Li}\email{gpliphys@yeah.net}
\affiliation{School of Physics and Astronomy, China West Normal University, Nanchong 637000, People's Republic of China}
\author{Chen-Yu Yang}\email{chenyuyang2024@163.com}
\affiliation{Department of Mechanics, Chongqing Jiaotong University, Chongqing 400000, People's Republic of China}
\author{Xiao-Xiong Zeng}\thanks{Corresponding author: xxzengphysics@163.com}
\affiliation{College of Physics and Electronic Engineering, Chongqing Normal University, Chongqing 401331, People's Republic of China}
\affiliation{Department of Mechanics, Chongqing Jiaotong University, Chongqing 400000, People's Republic of China}

\begin{abstract}
{ By employing ray-tracing techniques, we investigate the shadow images of rotating Bardeen black holes surrounded by perfect fluid dark matter. In this work, two models are considered for the background light source, namely the celestial light source model and the thin accretion disk model. Regarding the celestial light source, the investigation focuses on the impact of variations in relevant parameters and observed inclination on the contour and size of the shadow. For the thin accretion disk model, the optical appearance of a black hole is evidently contingent upon  the radiative properties exhibited by the accretion disk, as well as factors such as observed inclination and relevant parameters governing spacetime. With an increasing observation inclination, the observed flux of direct and lensed images of the accretion disk gradually converge towards the lower region of the image, while an increase in the dark matter parameter $a$ significantly expands the region encompassing both direct and lensed images. Furthermore, the predominant effect is redshift at lower observation angles, whereas the blueshift effect only becomes apparent at higher observation angles. Simultaneously, the increase in the  observation inclination will amplify the redshift effect, whereas an increase in the magnetic charge $\mathcal{G}$, rotation parameter $a$ and the absolute value of dark matter parameter $\alpha$ will attenuate the redshift effect observed in the image. These observations of a rotating Bardeen black hole surrounded by perfect fluid dark matter could provide a convenient way to distinguish it from other black hole models.}
\end{abstract}

\maketitle
 \newpage
\section{Introduction}
\label{sec:intro}
Since the inception of General Relativity (GR), it has indubitably entrenched its foundational status within the domain of modern physics. Among them, black holes, as one of the fundamental predictions of GR, have continued to attract the keen interest of physicists and astronomers. With the advancement of scientific exploration, the groundbreaking detection of gravitational waves and subsequent release of awe-inspiring images portraying the supermassive black hole situated at the core of the colossal elliptical galaxy M87$^*$\cite{EventHorizonTelescope:2019dse,EventHorizonTelescope:2019uob,EventHorizonTelescope:2019jan,EventHorizonTelescope:2019ths,EventHorizonTelescope:2019pgp,EventHorizonTelescope:2019ggy}, along with the black hole residing at Sagittarius A in the Milky Way \cite{EventHorizonTelescope:2022wkp,EventHorizonTelescope:2022vjs,EventHorizonTelescope:2022wok,EventHorizonTelescope:2022exc,EventHorizonTelescope:2022urf,EventHorizonTelescope:2022xqj}, have unequivocally substantiated the existence of black holes as tangible celestial entities within our universe, supported by unparalleled empirical evidence. These images of black holes not only offer a opportunities to directly observe and experimentally verify gravity in immensely strong magnetic fields, but also unveil the intricate electromagnetic interactions, precise matter distribution patterns, and  accretion processes in the vicinity of black holes. Most significantly, they serve as compelling evidence to robustly validate the accuracy and efficacy of GR under extremely intense gravitational field conditions\cite{Synge:1966okc,Bar,Cunha:2018acu,Perlick:2021aok,Liu:2022ruc,
Wang:2022kvg,Chen:2022scf,Vagnozzi:2022moj,Zhong:2021mty,Zhang:2022osx,Huang:2024wpj,Li:2020drn}, further solidifying its fundamental position in modern physics.

It is widely acknowledged that black holes themselves do not emit light; however, they possess the capability to absorb the  hot, magnetized plasma in their vicinity, thereby giving rise to a luminous accretion disk encircling the black hole. The light emitted by the thermally synchronized electrons in this disk is considered to be the most probable source of radiation within the frequency range corresponding to the black hole image. Additionally, in the case of black holes exhibiting extremely high rates of rotation (i.e., high-spin black holes), their electromagnetic energy can induce a phenomenon referred to as relativistic jet\cite{Blandford:1977ds}. The base of the jet is encompassed by a distinctive funnel wall structure, which not only shapes the morphology of the jet but also potentially triggers substantial amounts of thermal synchrotron radiation through intricate interactions with its surrounding environment. Therefore, both the accretion disk and  jet play a crucial role as the background light source providing an indispensable contribution to the horizon-scale imaging process of the black hole. In numerous investigations on black hole shadows, researchers have extensively examined the specific impacts of various accretion models on the black hole images. These models include, but are not limited to, spherical accretion models\cite{Narayan:2019imo,Zeng:2020dco,Heydari-Fard:2023ent}, optically  and geometrically thin accretion disk models\cite{Gralla:2019xty,Zeng:2020vsj,Peng:2020wun,Zhang:2023bzv,Hou:2022eev,He:2022yse,Li:2021ypw,Zeng:2021mok,Guo:2021bwr,Zeng:2021dlj,Li:2021riw,He:2021htq,
Gao:2023mjb,Cui:2024wvz,Wang:2023fge,Hu:2023pyd,Meng:2024puu,Zeng:2022pvb,Zeng:2022fdm}, and geometrically thick accretion disk models\cite{Zhang:2024jrw,Zhang:2024lsf,Gjorgjieski:2024csb}.  By conducting a meticulous analysis of these models, one can acquire a more profound comprehension of the attributes exhibited by black hole shadows and their manifestations under different conditions.

The current observations have not yet definitively ruled out the possibility of deviations from GR, thereby providing ample scope for the validation of alternative gravitational theories. The singularity problem in GR is regarded as one of the most fundamental issues, reflecting the incompleteness of GR. Consequently, numerous endeavors have been made to address this problem, encompassing quantum corrections and modifications to the gravitational theory\cite{Cognola:2013fva,Zhang:2019dgi}. In 1968, Bardeen proposed a black hole solution with regular non-singular geometry, i.e., the Bardeen black hole\cite{bardeen1968non}, and subsequent comprehensive analysis revealed that singularities can be circumvented through topological changes within this class of black holes\cite{Borde:1994ai,Borde:1996df}. It has been demonstrated that the physical source of such solutions is the gravitational collapse of magnetic monopole generated in a particular form of nonlinear electrodynamics\cite{Ayon-Beato:2000mjt}. Subsequently, Hayward  proposed another static spherically symmetric black hole solution that has the limitations and regularity of curvature invariants, while also eliminating the singularity problem of black holes\cite{Hayward:2005gi}.  Other regular black hole have also been obtained, such as the Ay\'{o}n-Beato-Garc\'{\i}a black hole\cite{Ayon-Beato:1998hmi}, Berej-Matyjasek-Trynieki-Wornowicz black hole and rotating Bardeen black hole black hole\cite{Berej:2006cc}. The pertinent physical properties of these regular black holes have been extensively investigated, encompassing dynamic stability, quasi-normal modes of scalar field perturbations, weak gravitational lensing, and geodesic motion of test particles\cite{Khan:2021tzv,Fernando:2012yw,Ghaffarnejad:2014zva,Zhou:2011aa,Guo:2021wcf}.

On the other hand, based on observations of galaxy cluster dynamics, Zwicky analyzed the mass of these clusters within the universe and inferred the existence of a significant amount of invisible matter, commonly known as dark matter\cite{Zwicky:1933gu}. In the work of, Holmberg and Smith arrived at a similar conclusion subsequent to their investigations on the mass of the galaxy system and the Virgo Cluster\cite{Hol,Smith:1936mlg}. The gravitational influence of dark matter in the universe drives the aggregation of visible matter, resulting in the formation of galaxies, galaxy clusters, and supergalactic structures. However, the concept of dark matter implies its lack of interaction with electromagnetic radiation, resulting in its inability to reflect, absorb, or emit such radiation and thus making direct observation impossible. According to the observation of cosmic microwave background (CMB) radiation anisotropy and the standard cosmological model ($\Lambda$CDM model), the results obtained by the Planck observation satellite have revealed that the cosmic matter-energy composition is primarily comprised of dark energy (approximately 68$\%$) and dark matter (approximately 27$\%$), with baryonic matter contributing a mere 5$\%$\cite{Planck:2013pxb}. Then, the dark matter is thought to be a halo around a black hole\cite{Jusufi:2019nrn,Konoplya:2019sns}, while the black hole solution surrounded by quintessence matter was first proposed by Kiselev\cite{Kiselev:2002dx}. In addition, Rahaman et al. proposed an interesting model in which black holes were surrounded by perfect fluid dark matter (PFDM), i.e., dark matter that maintains the properties of a perfect fluid, such as isotropic pressure and mass density\cite{Rahaman:2010xs}. Subsequently, Li et al. proposed a model wherein a black hole is enveloped by PFDM within a spatially heterogeneous phantom field background\cite{Li:2012zx}. Furthermore, Xu et al. employ the Newman-Janis algorithm to derive solutions for rotating black holes surrounded by  PFDM in Kerr-like spacetimes, thereby extending the existing solution for Kerr-de Sitter/anti-de Sitter spacetime with a cosmological constant\cite{Xu:2017bpz}.

Given the aforementioned discussion, it is imperative to investigate the impact of dark matter on the visual characteristics exhibited by black hole shadows. The previous study \cite{M:2022vyd} solely focused on the  shadow shape of rotating Bardeen black hole surrounded by PFDM, neglecting the consideration of its observed characteristics within an accretion model.  In this work, we  focus on the rotating Bardeen black holes surrounded by PFDM, aiming to analyze the impact of variations in key physical parameters, such as dark matter and magnetic charge, on the black hole shadow. Similarly,  the influence of variations in the observed inclination and related parameter on the shadow shape is also examined through the introduction of a celestial light source. After that, the thin accretion disk model as the exclusive background light source, we further investigate the observed characteristics of black hole shadows from various observation angles within this spacetime background,  and explore the dependence on these characteristics and associated parameters. Furthermore, a comprehensive investigation has been conducted on the redshift factors associated with both direct and lensed images of accretion disks.

The organization of this work is as follows. In  Section 2, we will provide a concise introduction to the Bardeen rotating black hole solution surrounded by PFDM, followed by the derivation of geodesic equations governing the motion  particles. In  Section 3, we employ a ray-tracing method to simulate images of the black hole shadow, where  the celestial light source is the only background source. The investigation focuses on the impact of relevant parameters and observed inclination on the morphology of black hole shadow. In  Section 4, we consider an optically and geometrically thin accretion disk positioned on the equatorial plane of the black hole in order to obtain the visually observed appearance of the black hole within the observation plane of the ZAMO. Moreover, we meticulously scrutinize the impacts of parameters on the redshift factor and the
appearance of the inner shadow. Finally, we draw the conclusions and discussions in Section 5.

\section{The rotating Bardeen black holes surrounded by PFDM}
\label{sec2}
In this section, we will provide a concise introduction to the Bardeen rotating black hole solution surrounded by PFDM, which serves as the underlying framework throughout our study. In the Bardeen spacetime, considering the coupling of gravity and non-linear electromagnetic field, the corresponding Einstein-Maxwell equations should be modified as\cite{Zhang:2020mxi}
\begin{align}
\mathrm{G}^\nu_\mu=2\left(\frac{\partial \mathcal{L}(\mathcal{F})}{\partial \mathcal{F}} \mathcal{F}_{\mu\lambda} \mathcal{F}^{\nu\lambda}-\delta^\nu_\mu \mathcal{L} \right)+8 \pi \mathcal{T}^\nu_\mu,\label{EME1}
\end{align}
\begin{align}
\nabla_\mu \left(\frac{\partial \mathcal{L}(\mathcal{F})}{\partial \mathcal{F}} \mathcal{F}^{\nu\mu} \right)=0,\label{EME2}
\end{align}
\begin{align}
\nabla_\mu \left( \ast \mathcal{F}^{\nu\mu} \right)=0.\label{EME3}
\end{align}
The term $\mathcal{F}^{\nu\mu}$ in the above equation represents $\mathcal{F}^{\nu\mu}=2 \nabla_{[\mu \mathbf{A}_\nu]}$, and $\mathcal{L}(\mathcal{F})$  denotes the Lagrangian for a nonlinear electrodynamics source\cite{Ayon-Beato:2000mjt}, that is
\begin{align}\label{Lag1}
\mathcal{L}(\mathcal{F})=\frac{3 M}{\mid \mathcal{G} \mid ^3 } \left(\frac{\sqrt{2 \mathcal{G} ^2 \mathcal{F}}}{1+\sqrt{2 \mathcal{G} ^2 \mathcal{F}}}\right)^{5/2},
\end{align}
and
\begin{align}\label{Lag2}
\mathcal{F} \equiv \frac{1}{4} \mathcal{F}_{\mu\nu}\mathcal{F}^{\mu\nu}.
\end{align}
Here, the parameters of $M$ in this context are integral constants that correspond to the mass of the black hole, while the term $\mathcal{G}$ represents an integration constant associated with magnetic charges, specifically referring to the magnetic charge. Considering a black hole enveloped by PFDM, the energy-momentum tensor $\mathcal{T}^\mu_\nu$ can be expressed as\cite{Li:2012zx}
\begin{align}
\mathcal{T}^\mu_\nu=\text{diag} (-\rho, p_r, p_\theta, p_\varphi),\label{EMT}
\end{align}
in which
\begin{align}
-\rho=p_r=-\frac{\alpha }{8 \pi r^3}, \qquad  p_\theta=p_\varphi=-\frac{\alpha }{16 \pi r^3}.\label{ro1}
\end{align}
The parameters of $\rho$ is the total energy density, and $(p_r, p_\theta, p_\varphi)$ is the pressure of dark energy. The solution for the Bardeen black hole surrounded by PFDM can be obtained by solving the Einstein-Maxwell equations and energy-momentum tensor, resulting in a static spherically symmetric black hole solution, which is
\begin{align}\label{metric1}
{ds}^2=-f(r){dt}^2+\frac{d r^2}{f(r)}+r^2 d \Omega^2,
\end{align}
and
\begin{align}\label{metric2}
f(r)=1-\frac{2 M r^2}{(r^2+\mathcal{G}^2)^{3/2}}+\frac{\alpha}{r} \ln \frac{r}{\alpha}.
\end{align}
Here, $d \Omega^2$  stands for the standard element on the unit 2-sphere $d \Omega ^2= {d \theta }^{2}+ \text{sin}^2(\theta) d\varphi^2$, and $\alpha$ is the dark matter parameter describing the intensity of PFDM. It is worth mentioning that in the limit of $\alpha \rightarrow 0$, the metric reverts to the Bardeen spacetime; in the limit of $\mathcal{G} \rightarrow 0$, it reduces to the Schwarzschild black hole surrounded by PFDM\cite{Li:2012zx}; and in the combined limit of $\alpha \rightarrow 0$ and $\mathcal{G} \rightarrow 0$, one can obtain the Schwarzschild black hole. Then, the line element of  rotating Bardeen black hole surrounded by PFDM was derived by\cite{Azreg-Ainou:2014pra,Azreg-Ainou:2014aqa},
\begin{align}\label{metric3}
{ds}^2=-\left(1-\frac{2 \rho r }{\Sigma}\right)d t^2+\frac{\Sigma}{\Delta_r}dr^2+\Sigma d \varphi^2+\sin^2 \theta \left(r^2+a^2+\frac{2 r a^2 \rho \sin^2 \theta}{\Sigma} \right)d\varphi^2
-4 a r\rho \frac{\sin^2 \theta}{\Sigma}dt d\varphi,
\end{align}
and
\begin{align}\label{metric4}
2\rho=\frac{2 M r^3}{(r^2+\mathcal{G}^2)^{3/2}}-\alpha \ln \frac{r}{|\alpha|},
\end{align}
\begin{align}\label{metric5}
\Delta_r=r^2+a^2-\frac{2 M r^4}{(r^2+\mathcal{G}^2)^{3/2}}+{\alpha}{r} \ln \frac{r}{|\alpha|},
\end{align}
\begin{align}\label{metric6}
\Sigma=r^2+a^2 \cos^2\theta.
\end{align}
Here, the parameters of $a$  represents the spin per unit mass of the black hole. It can be observed that the aforementioned line element converges to the rotating Bardeen metric in the case of dark matter parameters $\alpha=0$, and converges to the Kerr metric when $\mathcal{G}=\alpha=0$.  In this system, the curvature scalar is\cite{Zhang:2020mxi}
\begin{align}\label{CS}
\mathcal{R}=\frac{6 M \mathcal{G}^2 (4 \mathcal{G}^2-r^2)}{(\mathcal{G}^2+r^2)^{7/2}}-\frac{\alpha}{r^3},
\end{align}
and
\begin{align}\label{CS2}
\mathcal{R}_{\mu\nu} \mathcal{R}^{\mu\nu}=\frac{18 M^2 \mathcal{G}^4 (8\mathcal{G}^4-4\mathcal{G}^2 r^2+13 r^4)}{(\mathcal{G}^2+r^2)^{7}}+\frac{5 \alpha ^2}{2 r^6}-\frac{6M \mathcal{G}^2 (2\mathcal{G}^2+7 r^2)\alpha}{r^3(\mathcal{G}^2+r^2)^{7/2})}.
\end{align}
The results obtained from the above equation at $r = 0$ do not diverge, indicating that the presence of PFDM has no impact on the regularity of a Bardeen black hole. In other words, a Bardeen black hole surrounded by PDFM still has no singularity. The  horizon of  black hole can be given by the condition that
\begin{align}\label{HR}
\Delta_r=r^2+a^2-\frac{2 M r^4}{(r^2+\mathcal{G}^2)^{3/2}}+{\alpha}{r} \ln \frac{r}{|\alpha|}=0.
\end{align}
By considering equation $(\ref{HR})$, one can determine the Cauchy horizon $r_c$ (inner horizon)and event horizon $r_h$ (outer horizon)of the rotating Bardeen black hole enveloped by PDFM. The investigation of photon behavior in the vicinity of a  rotating Bardeen black hole enveloped by PDFM is a prerequisite for conducting a comprehensive study on the shadow associated with it. In this spacetime, the dynamics of photons can be effectively described by the Hamilton-Jacobi equation, which can be written as
\begin{align}\label{HJ}
\frac{\partial \mathcal{S}}{\partial \tau}=-\frac{1}{2} g^{\mu\nu} \frac{\partial \mathcal{S}}{\partial x^\mu} \frac{\partial \mathcal{S}}{\partial x^\nu}.
\end{align}
In the equation (\ref{HJ}), the symbol $\tau$ represents the affine parameter, while $\mathcal{S}$ denotes the Jacobi action, that is
\begin{align}\label{HJ2}
\mathcal{S}=\frac{1}{2}m^2 \tau - \vec{E}t + \vec{L} \varphi + \mathcal{S}_{r}(r)+ \mathcal{S}_{\theta}(\theta),
\end{align}
In which, $m$ is the rest mass. Since the  function (\ref{HJ}) does not depend on time $t$ and azimuthal angle $\varphi$, two conserved quantities, namely energy $\vec{E}$ and angular momentum $\vec{L}$, can be defined.
From equations (\ref{HJ})-(\ref{HJ2}), one can get that\cite{M:2022vyd}
\begin{align}\label{EN}
&\Delta_r \left(\frac{\partial \mathcal{S}_r}{\partial r}\right)^2+\left(\frac{\partial \mathcal{S}_\theta}{\partial \theta}\right)^2+\frac{\vec{L}^2}{\sin^2\theta}-\vec{E}^2 a^2 \sin^2\theta-\frac{1}{\Delta_r}(a^2 \vec{L}^2+\vec{E}^2(r^2+a^2) \nonumber\\
&-2a \vec{L}\vec{E}(r^2+a^2))-2a \vec{L}\vec{E}\frac{\chi}{\Delta_r}=0.
\end{align}
The expression for $\chi$ is given by $\chi=r^2+a^2-2 \rho a^2 r$, while $m^2=1,0,-1$ corresponds to time-like, null and space-like geodesic respectively. In addition, the term of $\mathcal{S}_{r}$ and $\mathcal{S}_{\theta}$ can be express as
\begin{align}\label{HR}
\Delta_r \left(\frac{\partial \mathcal{S}_r}{\partial r}\right)^2=\mathbf{R}(r)=(a \vec{L} -(a^2+r^2)\vec{E})^2-(\kappa+(a \vec{E}-\vec{L})^2)\Delta_r,
\end{align}
and
\begin{align}\label{HT}
\left(\frac{\partial \mathcal{S}_\theta}{\partial \theta}\right)^2=\Theta(\theta)=\kappa-\left(\frac{\vec{L}^2}{\sin^2 \theta}- a^2 \vec{E}^2\right)\cos \theta.
\end{align}
where $\kappa$ is a constant derived from the geometric structure of this spacetime. The equation governing geodesic motion can be expressed as
\begin{align}\label{GE1}
\Sigma \dot{t}=\frac{1}{\Delta_r}\left(\vec{E}(r^2+a^2)-a\vec{L}\right)(r^2+a^2)-a\left(a\vec{E} \sin^2 \theta-\vec{L}\right),
\end{align}
\begin{align}\label{GE2}
\Sigma \dot{\varphi}=\frac{a}{\Delta_r}  \left(\vec{E}(r^2+a^2)-a\vec{L}\right)-\frac{1}{\sin^2 \theta}(a\vec{E} \sin \theta^2-\vec{L}),
\end{align}
\begin{align}\label{GE3}
\Sigma \dot{r}=\sqrt{\mathbf{R}(r)},
\end{align}
\begin{align}\label{GE4}
\Sigma \dot{\theta}=\sqrt{\Theta(\theta)}.
\end{align}
Among them, symbols $(\cdot)$ represents the derivative of the affine parameter $\lambda$.
Therefore, the behavior of photons in close proximity to a black hole can be accurately described by equations (\ref{GE1})-(\ref{GE4}), and subsequent discussions on black hole shadows rely on these four  differential equations.

\section{ Shadow of the rotating Bardeen black holes surrounded by PFDM}
In order to obtain the shadow  cast by a rotating Bardeen black hole  surrounded by PFDM, it is necessary to examine the photon sphere of the black hole. The boundary and size of the black hole shadow is dictated by the dimensions of the photon sphere $r_p$; thus, one can derive the conditions for the photon sphere by imposing $\dot{r}=0$ and $\ddot{r}=0$, that is
\begin{align} \label{PS}
\mathbf{R}(r)=0, \qquad  \mathbf{R}'(r)=0.
\end{align}
In addition, there are two impact parameter $\xi=\vec{L}/\vec{E}$ and $\varsigma=\kappa/\vec{E}^2$  is introduced to describe the motion behavior of the photon, and the  form of $\xi$ and $\varsigma$ can be expressed as
\begin{align} \label{XI}
\xi=\frac{1}{a \Delta'_r}\left(a^2 \Delta'_r + r^2 \Delta'_r -4r \Delta_r \right),
\end{align}
and
\begin{align} \label{VS}
\varsigma=\frac{r^2}{a^2 \Delta'^2_r} \left(16 a^2 \Delta_r-16\Delta^2_r +8r\Delta_r \Delta'_r-r^2\Delta'^2_r \right).
\end{align}
The motion of photons within the photon sphere can be described by $\varsigma$ and $\xi$, with these two conserved quantities naturally determining the boundary of the black hole shadow.
In the work of\cite{M:2022vyd}, they provides a discussion on the distinctive characteristics of the  shape of black hole shadow under various conditions. Therefore, we propose employing  the backward ray-tracing technique\cite{Cunha:2015yba} to validate the findings obtained in \cite{M:2022vyd} from a numerical perspective. The crucial aspect is to determine the portion of light emitted from the background light source that can reach the observer, as well as discerning which part is being absorbed by the black hole. Due to the reversibility of light, it can be evolve that the light propagates backwards in time from the observer and then the position of each pixel in the final image can be determined by numerically solving the  null geodesic equations. Among them, the shadow image in the observer's field of view is composed of pixels connected to  light rays falling into the  black hole.

To calculate the observer's obtainable shadow of the black hole, we also establish a local basis coordinate system at the distance $r \rightarrow r_o$ from the black hole. The coordinates of the observer's local basis  $\{e_{\tilde{t}}, e_{\tilde{r}},e_{\tilde{\theta}},e_{\tilde{\varphi}}\}$ can be transformed into the  coordinate basis of black hole spacetime  $e_{\tilde{\mu}}\{\partial_t, \partial_r, \partial_\theta, \partial_\varphi\}$ through the relationship $e_{\tilde{\mu}}=e^\nu_{\tilde{\mu}}\partial_\nu$, where $e^\nu_{\tilde{\mu}}$ is the transformation matrix, that is
\begin{align} \label{TM}
g_{\mu\nu}e^\mu_{\tilde{\alpha}}e^\nu_{\tilde{\beta}}=\eta_{\tilde{\alpha}\tilde{\beta}}.
\end{align}
Here, $\eta_{\tilde{\alpha}\tilde{\beta}}$  is the Minkowski spacetime metric.  It is worth mentioning that the choice of the transformation matrix $e^\nu_{\tilde{\mu}}$ is not unique, as it allows for both spatial rotations and Lorentz transformations. We adopt the selection in Ref.\cite{Long:2020wqj}, which is
\begin{align} \label{TM}
e^\nu_{\tilde{\mu}}=\left(
\begin{array}{cccc}
\varpi & 0 &0 & \sigma \\
0 & \mathbf{A}^r &0 &0\\
0 & 0 &\mathbf{A}^\theta &0\\
0 & 0 &0 &\mathbf{A}^\varphi
\end{array}\right).
\end{align}
In particular, the reference frame of the observer at infinity in space has zero axial angular momentum, which is also called the zero angular momentum observer (ZAMO). And, the term of ($\varpi, \sigma, \mathbf{A}^r, \mathbf{A}^\theta, \mathbf{A}^\varphi $) are real coefficients, which can be expressed as
\begin{align} \label{TM2}
&\mathbf{A}^r=\frac{1}{\sqrt{g_{rr}}}, \qquad \mathbf{A}^\theta=\frac{1}{\sqrt{g_{\theta \theta}}}, \qquad \mathbf{A}^\varphi=\frac{1}{\sqrt{g_{\varphi \varphi}}}, \nonumber\\
&\varpi=\sqrt{\frac{g_{\varphi \varphi}}{g^2_{t \varphi}-g_{tt}g_{\varphi\varphi}}}, \qquad \sigma=-\frac{g_{t \varphi}}{g_{\varphi \varphi}}\sqrt{\frac{g_{\varphi \varphi}}{g^2_{t \varphi}-g_{tt}g_{\varphi\varphi}}}.
\end{align}
Through the  above transformation, one can obtain the 4-momentum of a photon  measured in the observer's local inertial coordinate system, which is  defined as
\begin{align} \label{TM3}
p^{\tilde{t}}=- p_{\tilde{t}}=- e^\nu_{\tilde{t}} p_\nu, \qquad  p^{\tilde{i}}=- p_{\tilde{i}}=- e^\nu_{\tilde{i}} p_\nu ,
\end{align}
and $\tilde{i}=(r, \theta, \varphi)$. The locally measured 4-momentum  $ p^{\tilde{\mu}}$  in the spacetime of a rotating Bardeen black hole  surrounded by PFDM can  be  rewritten as
\begin{align} \label{TM3}
p^{\tilde{t}}=\varpi \vec{E}- \sigma \vec{L}, \qquad  p^{\tilde{\varphi}}=\frac{\vec{L}}{\sqrt{g_{\varphi\varphi}}}, \qquad
p^{\tilde{\theta}}=\frac{p_\theta}{\sqrt{g_{\theta \theta}}}, \qquad  p^{\tilde{r}}=\frac{p_r}{\sqrt{g_{r r}}},
\end{align}
Where $p_\theta= g_{\theta \theta} \frac{d\theta}{d \sigma}$ and $p_r = g_{rr} \frac{d r}{d \sigma}$ are the components of momentum of the photon, respectively. Therefore, in the ZAMO framework, it enables the definition of each light ray using celestial coordinates ($X$, $Y$). 
In accordance with the convention outlined in\cite{Hu:2020usx}, the relationship between  4-momentum of photon $p_{\tilde{\mu}}$ and celestial coordinates $(X, Y)$ is established as
\begin{align} \label{TM3}
\cos X=\frac{p^{\tilde{r}}}{p^{\tilde{t}}}, \qquad \tan Y= \frac{p^{\tilde{\varphi}}}{p^{\tilde{\theta}}}.
\end{align}
The acquisition of an image of a black hole necessitates the mapping of celestial coordinates $(X, Y)$ onto  the observation plane $(x, y)$. On the observation plane, a standard Cartesian coordinate system can be established, and the relationship between the observation plane and the celestial coordinate  follows
\begin{align} \label{TM3}
x=-2\tan\frac{X}{2}\sin Y, \qquad y= -2 \tan \frac{X}{2}\cos Y.
\end{align}
In essence, it determines the initial momentum value of the photon at the ZAMO as well as its initial position. In this way, the shadow image of rotating Bardeen black holes surrounded by PFDM can be obtained in celestial coordinates using the  backward ray-tracing technique.

\begin{figure}
    \centering
    \subfigure[$a=0.001$]{
        \includegraphics[width=2in]{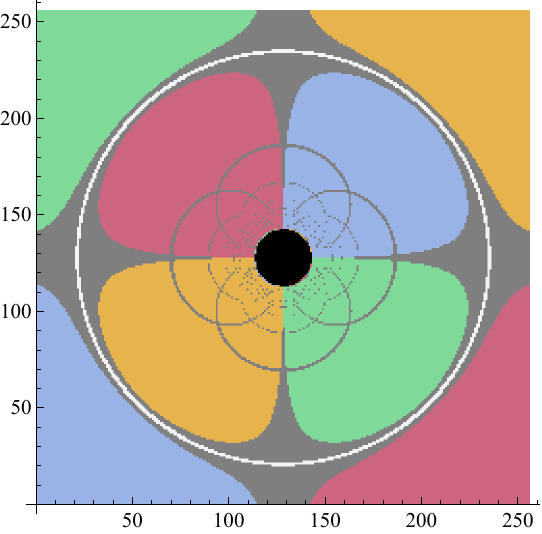}
    }
	\subfigure[$a=0.5$]{
        \includegraphics[width=2in]{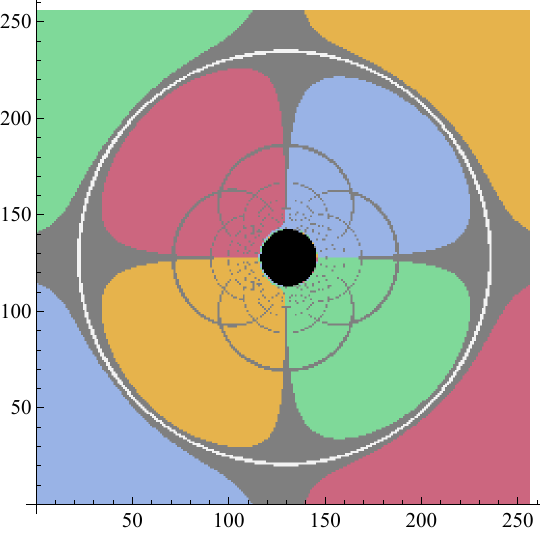}
    }
    \subfigure[$a=0.95$]{
        \includegraphics[width=2in]{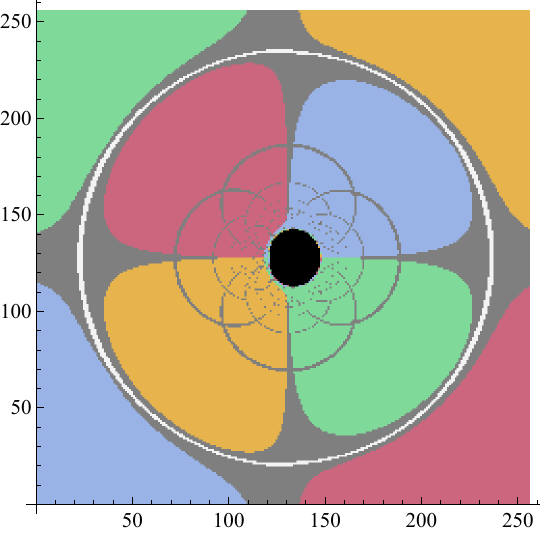}
    }
   \caption{The shadow of the rotating Bardeen black hole surrounded by PFDM is observed for different values of the spin parameter $a$, while keeping other relevant parameters fixed at $\mathcal{G}=0.3$, $\alpha=-0.5$ and $\theta_o = 90^{\circ} $.}\label{figbl1}
\end{figure}

\begin{figure}
    \centering
    \subfigure[$\mathcal{G}=0.01$]{
        \includegraphics[width=2in]{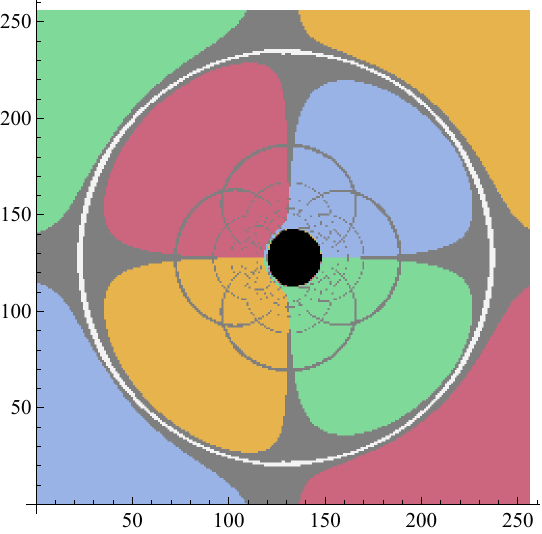}
    }
	\subfigure[$\mathcal{G}=0.1$]{
        \includegraphics[width=2in]{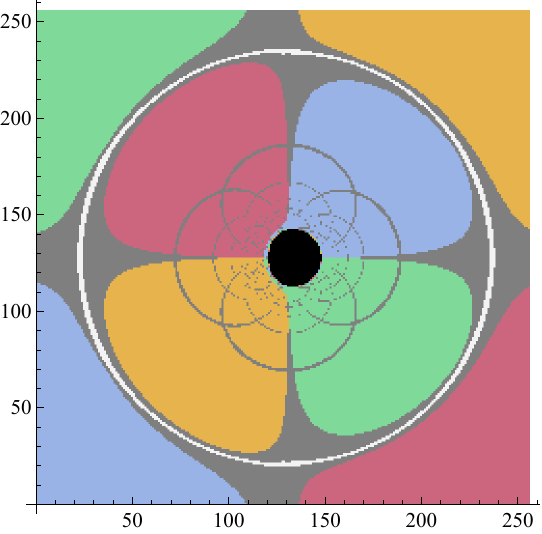}
    }
    \subfigure[$\mathcal{G}=0.2$]{
        \includegraphics[width=2in]{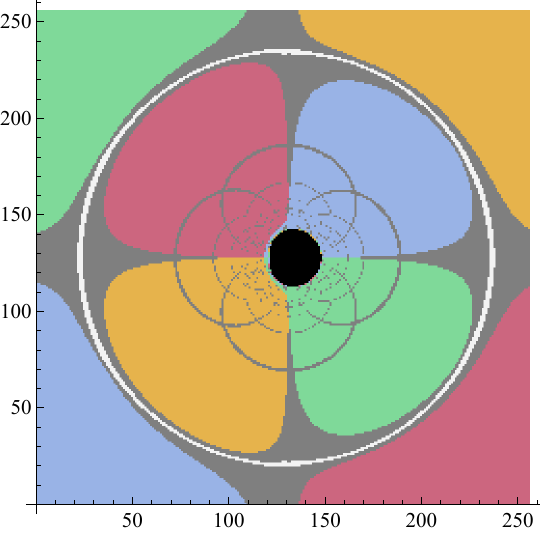}
    }
   \caption{The shadow of the rotating Bardeen black hole surrounded by PFDM is observed for different values of the magnetic charge $\mathcal{G}$, while keeping other relevant parameters fixed at $a=0.99$, $\alpha=-0.5$ and $\theta_o = 90^{\circ} $.}\label{figbl2}
\end{figure}

\begin{figure}
    \centering
    \subfigure[$\alpha=-0.9$]{
        \includegraphics[width=2in]{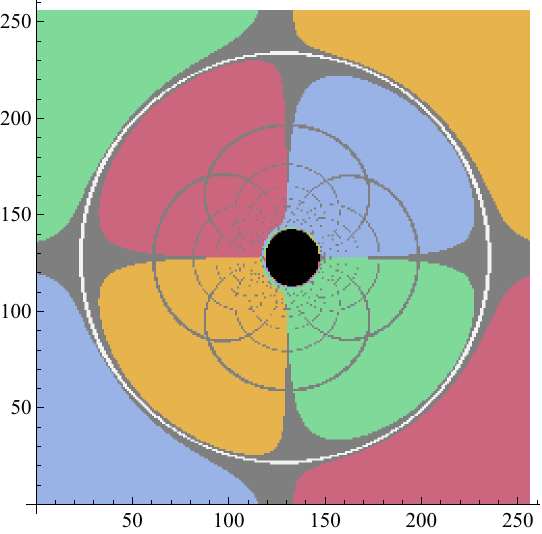}
    }
	\subfigure[$\alpha=-0.5$]{
        \includegraphics[width=2in]{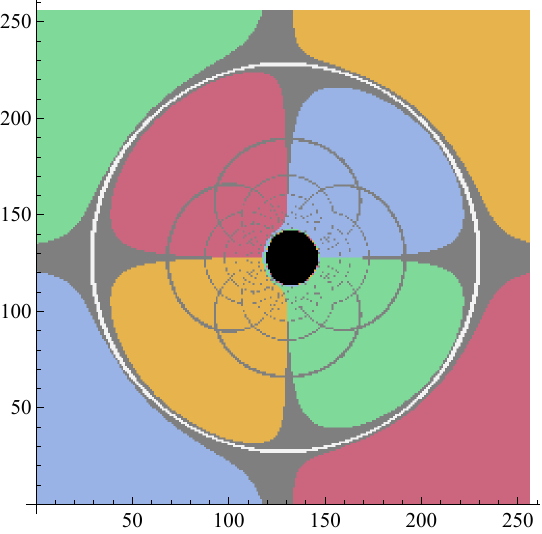}
    }
    \subfigure[$\alpha=-0.01$]{
        \includegraphics[width=2in]{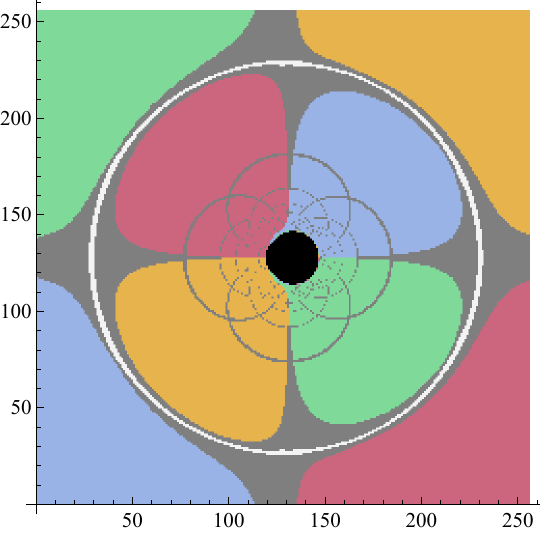}
    }
   \caption{The shadow of the rotating Bardeen black hole surrounded by PFDM is observed for different values of the dark matter parameter $\alpha$, while keeping other relevant parameters fixed at $a=0.99$, $\mathcal{G}=0.3$ and $\theta_o = 90^{\circ}$.}\label{figbl3}
\end{figure}

In order to enhance the demonstrative impact of curved spacetime on light ray propagation, we partition the spherical background into four quadrants and designate them with distinct colors (red, blue, yellow, and green), while using brown markers at 10 intervals along the spherical latitude and longitude lines. In addition, the black hole is situated precisely at the center of the sphere, while the observer is positioned at an intersection point where all four quadrants meet. Similarly to the approach employed in previous studies, we designate the other intersection point with the observer, located in the opposite quadrant, as white color to serve as a reference light source for studying the strong gravitational lensing effects of Einstein rings. Considering the impact of relevant parameters on the spatiotemporal structure, we examined the influence of spin parameter $a$, magnetic charge $\mathcal{G}$ and dark matter parameter $\alpha$ on the black hole shadow, as depicted in Fig.\ref{figbl1}-Fig.\ref{figbl3}. The images consistently exhibit a central dark region, encompassed by a vibrant background light source, and accompanied by a white ring demonstration of the Einstein ring. Clearly, these images show the warping of space by a black hole and the gravitational lensing effect of a black hole.

The shadows of a rotating Bardeen black hole surrounded by PFDM  obtained by varying the spin parameter $a$ are depicted in Fig.\ref{figbl1}, while keeping other relevant parameters fixed at the dark matter parameter $a=-0.5$, magnetic charge $\mathcal{G}=0.3$ and observation angle $\theta_{o}=\pi/2$ (the equatorial plane). The increase in the spin parameter $a$ is observed to cause deformation in the shape of the black hole shadow. For small values of $a$, the shadow maintains a disk-like shape, while for large values of $a$, it evolves into a D-shaped form, deviating from its perfect disk shape which is similar to the change in the shadow of  the usual rotating (Kerr) black hole. The size of the shadow on the vertical axis remains relatively constant, while its position gradually shifts towards the right along the horizontal axis as parameter $a$ increases. In addition, the increase in parameter a also causes the color image near the shadow to distort, demonstrating the space drag effect caused by the rotation in the spacetime of Bardeen black hole surrounded by PFDM.

The correlation effect of a change in the magnetic charge $\mathcal{G}$ on the black hole shadow is depicted in Fig.\ref{figbl2},  where the  parameters $a=0.99$, $\alpha=-0.5$ and $\theta_o=\pi/2$ are considered. Similarly, for a significantly large spin parameter $a$, irrespective of the magnetic charge value $\mathcal{G}$, the black hole shadow will undergo deformation to a certain extent and exhibit noticeable drag effects in its vicinity. However, as the magnetic charge value increases, there is a significant increase in the degree of deformation of the shadow, resulting in a shape that closely resembles a D-shape when  $\mathcal{G}=0.2$. Meanwhile, the fluctuation of magnetic charge $\mathcal{G}$ does not exert any discernible impact on the dimensions or location of the shadow. Subsequently, the impact of the dark matter parameter $\alpha$ on the  the black hole shadow  is  investigated, see Fig.\ref{figbl3}. In the case of other parameters is set as $a=0.99$ and $\mathcal{G}=0.3$, the result show that the influence of the change of  dark matter parameter $\alpha$ on the shadow is mainly reflected in the size and deformation degree. Specifically, the deformation degree of the shadow decreases with the increase of the absolute value of the dark matter parameter $\alpha$, but its size increases significantly. The size of the shadow cast by the black hole is significantly larger in the case of $\alpha=-0.9$ compared to $\alpha=-0.01$, and it also appears as a slightly distorted circular shape without excessive deformation. Hence, the impact of the dark matter parameter $a$ on the black hole shadow is inversely proportional to the influence of the magnetic charge $\mathcal{G}$.

\begin{figure}[h]
\centering 
\subfigure[$\theta_o = 0^{\circ}$]{\includegraphics[scale=0.55]{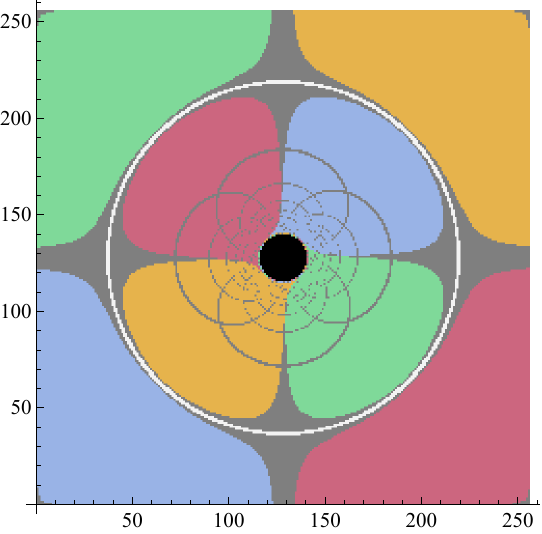}}
\subfigure[$\theta_o = 17^{\circ}$]{\includegraphics[scale=0.55]{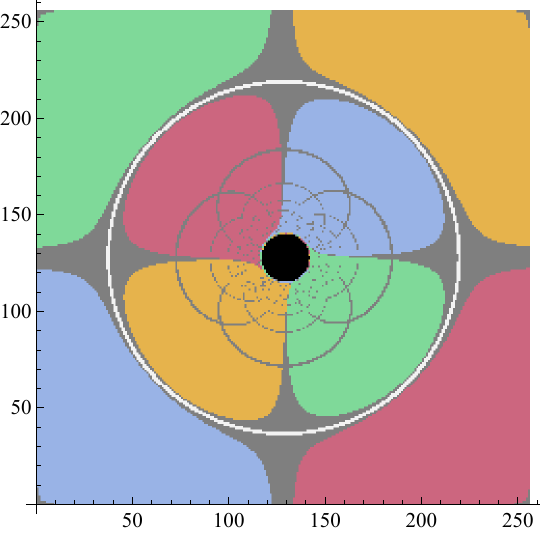}}
\subfigure[$\theta_o = 30^{\circ}$]{\includegraphics[scale=0.55]{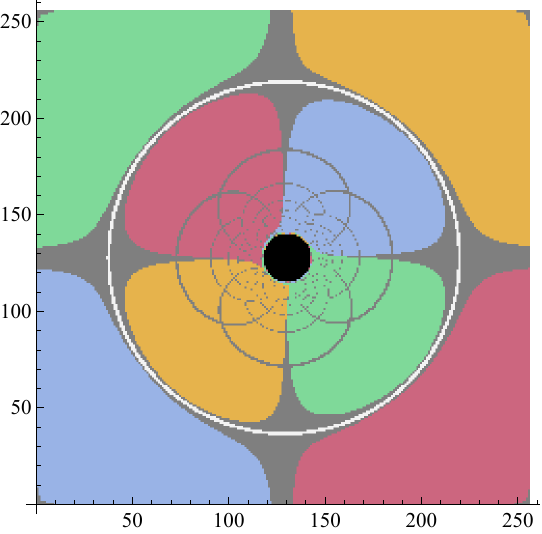}}
\subfigure[$\theta_o = 60^{\circ}$]{\includegraphics[scale=0.55]{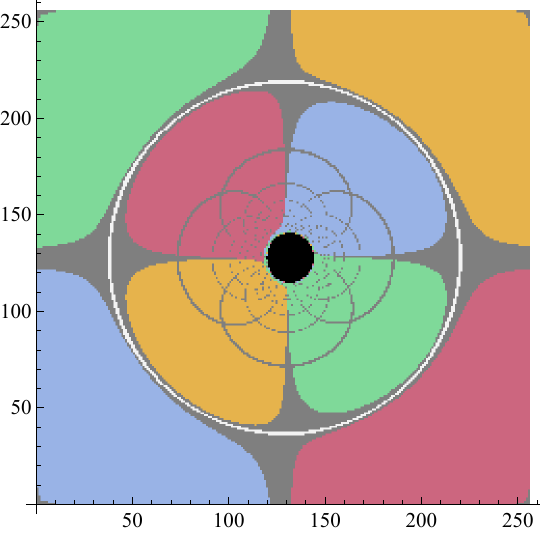}}
\subfigure[$\theta_o = 75^{\circ}$]{\includegraphics[scale=0.55]{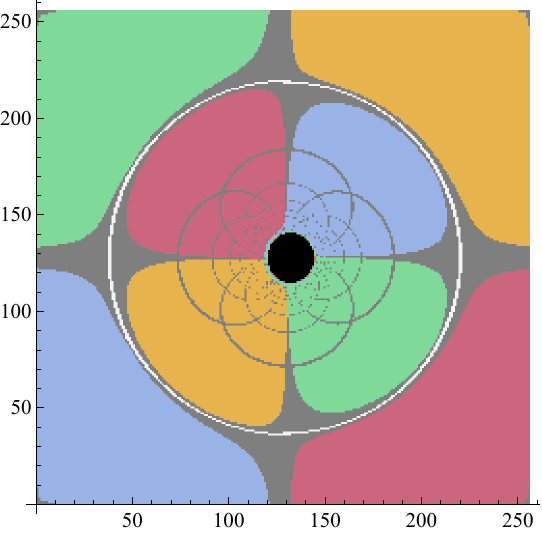}}
\subfigure[$\theta_o = 90^{\circ}$]{\includegraphics[scale=0.55]{figFG5.pdf}}
\caption{\label{figbl4} The shadow of the rotating Bardeen black hole surrounded by PFDM is observed for different values of the observation inclination angles $\theta_o$, while keeping other relevant parameters fixed at $a=0.99$, $\mathcal{G}=0.3$ and $\alpha=-0.5$. }
\end{figure}

In order to gain a deeper understanding of the geometric structure of spacetime, we also investigate the apparent characteristics of the black hole shadow from various observation inclination angles ($\theta_o=0^{\circ}, 17^{\circ}, 30^{\circ}, 60^{\circ}, 75^{\circ}, 90^{\circ}$), while keeping other relevant parameters fixed at $a=0.99$, $\mathcal{G}=0.3$ and $\alpha=-0.5$, as shown in Fig. \ref{figbl4}.  When the observation inclination is  set to $\theta_o=0$, i.e., the observer's position aligns with the rotation axis of the black hole, the contour shape of the black hole shadow exhibits a perfect circular form,  and the space drag effect caused by the rotation of the black hole is also obviously displayed, which is in line with the expected result. As the angle of observation increases, the shadow shape of the black hole gradually deviates from a perfect circle and ultimately manifests as a D-shape. Additionally, there is a significant increase in the size of the shadow along the vertical axis. Consequently, an observer situated on the equatorial plane would perceive larger and more distorted black hole shadows.

\section{Optical appearance of the black hole within a thin accretion disk model}
In the preceding chapter, employing the model of a spherical background light source, we conducted numerical simulations to investigate the shadow characteristics of the black hole, encompassing its apparent shape and size, and the obtained results align with those reported in\cite{M:2022vyd}. In the real universe, black holes are typically enveloped by a substantial amount of matter that undergoes gravitational acceleration and emits high-energy radiation, thereby giving rise to luminous accretion disks encircling the black hole. In this section, we will further explore the significance of bright accretion disks as the primary source of background light, given their crucial role in black hole imaging. In addition, we continue to employ  the backward ray-tracing technique , while the accretion  model is regarded as  an optically thin, geometrically thin accretion disk.
\subsection{The configuration of the thin accretion disk model}
It is worth noting that there are several specific aspects that we must account for in this accretion model. Firstly, the accretion disk can be regarded as a free electrically neutral plasma in motion along an equatorial timelike geodesic, due to its placement on the equatorial plane and its geometrically thin nature. Furthermore, the width of the accretion disk extends beyond the inner stable circular orbit (ISCO), while its inner radius reaches towards the  black hole horizon. In astrophysics, the viscosity of the accretion disk induces outward transfer of angular momentum by the accretion material as it orbits around the black hole, leading to gradual migration of particles in Keplerian region towards the ISCO. Subsequently, these particles undergo acceleration and spiral towards the event horizon until they reach the black hole, and the particle motion mechanism during this process has been verified to be in accordance with astrophysical observations\cite{Chael:2021rjo}. Hence, the scheme proposed in\cite{Hou:2022eev} can be employed to elucidate the particle motion behavior within the accretion disk. Specifically,  the ISCO represents a demarcation line, beyond which the particles of the accretion disk will maintain a stable circular orbits, while within this boundary they will undergo a critical plunging orbits. In other words, the radial motion of the particles in the accretion disk within the range $r \geq r_{i}$  should satisfy
\begin{align} \label{RM1}
V_{eff}=\partial_r V_{eff}=0.
\end{align}
And, within the range $r < r_{i}$, the radial motion $\mathcal{U}_r$ is governed by
\begin{align} \label{RM2}
\mathcal{U}_r=-\sqrt{-\frac{V_{eff}}{g_{rr}}},
\end{align}
where the minus sign preceding the square root denotes inward motions, while the term $V_{eff}=\frac{\vec{E}^2-1}{2}-\frac{\dot{r}^2}{2}$ signifies the effective potential. When retracing the trajectory of light rays, multiple intersections with the accretion disk in the equatorial plane may occur, and each intersection point may vary radially. The radius at which light intersects with the equatorial plane for the $n^{th}$ time is denoted as $r_n(x,y)\mid_{n=1,2,3...N_m(x, y)}$, a radiative transfer function governed by the Boltzmann equation for photons, where $N(x, y)$ represents the maximum number of intersection points.
In fact, the function $r_n(x,y)$ generates the shape of the n$^{th}$ image of the disk. Specifically, the direct image corresponds to the case of $n=1$, while the lens image produced on the screen corresponds to $n=2$. The radiative transfer function is closely dependent on the observation angle $\theta_o$ and the algorithm employed for image generation. And, each pixel in the image corresponds to a wave vector.

The intensity of light ray undergoes variations at each interaction with the accretion disk in a complete light path, originating from the accretion disk and reaching the screen within the ZAMO frame, due to emission and absorption effects. The refraction effect of the disk medium will be disregarded for the sake of convenience, thereby enabling the change in light intensity to be denoted as\cite{Lindquist:1966igj}
\begin{align} \label{II1}
\frac{d}{d\tau}\left(\frac{\mathcal{I}_\nu}{\nu^3}\right)=\frac{1}{\nu^2}(\mathcal{J}_\nu-\kappa_\nu \mathcal{I}_\nu).
\end{align}
In this context, the parameter $\tau$ represents the affine parameter of null geodesics, while $\mathcal{I}_\nu$ (erg cm$^{-2}$ s$^{-1}$ sr$^{-1}$ Hz$^{-1}$), $\mathcal{J}_\nu$ (erg cm$^{-3}$ s$^{-1}$ sr$^{-1}$ Hz$^{-1}$), and $\kappa_\nu$ (erg cm$^{-1}$ )represent the specific intensity, emissivity, and absorption coefficient at frequency $\nu$, respectively. When the beams of light propagate in different directions at a single point in space, both $\mathcal{J}_\nu$ and $\kappa_\nu$ tend towards zero, resulting in the conservation of the quantities $\mathcal{I}_\nu /{\nu^3}$ along the geodesic. The accretion disk should exhibit stable, axisymmetric properties and possess $Z_2$ symmetry at the equatorial plane. The emissivity and absorption coefficient of the accretion disk remain constant as light passes through it, owing to its geometrically thin. The integration of equation (\ref{II1}) along the path of light yields an intensity formula for each position on the observer screen, that is
\begin{align} \label{II2}
\mathcal{I}_{\nu_{o}}=\sum^{N_m}_{n=1}\left(\frac{\nu_o}{\nu_n}\right)^3 \frac{\mathcal{J}_n}{\varrho_{n-1}} \left(\frac{1-e^{-\kappa_n \mathbf{F_n}}}{\mathbf{F_n}}\right).
\end{align}
The equation above defines $\nu_{o}$ as the observed frequency  by the observer, and $\nu_{n}$ as the frequency observed in the stationary coordinate system moving relative to the radiation profile.  Meanwhile, one can designate the class of frames as $\{{\mathfrak{F}_n}$\}, where $n = 1...N_m$ represents the number of times the ray intersects the equatorial plane. Additionally, we employ subscript $n$ to indicate the corresponding measurements in their respective local rest frames ${\mathfrak{F}_n}$. The parameter $\varrho_n$ represents the optical depth of photons, and its value varies accordingly upon emission from a designated position denoted as $\mathrm{N}$, that is
\begin{align} \label{OD}
\varrho_n=\begin{cases} \exp \left(\Sigma^{\mathrm{N}}_{n=1} \kappa_n \mathbf{F}_n\right)
    , &  \mathrm{N}\geq 1,  \\
    1,  & \mathrm{N}=0.
    \end{cases}
\end{align}
Here, the term $\mathbf{F}_n=\nu_n \Delta \lambda_n $ is commonly known as the fudge factor that governs the luminosity of the higher-order photon ring, necessitating further elucidation contingent upon the specific model of the accretion disk. In the context of fudge factor, $\Delta \lambda_n$ represents the variation in the affine parameter as a specific ray traverses through the accretion disk medium at ${\mathfrak{F}_n}$. Given the optically thin properties of the accretion disk, one can disregard any absorption effects. Consequently, equation (\ref{II2})  is simplified as\cite{Chael:2021rjo,Hadar:2020fda}
\begin{align} \label{II3}
\mathcal{I}_{\nu_{o}}=\sum^{N_m}_{n=1} \mathbf{F}_n g^3_n (r_n, x, y) \mathcal{J}_n,
\end{align}
where $r_n$ represents the radius at which the ray intersects the equatorial plane for the $n^{th}$ time. In addition, the term $g^3_n(r_n,x,y)$ represents the redshift factor, which is defined as the ratio of the observed frequency to the emission frequency at radius $r_n$, that is
\begin{align} \label{RF}
g_n=\frac{\nu_{o}}{\nu_n},
\end{align}
The factor $g_n^3$ applies to the intensity of a specific frequency, while  $g_n^4$ is suitable for integrated intensity\cite{Wang:2023fge}.  Since the behavior of particles in the accretion disk can be classified into two distinct categories based on the ISCO, the redshift factor of the accretion disk within and beyond the ISCO exhibits discernible different. The particles in the accretion disk outside the ISCO move along a circular  orbits with an angular velocity $\mathbf{W}$, and the redshift factor can be expressed as
\begin{align} \label{RF2}
g_{circular}=\frac{e}{(1-\mathbf{W}_n b)\upsilon}, \qquad r_n\geq r_i,
\end{align}
where
\begin{align} \label{RFw}
\mathbf{W}_n(r)=\frac{\mathcal{U}^\varphi}{\mathcal{U}^t}\mid _{r=r_n}.
\end{align}
The equation above defines $b$ as the impact parameter, i.e., the ratio of the energy of the photon along null geodesic wire to the angular momentum, $\upsilon$ as the angular velocity function, and $e$ as the ratio of observed energy on the screen to conserved energy  along  null geodesic. These components are expressed as
\begin{align} \label{RFw2}
b=\frac{\vec{L}_n}{\vec{E}_n},  \qquad e=\frac{\vec{E}_n}{\vec{E}^o_n}=\varpi+b\sigma, \qquad \upsilon=\sqrt{\frac{-1}{g_{tt}+2g_{t \varphi}\mathbf{W}_n+g_{\varphi \varphi}\mathbf{W}_n^2}}\mid_{r=r_n}.
\end{align}
In the case of $r_o\rightarrow \infty$,  the value of $e$ is $e=0$ for the asymptotically flat spacetimes. In addition, in the region within ISCO, the particles are moving along the critical plunge orbit,  and its redshift is expressed in the form
\begin{align} \label{RFP}
g_{plunge}=-\frac{e}{\mathcal{U}_r k_r /\vec{E}_n +\vec{E}_I(g^{tt}-b g^{t \varphi})+\vec{L}_I(b g^{\varphi \varphi}-g^{t \varphi})}\mid_{r=r_n}, \qquad r_n < r_i,
\end{align}
where the terms $\vec{E}_I$and $\vec{L}_I$ are the energy and angular momentum of the particle on ISCO, respectively. Given that the observed wavelength (1.3 mm) is consistent with the image of M87 and Sgr A* at 230 GHz, we adhere to the definition of emissivity as stated in\cite{Hou:2022eev}, which adopts a  second-order polynomial in log-space as
\begin{align} \label{EMI}
\log[\mathcal{J}(r)]=(\eta_1 \varepsilon^2+ \eta_2 \varepsilon).
\end{align}
In which,  the component $\varepsilon$ is $\varepsilon=\log(r/r_h)$. The values of $\eta_1$ and $\eta_2$ are set to $\eta_1=-1/2$ and $\eta_2=-2$, respectively, in order to achieve a more visually appealing effect that aligns with the 230 GHz image. The fudge factor $\mathbf{F}_n$ has various options in\cite{Chael:2021rjo}, whereas in this paper, one can normalize all the fudge factor to $1$, i.e., $\mathbf{F}_n \rightarrow 1 $\cite{Hou:2022eev}. The primary focus of our study lies in the impact of the correlation parameteron the emission profile, rendering the optical manifestation of the photon ring inconsequential due to variations in $\mathbf{F}_n$ value, thereby exerting a minimal influence on the overall image.

\subsection{Observational appearance of the black hole}
Within the framework of the accretion thin disk model, one can explicitly simulate the visual representation of a rotating Bardeen black hole enveloped by PFDM on the display using equation (\ref{II3}). The subsequent simulation was performed using a constant observation distance of $r_o=100M$, and the outer radius of the accretion disk was defined as $r_{d1}=20M$, while its extension towards the event horizon was achieved by setting the inner radius as $r_{d2}=r_h$. In the spacetime of (\ref{metric3}), the geometry structure  is evidently influenced by various parameters, including the rotation parameter $a$, dark matter parameter $\alpha$, and magnetic charge $\mathcal{G}$.  Therefore, we conducted a specific investigation on the impact of these parameters on the imaging of a rotating Bardeen black hole surrounded by PFDM at various observation angles  as $\theta_o=0^{\circ}, 17^{\circ}, 60^{\circ}$ and $90^{\circ}$.

\begin{figure}[!t]
\centering 
\subfigure[$a=0$, $\theta_o = 0^{\circ}$]{\includegraphics[scale=0.35]{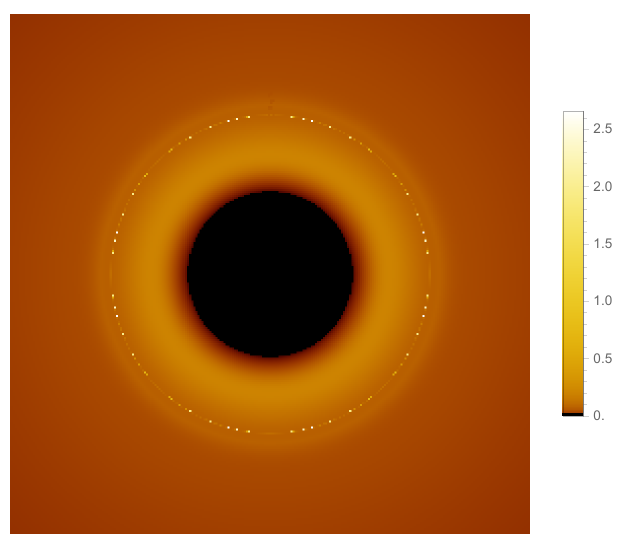}}
\subfigure[$a=0.1$, $\theta_o = 0^{\circ}$]{\includegraphics[scale=0.35]{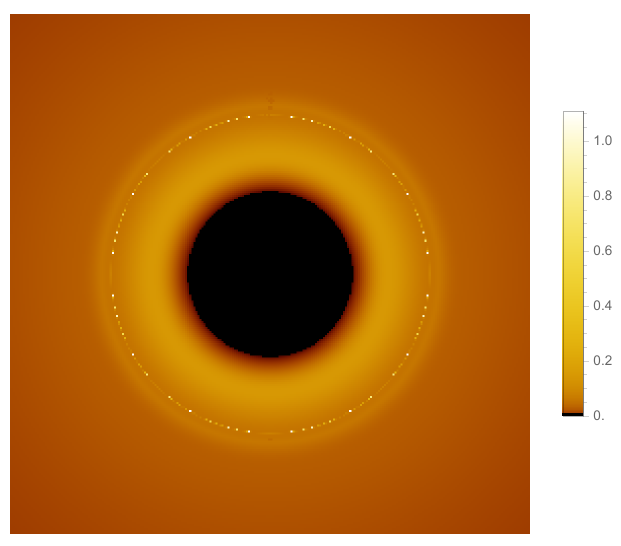}}
\subfigure[$a=0.5$, $\theta_o = 0^{\circ}$]{\includegraphics[scale=0.35]{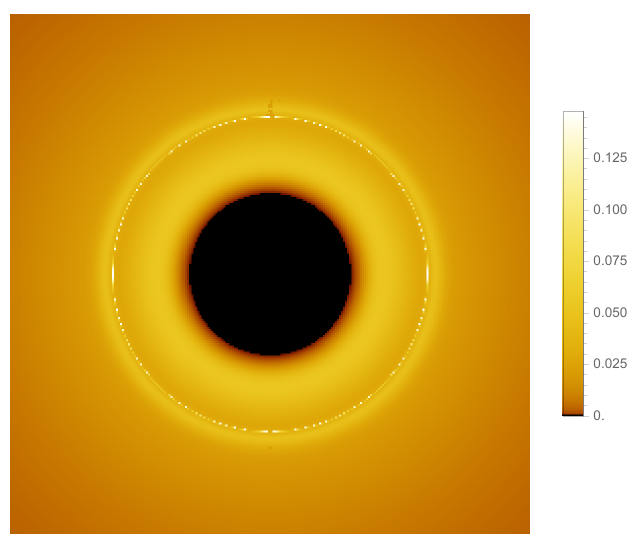}}
\subfigure[$a=0.99$, $\theta_o = 0^{\circ}$]{\includegraphics[scale=0.35]{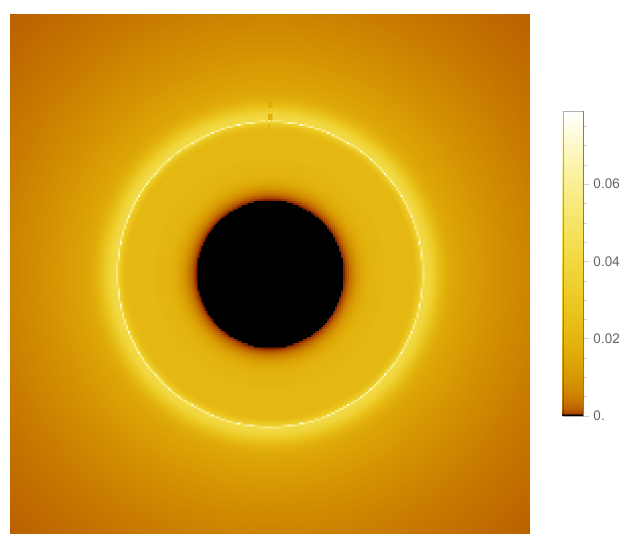}}
\subfigure[$a=0$, $\theta_o = 17^{\circ}$]{\includegraphics[scale=0.35]{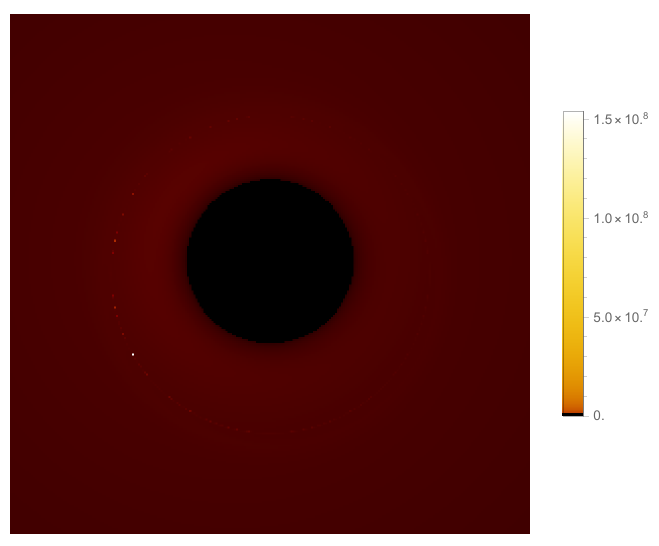}}
\subfigure[$a=0.1$, $\theta_o = 17^{\circ}$]{\includegraphics[scale=0.35]{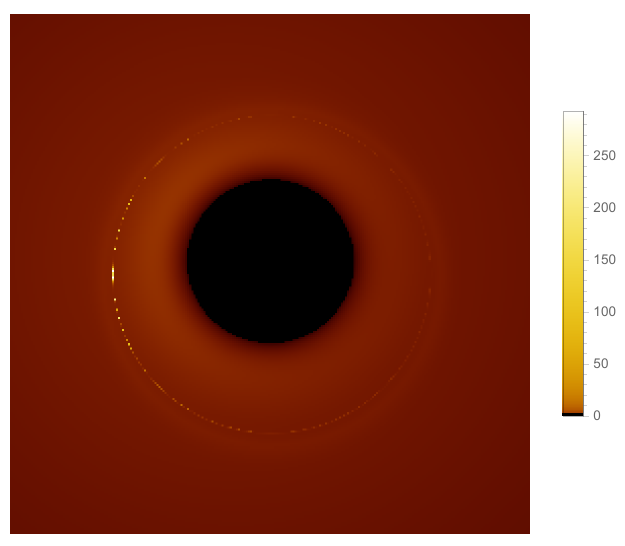}}
\subfigure[$a=0.5$, $\theta_o = 17^{\circ}$]{\includegraphics[scale=0.35]{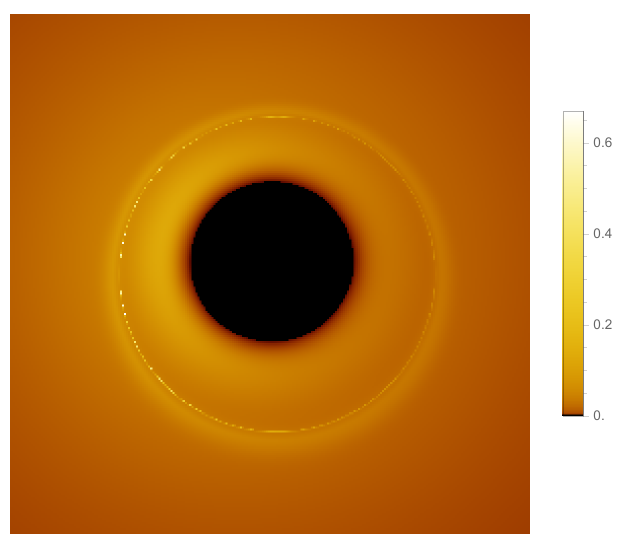}}
\subfigure[$a=0.99$, $\theta_o = 17^{\circ}$]{\includegraphics[scale=0.35]{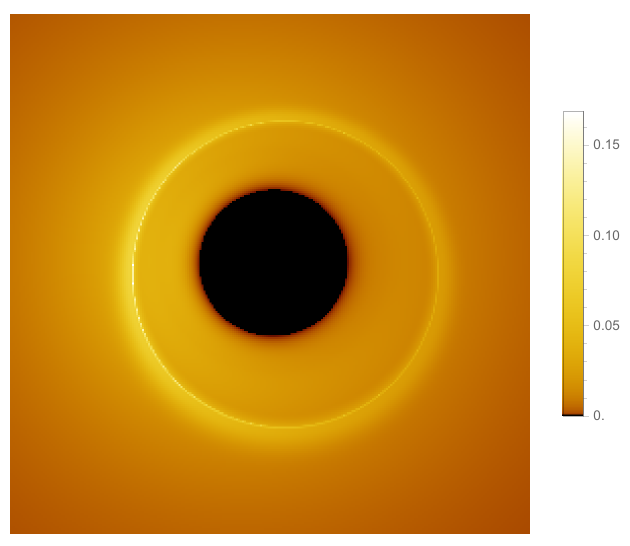}}
\subfigure[$a=0$, $\theta_o = 60^{\circ}$]{\includegraphics[scale=0.35]{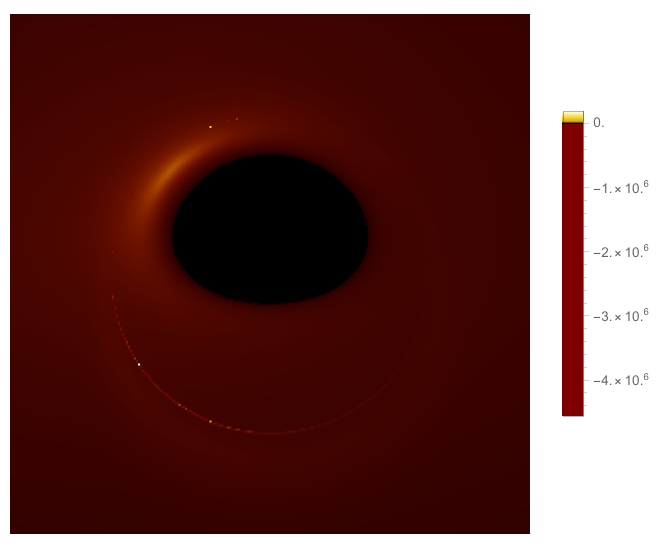}}
\subfigure[$a=0.1$, $\theta_o = 60^{\circ}$]{\includegraphics[scale=0.35]{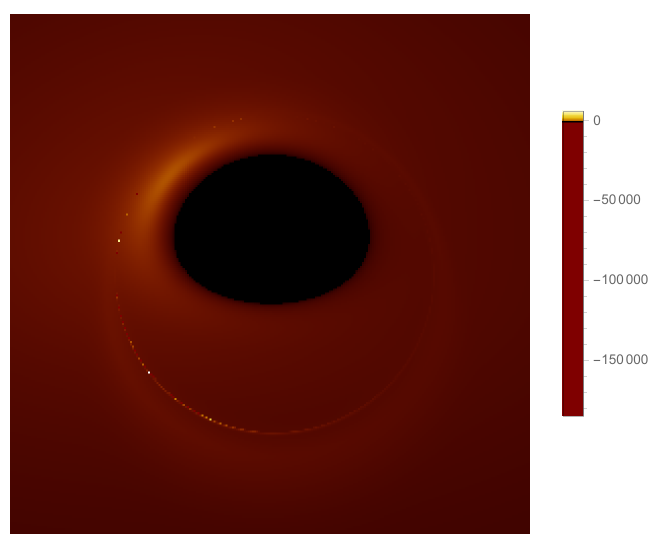}}
\subfigure[$a=0.5$, $\theta_o =60^{\circ}$]{\includegraphics[scale=0.35]{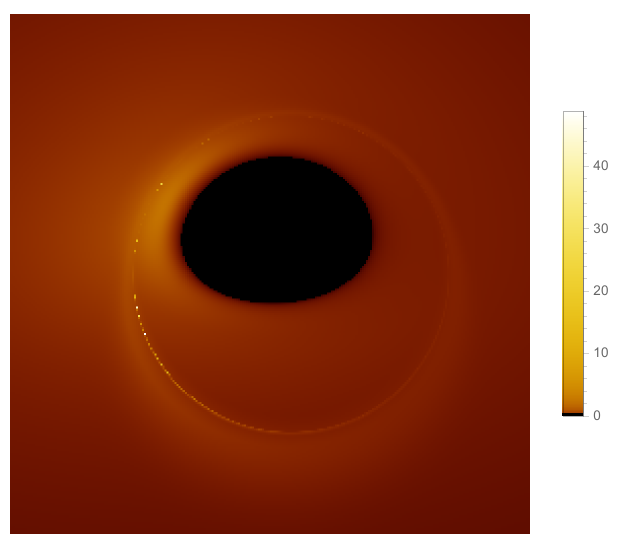}}
\subfigure[$a=0.99$, $\theta_o = 60^{\circ}$]{\includegraphics[scale=0.35]{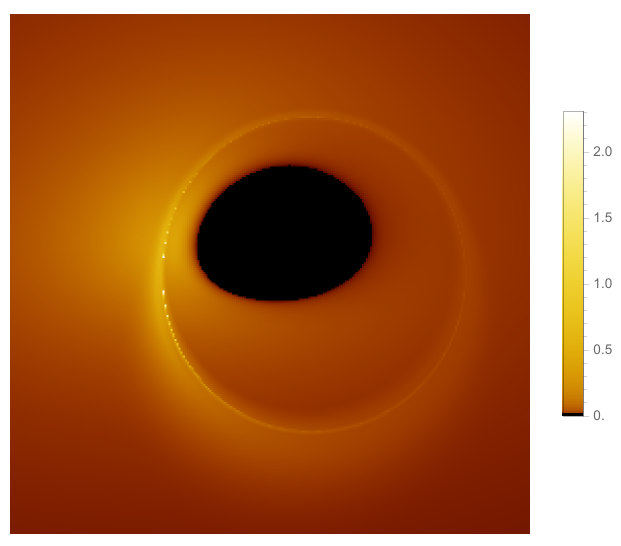}}
\subfigure[$a=0$, $\theta_o = 83^{\circ}$]{\includegraphics[scale=0.35]{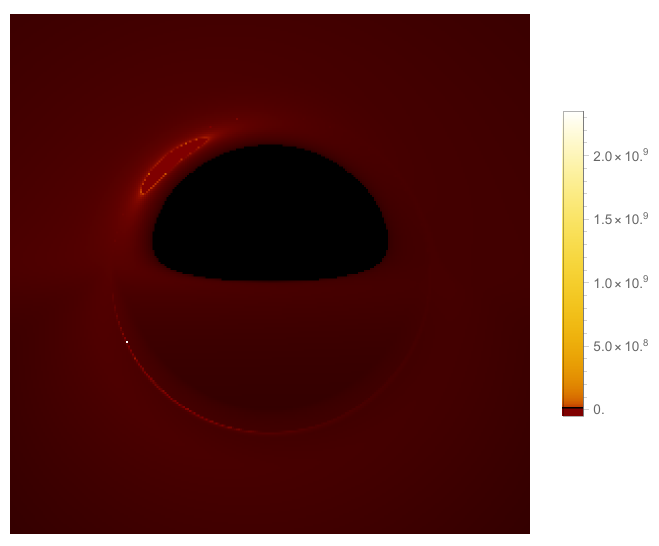}}
\subfigure[$a=0.1$, $\theta_o = 83^{\circ}$]{\includegraphics[scale=0.35]{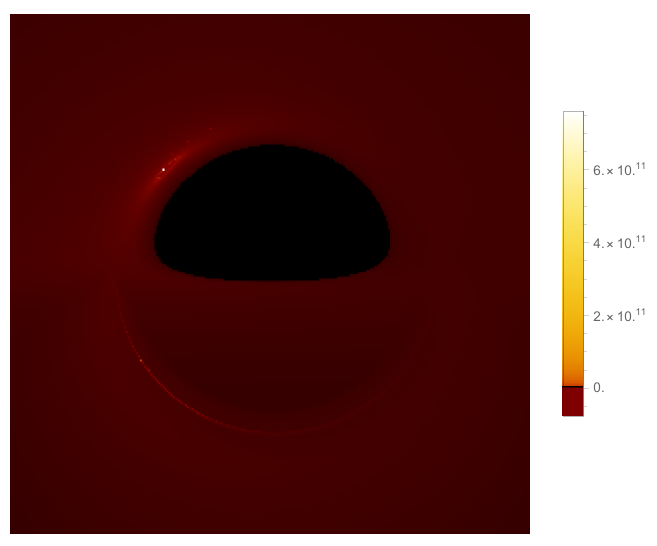}}
\subfigure[$a=0.5$, $\theta_o = 83^{\circ}$]{\includegraphics[scale=0.35]{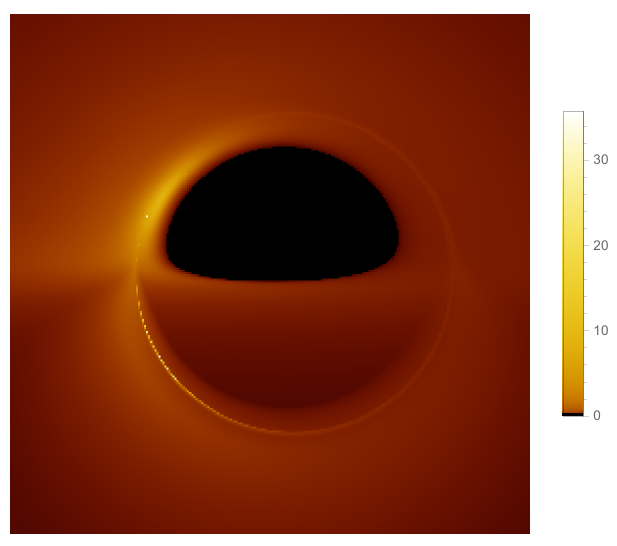}}
\subfigure[$a=0.99$, $\theta_o = 83^{\circ}$]{\includegraphics[scale=0.35]{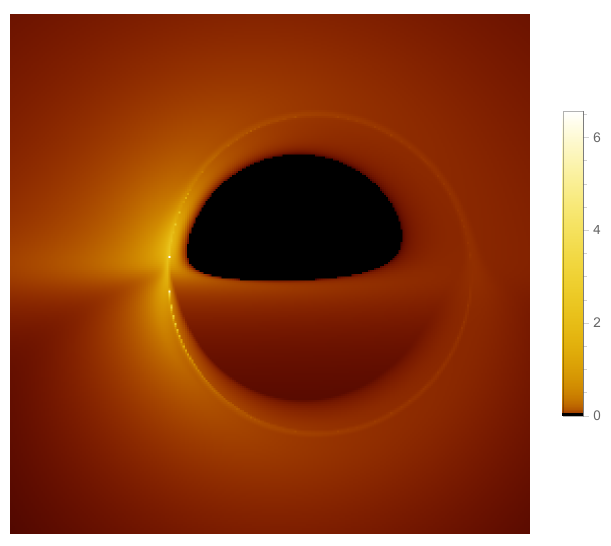}}
\caption{\label{figbpA1} In the thin accretion disk model, the images of a rotating Bardeen black holes surrounded by PFDM at 230 GHz across different parameter spaces. The observation angles in the first to fourth rows are $\theta_o = 0^{\circ}, 17^{\circ}, 60^{\circ}$ and $83^{\circ}$. And, the rotation parameters in the first to fourth columns are $\alpha=0, 0.1, 0.5$ and $0.99$, respectively. In which, the other relevant parameters are fixed as $\mathcal{G}=0.1$ and $\alpha=-05$.}
\end{figure}

The black hole image with the different value of rotation parameter is depicted in Fig.\ref{figbpA1}, where the values of $a$ are taken as $a=0, 0.1, 0.5$, and $0.99$, while keeping the other parameters constant at $\mathcal{G}=0.1$ and $\alpha=-0.5$. In general, one can always observe a dark area, with its periphery invariably encompassed by a luminous ring. The central dark region  corresponds to the accretion disk image at $r = r_h$, commonly referred to as the inner shadow according to \cite{Hou:2022eev}, while this distinctive luminous ring, known as the photon ring, is closely aligned with the critical curve of black hole. When the observed inclination is $\theta_o = 0^{\circ}$ (the first row), the dark region in the center of the image appears as an axisymmetric circle and is arranged in concentric circles with the photon ring. This symmetry arises because, when the line of sight is perpendicular to the disk, the motion of the particles in the accretion disk has no component in the line of sight, resulting in a lack of a Doppler component in the redshift factor, including only gravitational redshift. When the observed inclination is $\theta_o = 17^{\circ}$ (the second row), as the rotation parameter $a$ increases, the central dark region gradually undergoes deformation, albeit with subtle changes. For a larger value of parameter $a=0.99$, the central dark region no longer exhibits concentric circle symmetry characteristic with the photon ring.  By adjusting the observed inclination $\theta_o$ to a larger value, such as $\theta_o = 60^{\circ}$ (the third row) and $\theta_o = 90^{\circ}$(the fourth row), the deformation in the central dark region can be clearly identified, and an increase in the parameter $a$ will result in a higher degree of deformation. The left side of the screen also displays a distinctively luminous crescent or eyebrow-shaped region, which exhibits a slight increase in both brightness and size as parameter $a$ is incremented. This characteristic becomes particularly prominent when $\theta_o = 90^{\circ}$, and  it can be attributed to the amplification of the Doppler effect with an increasing inclination angle $\theta_o$. It should be emphasized that when the observed inclination and parameter $a$ change, the position of the photon ring always remain the same, but its brightness will change slightly.

\begin{figure}[!t]
\centering 
\subfigure[$a=0$, $\theta_o = 0^{\circ}$]{\includegraphics[scale=0.375]{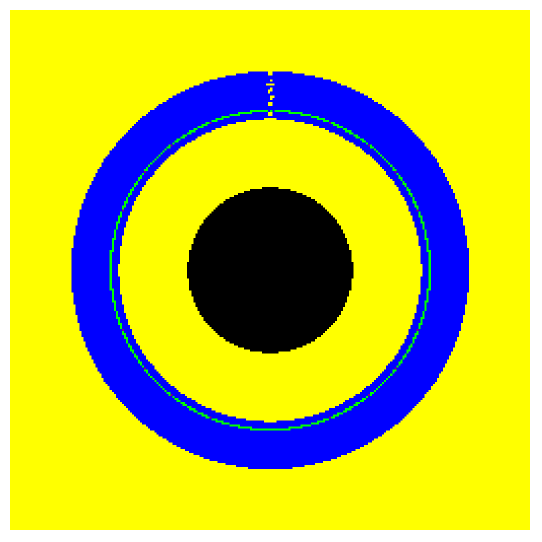}}
\subfigure[$a=0.1$, $\theta_o = 0^{\circ}$]{\includegraphics[scale=0.375]{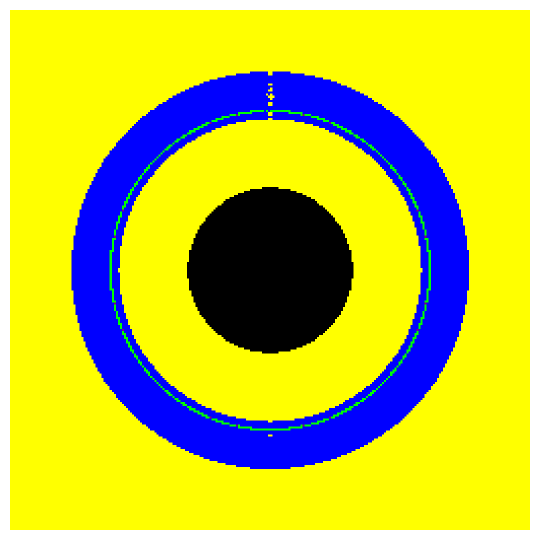}}
\subfigure[$a=0.5$, $\theta_o = 0^{\circ}$]{\includegraphics[scale=0.375]{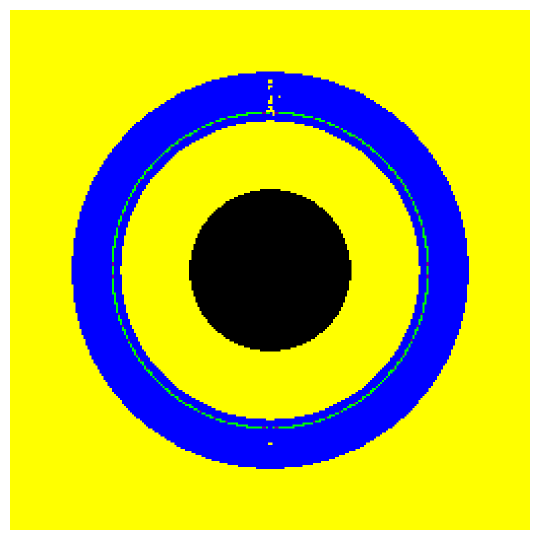}}
\subfigure[$a=0.99$, $\theta_o = 0^{\circ}$]{\includegraphics[scale=0.375]{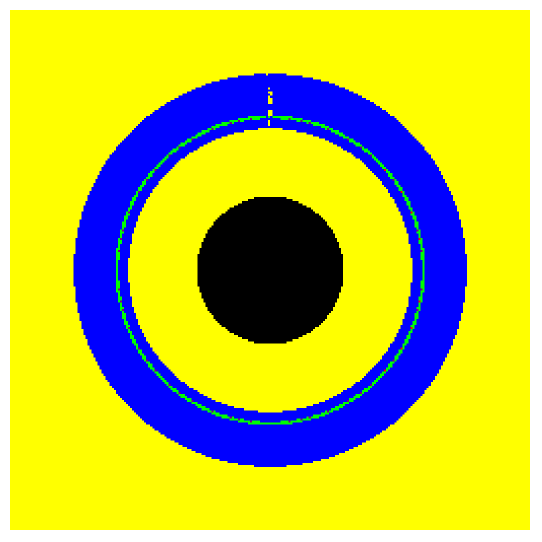}}
\subfigure[$a=0$, $\theta_o = 17^{\circ}$]{\includegraphics[scale=0.375]{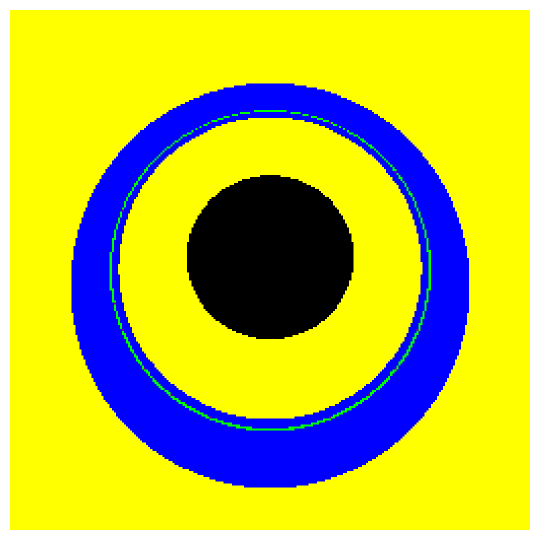}}
\subfigure[$a=0.1$, $\theta_o = 17^{\circ}$]{\includegraphics[scale=0.375]{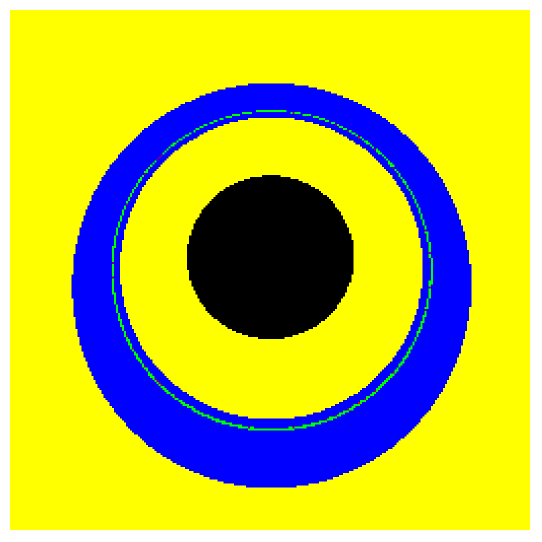}}
\subfigure[$a=0.5$, $\theta_o = 17^{\circ}$]{\includegraphics[scale=0.375]{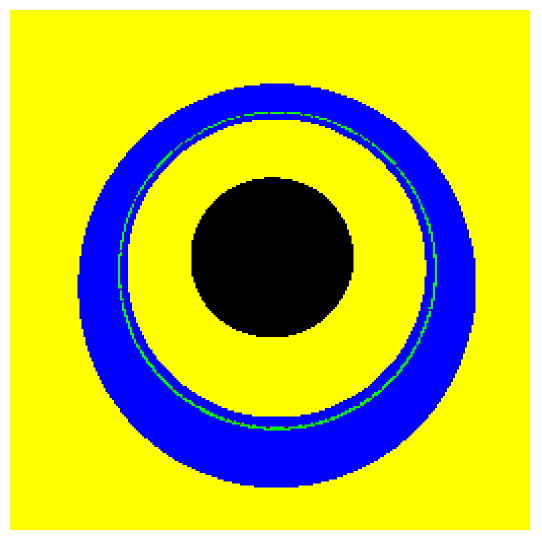}}
\subfigure[$a=0.99$, $\theta_o = 17^{\circ}$]{\includegraphics[scale=0.375]{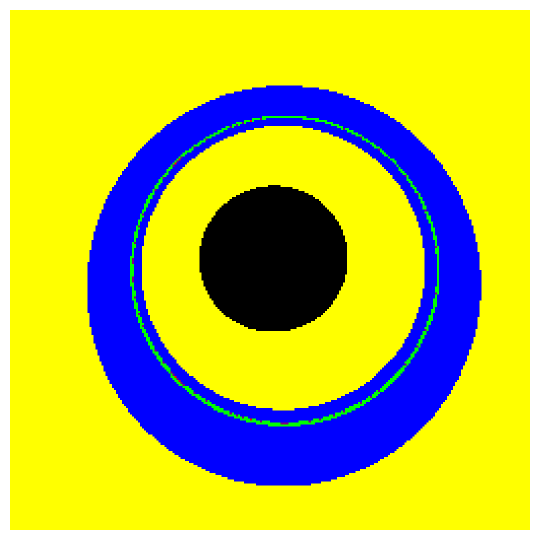}}
\subfigure[$a=0$, $\theta_o = 60^{\circ}$]{\includegraphics[scale=0.375]{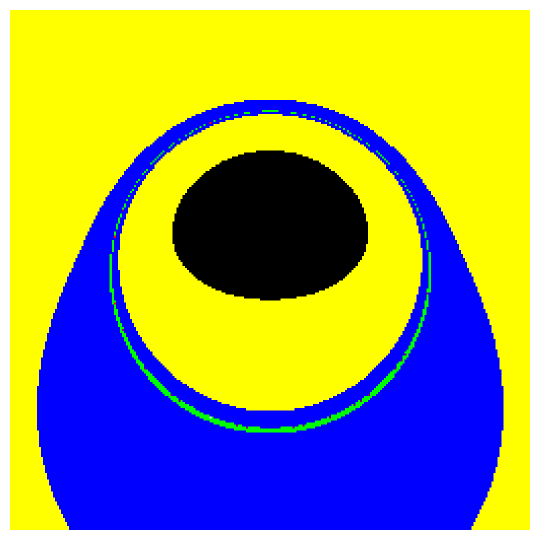}}
\subfigure[$a=0.1$, $\theta_o = 60^{\circ}$]{\includegraphics[scale=0.375]{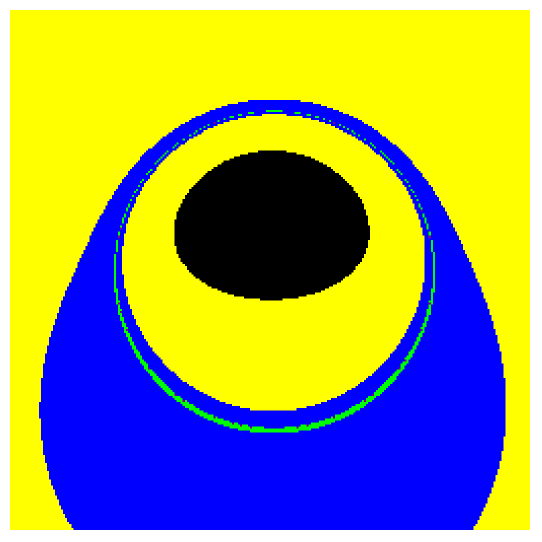}}
\subfigure[$a=0.5$, $\theta_o =60^{\circ}$]{\includegraphics[scale=0.375]{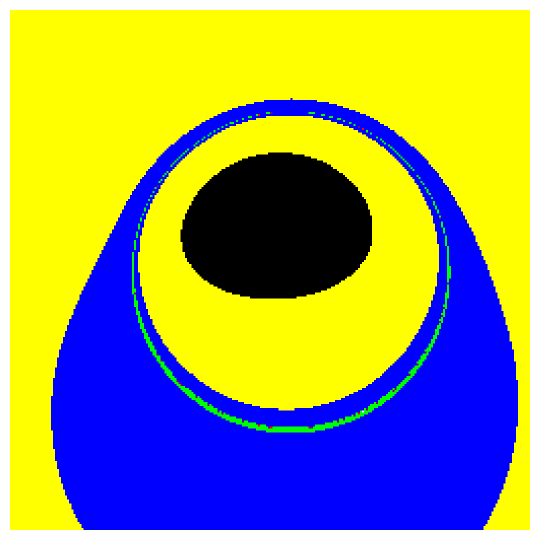}}
\subfigure[$a=0.99$, $\theta_o = 60^{\circ}$]{\includegraphics[scale=0.375]{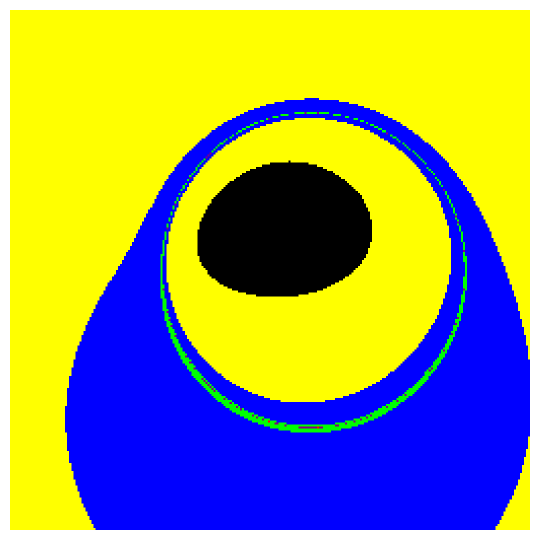}}
\subfigure[$a=0$, $\theta_o = 83^{\circ}$]{\includegraphics[scale=0.375]{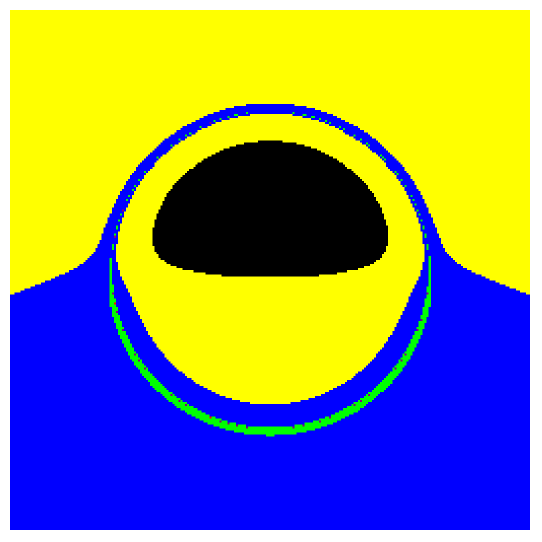}}
\subfigure[$a=0.1$, $\theta_o = 83^{\circ}$]{\includegraphics[scale=0.375]{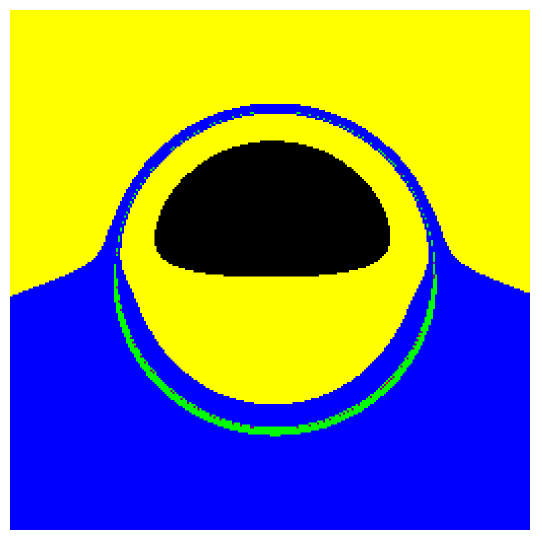}}
\subfigure[$a=0.5$, $\theta_o = 83^{\circ}$]{\includegraphics[scale=0.375]{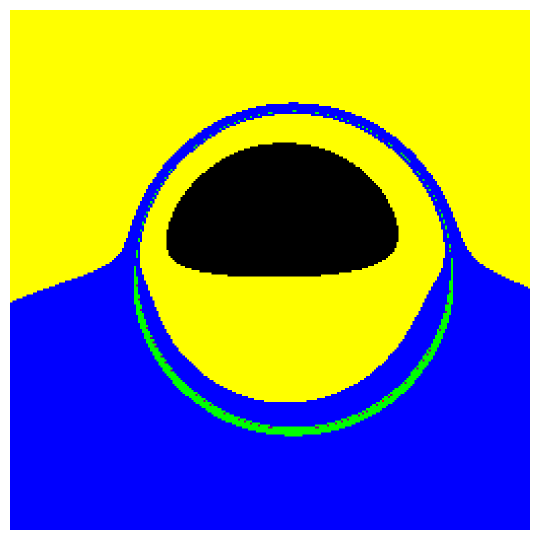}}
\subfigure[$a=0.99$, $\theta_o = 83^{\circ}$]{\includegraphics[scale=0.375]{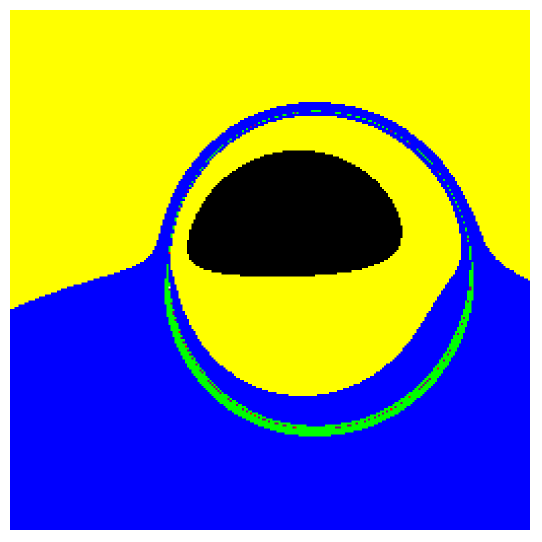}}
\caption{\label{figbpA2} The observed flux distribution of direct and lensed images of  the accretion disk. The colors yellow, blue and green correspond to the direct image, lensed image and photon ring, respectively. In addition, the rotation parameter $a$ increases gradually in each column, while keeping the other relevant parameters fixed at $\mathcal{G} = 0.1$ and $\alpha=-0.5$, and each row represents a fixed observation angle.}
\end{figure}

In order to enhance the distinction between direct  and lensed images of the thin accretion disk, we present their observed fluxes under relevant parameters in Fig. \ref{figbpA2}, where the parameter values correspond to those in Fig. \ref{figbpA1}.
In Figure 6, the colors yellow, blue and green  represent direct emission, lensed emission and photon ring regions respectively. These colors indicate the light that passes through the equatorial plane of the black hole multiple times: once for yellow, twice for blue, and three times for green. It can be observed that the position of the photon ring falls within the lensed image range, indicating a small amount of observed flux of the lensed emission inside the photon ring.
When the observed inclination angle $\theta_o = 0^{\circ}$, variations in the rotation parameter $a$ do not impact the appearance of direct and transparent images of the accretion disk, which consistently manifest as symmetrical rings with a specific width. When the observed inclination angle is a small value, i.e., $\theta_o=17^\circ$, both direct and lensed images appear as slightly deformed ring structures, and the radiation flux tends to converge towards the lower half of the observation plane. However, the observed intensity distribution in Fig.\ref{figbpA1} lacks the necessary sharpness to effectively discern between direct and lensed regions. For large observed inclination angles, i.e., $\theta_o = 60^{\circ}$ and $\theta_o = 83^{\circ}$, the direct and lensed images, as well as the inner shadow region of the accretion disk, exhibit significant deformations,  where an increase in parameter $a$ corresponds to a higher degree of deformation. Moreover, the majority of the observed flux of lensed emission is concentrated in the lower half region of the observation plane, with only a minor portion present in the upper half region. Meanwhile, one can also clearly distinguish between the the direct and lensed images in Fig.\ref{figbpA1}. The  observations suggest that variations in the observed inclination angle $\theta_o$ could potentially impact the discernible characteristics of black hole images, particularly those that undergo gravitational lensing.

\begin{figure}[!t]
\centering 
\subfigure[$\mathcal{G}=0.01$, $\theta_o = 0^{\circ}$]{\includegraphics[scale=0.375]{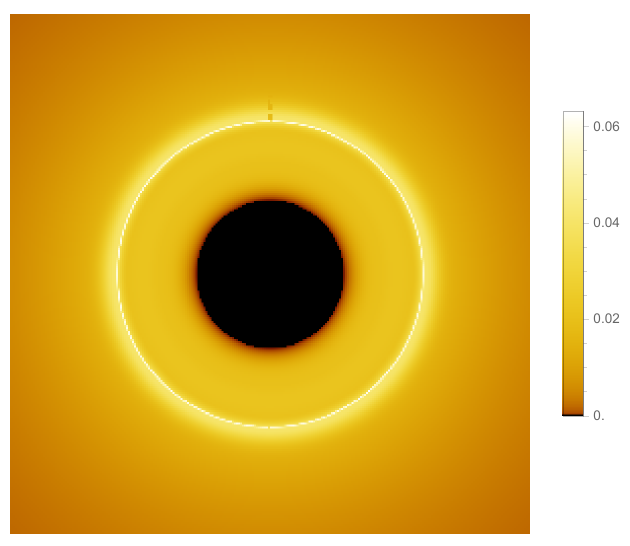}}
\subfigure[$\mathcal{G}=0.1$, $\theta_o = 0^{\circ}$]{\includegraphics[scale=0.375]{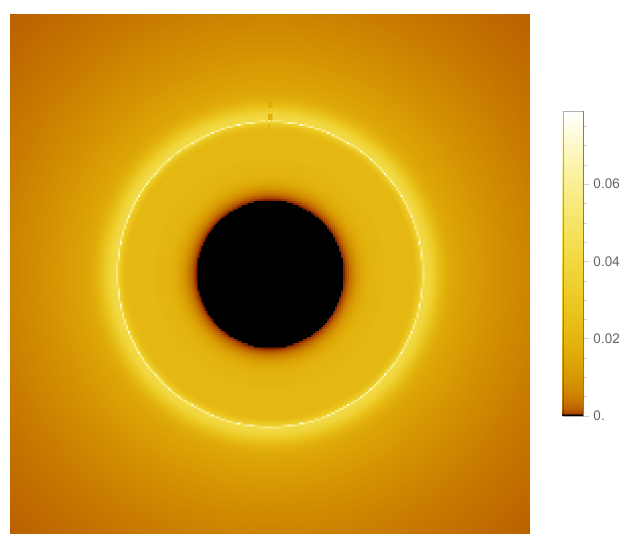}}
\subfigure[$\mathcal{G}=0.2$, $\theta_o = 0^{\circ}$]{\includegraphics[scale=0.375]{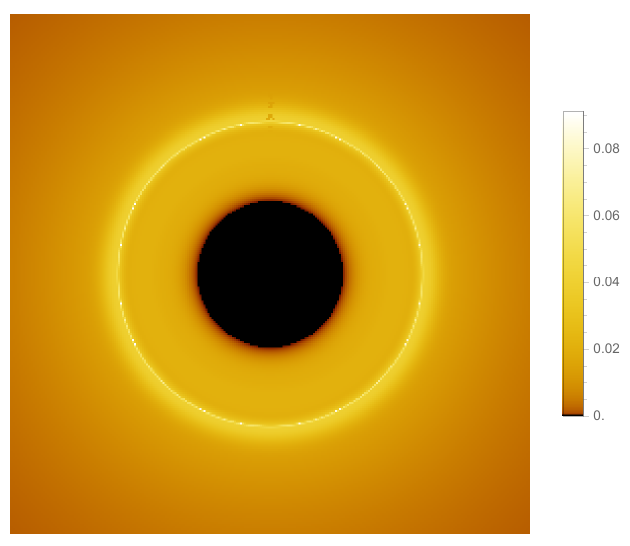}}
\subfigure[$\mathcal{G}=0.3$, $\theta_o = 0^{\circ}$]{\includegraphics[scale=0.375]{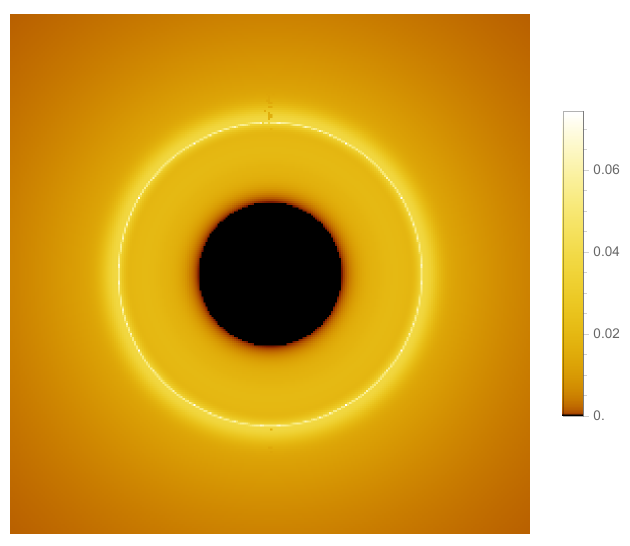}}
\subfigure[$\mathcal{G}=0.01$, $\theta_o = 17^{\circ}$]{\includegraphics[scale=0.375]{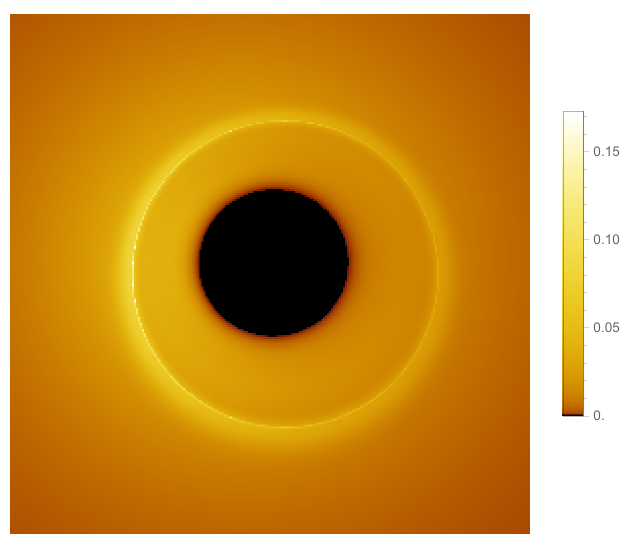}}
\subfigure[$\mathcal{G}=0.1$, $\theta_o = 17^{\circ}$]{\includegraphics[scale=0.375]{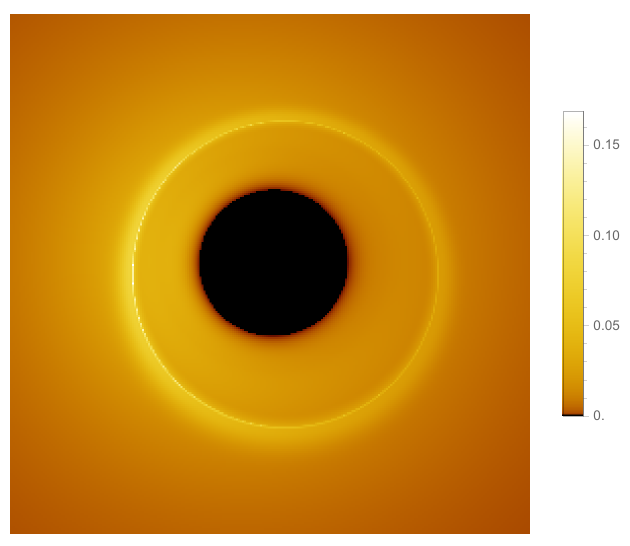}}
\subfigure[$\mathcal{G}=0.2$, $\theta_o = 17^{\circ}$]{\includegraphics[scale=0.375]{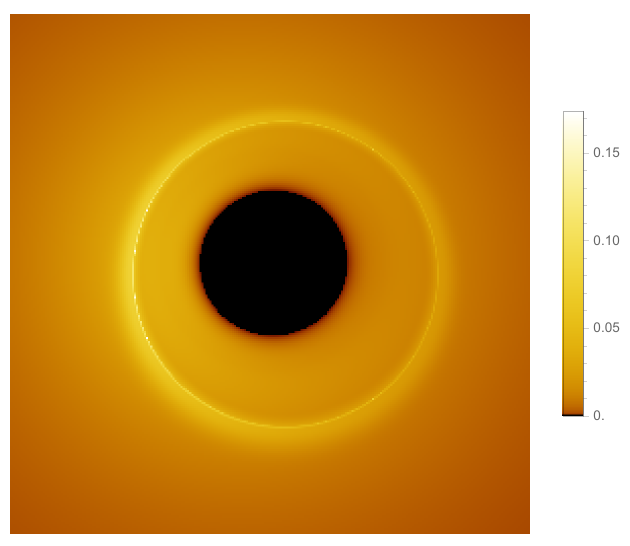}}
\subfigure[$\mathcal{G}=0.3$, $\theta_o = 17^{\circ}$]{\includegraphics[scale=0.375]{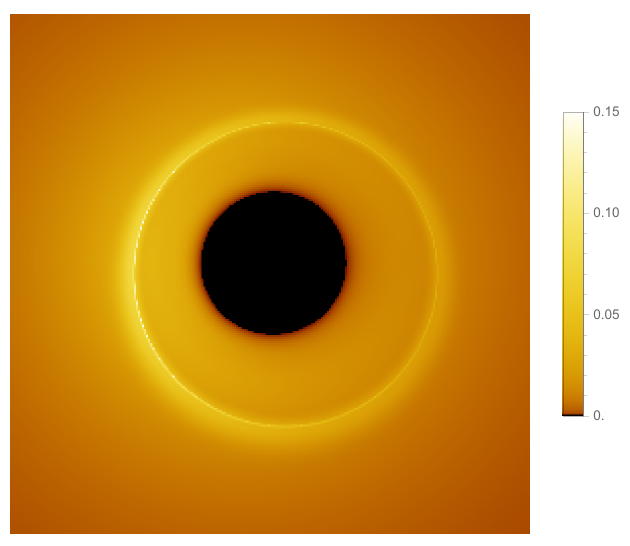}}
\subfigure[$\mathcal{G}=0.01$, $\theta_o = 60^{\circ}$]{\includegraphics[scale=0.375]{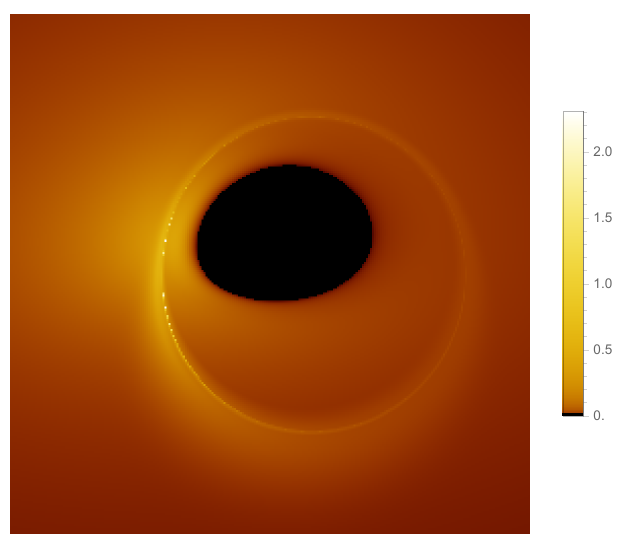}}
\subfigure[$\mathcal{G}=0.1$, $\theta_o = 60^{\circ}$]{\includegraphics[scale=0.375]{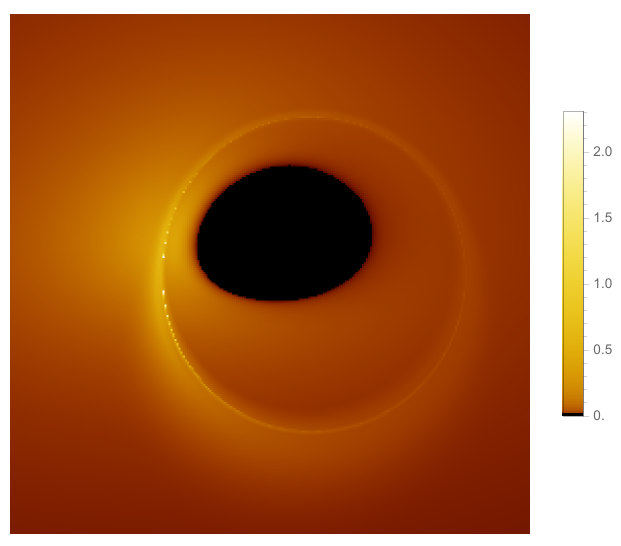}}
\subfigure[$\mathcal{G}=0.2$, $\theta_o =60^{\circ}$]{\includegraphics[scale=0.375]{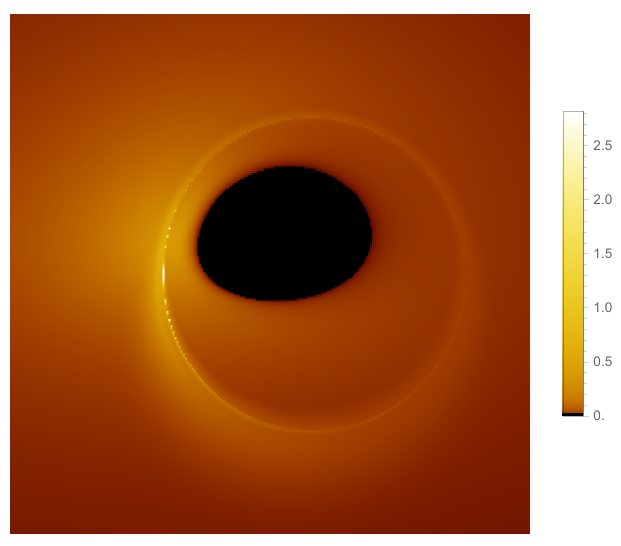}}
\subfigure[$\mathcal{G}=0.3$, $\theta_o = 60^{\circ}$]{\includegraphics[scale=0.375]{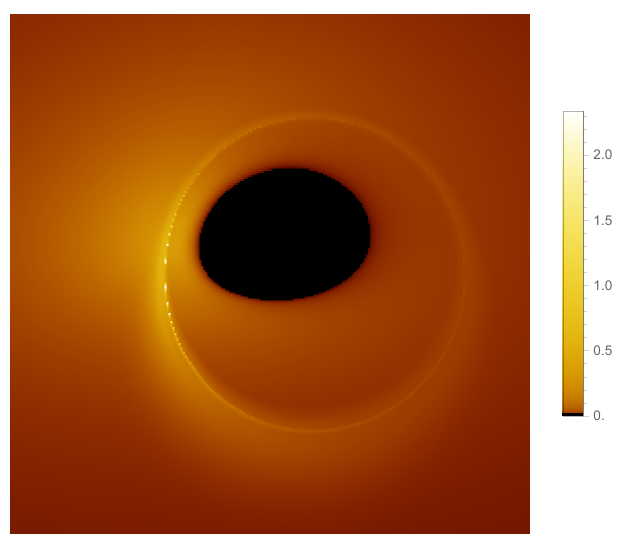}}
\subfigure[$\mathcal{G}=0.01$, $\theta_o = 83^{\circ}$]{\includegraphics[scale=0.385]{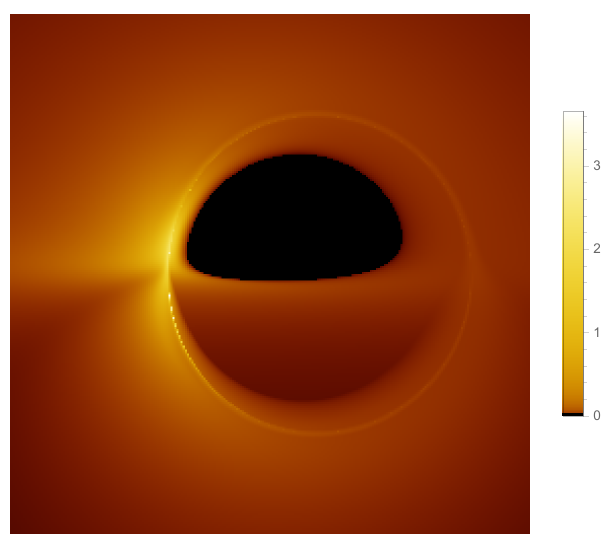}}
\subfigure[$\mathcal{G}=0.1$, $\theta_o = 83^{\circ}$]{\includegraphics[scale=0.385]{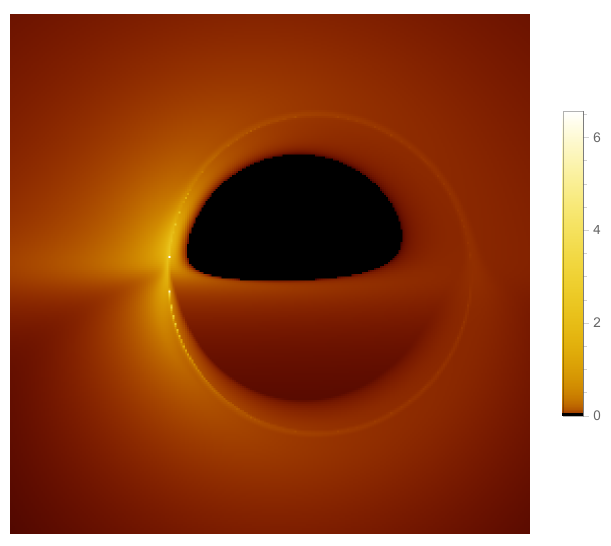}}
\subfigure[$\mathcal{G}=0.2$, $\theta_o = 83^{\circ}$]{\includegraphics[scale=0.385]{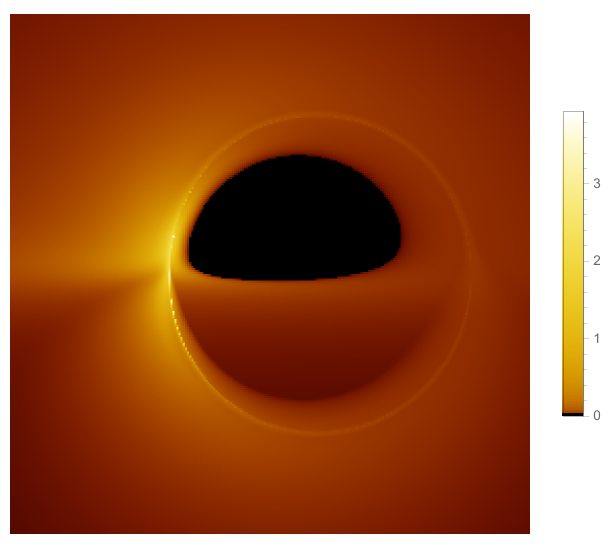}}
\subfigure[$\mathcal{G}=0.3$, $\theta_o = 83^{\circ}$]{\includegraphics[scale=0.385]{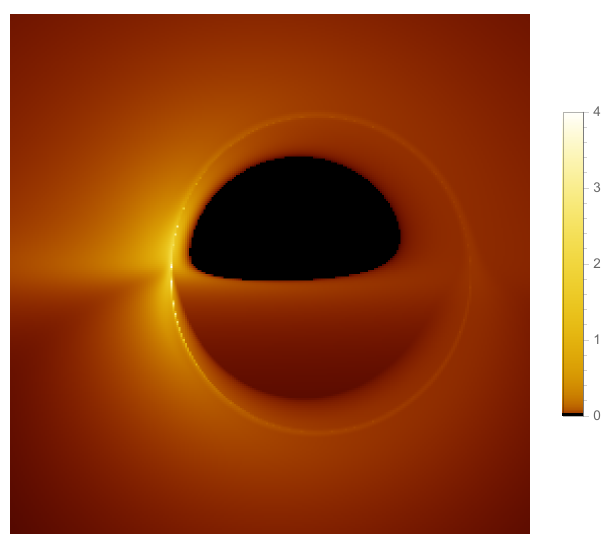}}
\caption{\label{figbpG1} In the thin accretion disk model, the images of a rotating Bardeen black holes surrounded by PFDM at 230 GHz across different parameter spaces. The observation angles in the first to fourth rows are $\theta_o = 0^{\circ}, 17^{\circ}, 60^{\circ}$ and $83^{\circ}$. And, the magnetic charge $\mathcal{G}$ in the first to fourth columns are $\mathcal{G}=0.01, 0.1, 0.2$ and $0.3$, respectively. In which, the other relevant parameters are fixed as $a=0.99$ and $\alpha=-05$.}
\end{figure}

\begin{figure}[!t]
\centering 
\subfigure[$\mathcal{G}=0.01$, $\theta_o = 0^{\circ}$]{\includegraphics[scale=0.38]{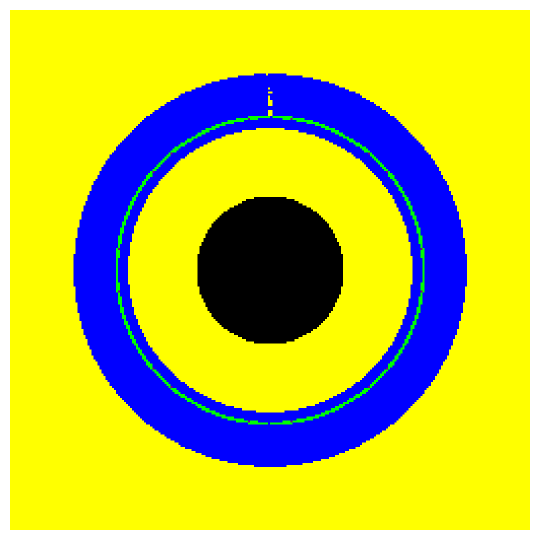}}
\subfigure[$\mathcal{G}=0.1$, $\theta_o = 0^{\circ}$]{\includegraphics[scale=0.38]{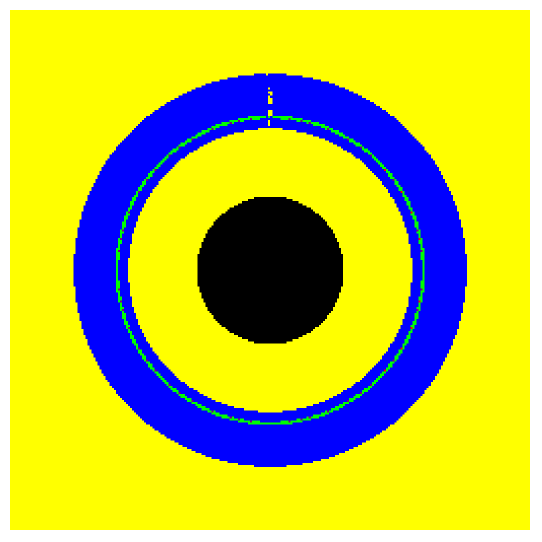}}
\subfigure[$\mathcal{G}=0.2$, $\theta_o = 0^{\circ}$]{\includegraphics[scale=0.38]{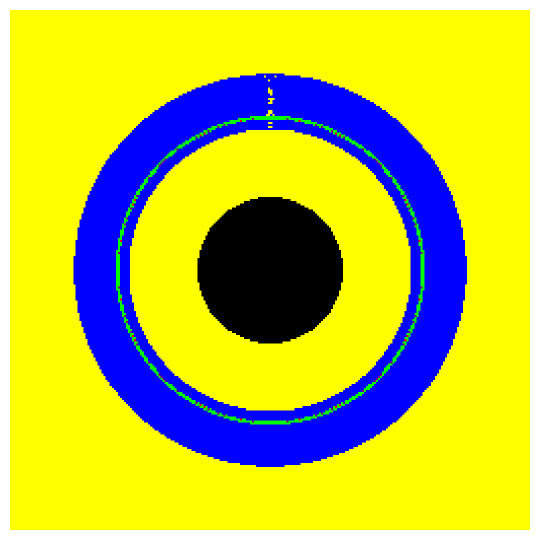}}
\subfigure[$\mathcal{G}=0.3$, $\theta_o = 0^{\circ}$]{\includegraphics[scale=0.38]{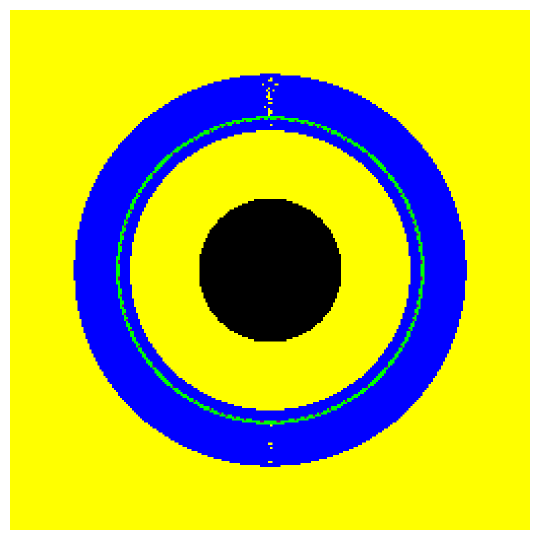}}
\subfigure[$\mathcal{G}=0.01$, $\theta_o = 17^{\circ}$]{\includegraphics[scale=0.38]{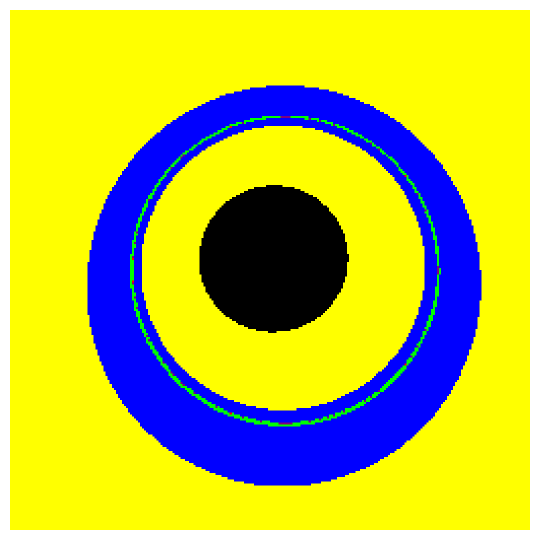}}
\subfigure[$\mathcal{G}=0.1$, $\theta_o = 17^{\circ}$]{\includegraphics[scale=0.38]{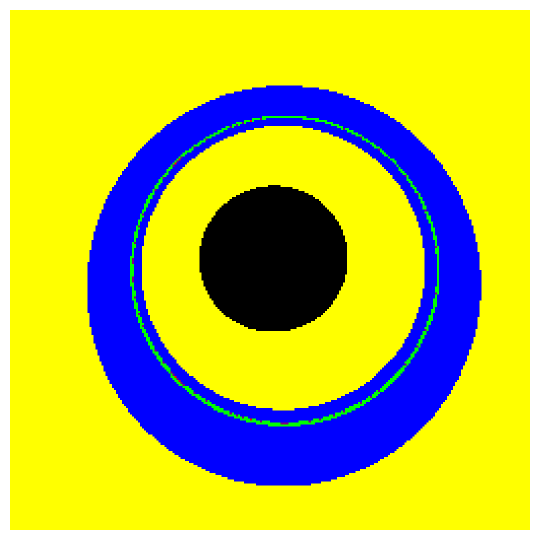}}
\subfigure[$\mathcal{G}=0.2$, $\theta_o = 17^{\circ}$]{\includegraphics[scale=0.38]{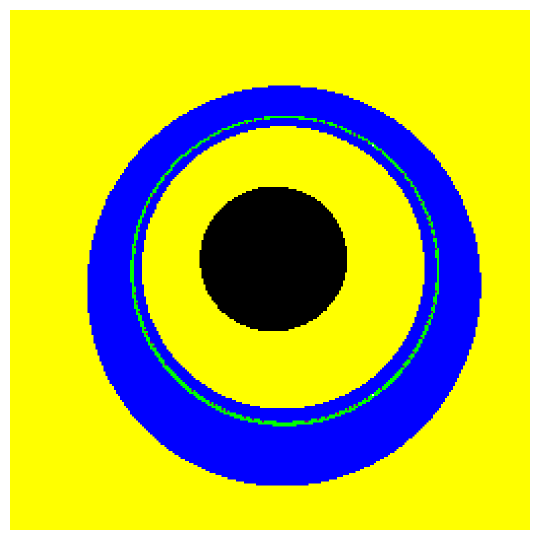}}
\subfigure[$\mathcal{G}=0.3$, $\theta_o = 17^{\circ}$]{\includegraphics[scale=0.38]{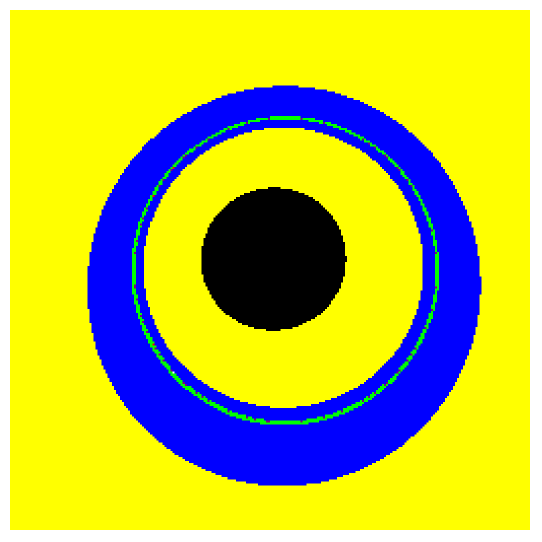}}
\subfigure[$\mathcal{G}=0.01$, $\theta_o = 60^{\circ}$]{\includegraphics[scale=0.38]{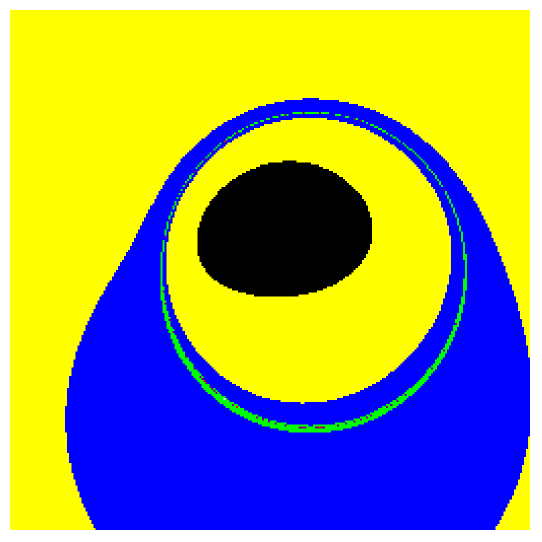}}
\subfigure[$\mathcal{G}=0.1$, $\theta_o = 60^{\circ}$]{\includegraphics[scale=0.38]{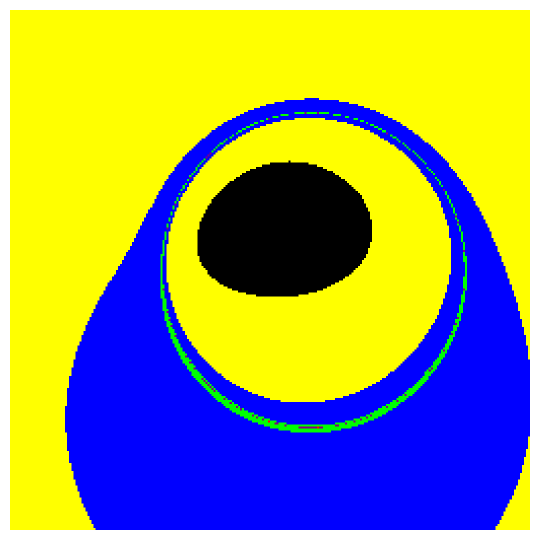}}
\subfigure[$\mathcal{G}=0.2$, $\theta_o =60^{\circ}$]{\includegraphics[scale=0.38]{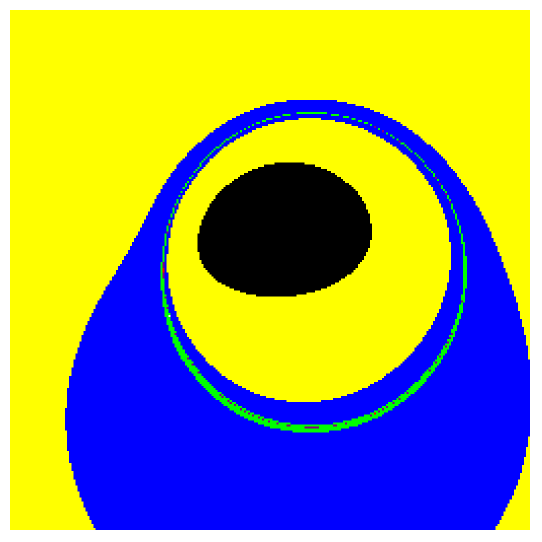}}
\subfigure[$\mathcal{G}=0.3$, $\theta_o = 60^{\circ}$]{\includegraphics[scale=0.38]{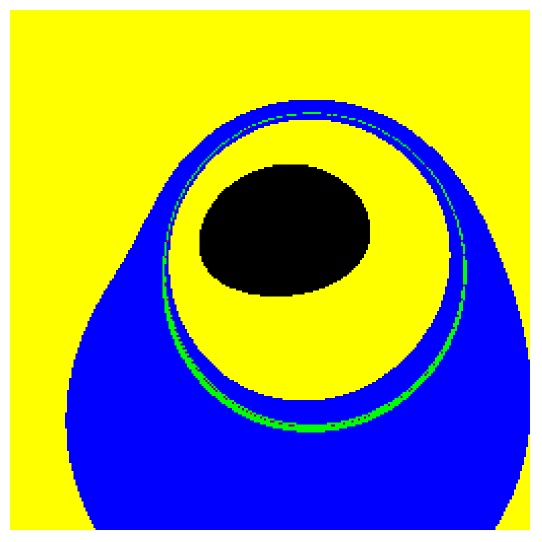}}
\subfigure[$\mathcal{G}=0.01$, $\theta_o = 83^{\circ}$]{\includegraphics[scale=0.38]{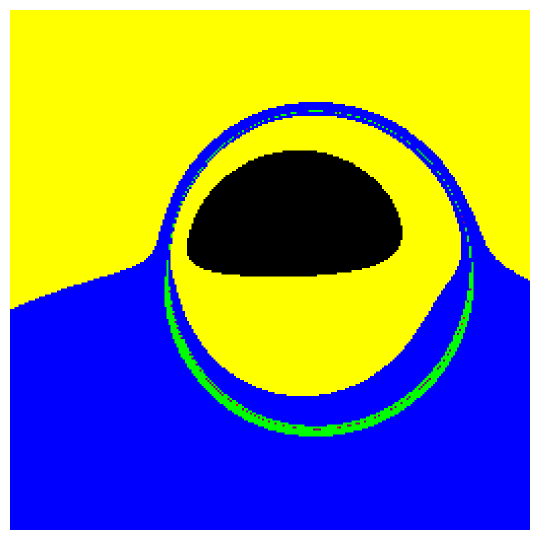}}
\subfigure[$\mathcal{G}=0.1$, $\theta_o = 83^{\circ}$]{\includegraphics[scale=0.38]{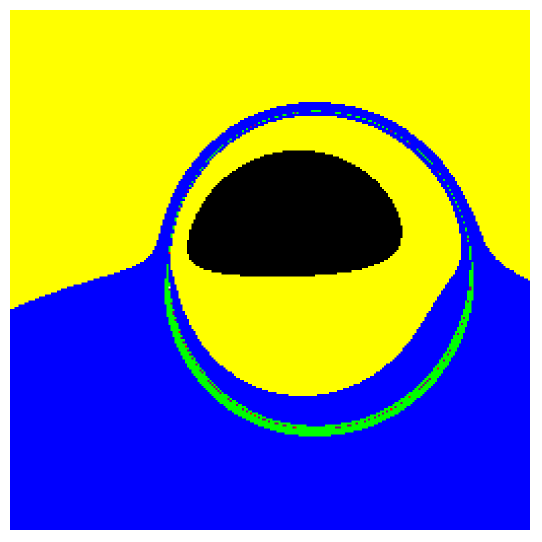}}
\subfigure[$\mathcal{G}=0.2$, $\theta_o = 83^{\circ}$]{\includegraphics[scale=0.38]{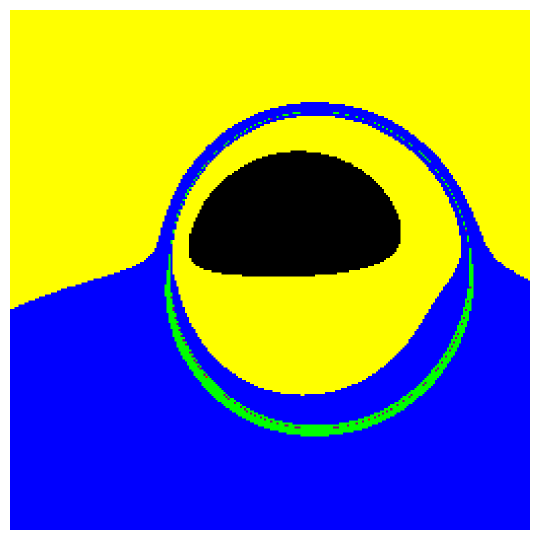}}
\subfigure[$\mathcal{G}=0.3$, $\theta_o = 83^{\circ}$]{\includegraphics[scale=0.38]{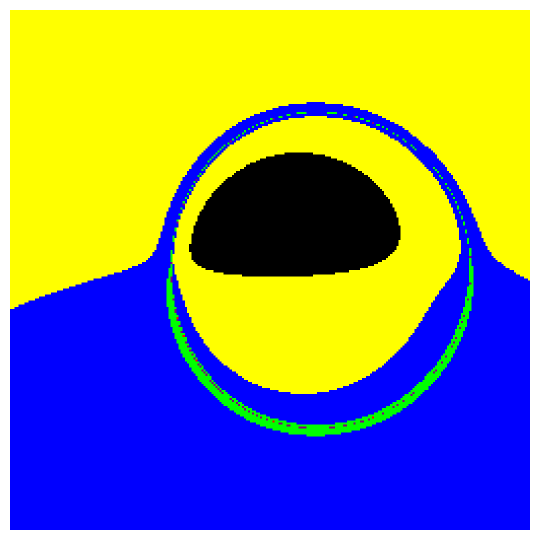}}
\caption{\label{figbpG2}  The observed flux distribution of direct and lensed images of  the accretion disk. The colors yellow, blue and green correspond to the direct image, lensed image and photon ring, respectively. In addition, the magnetic charge $\mathcal{G}$ increases gradually in each column, while keeping the other relevant parameters fixed at $a = 0.99$ and $\alpha=-0.5$, and each row represents a fixed observation angle.}
\end{figure}

The impact of variations in the magnetic charge $\mathcal{G}$ on the black hole image is illustrated in Fig. \ref{figbpG1}, while Fig. \ref{figbpG2} presents the observed flux distributions of the direct and lens images corresponding to these variations. Among them, the values of the magnetic charge are $\mathcal{G}=0.01,0.1,0.2$ and $0.3$, while the remaining relevant parameters are set to $a=0.99$ and $\alpha=-0.5$. From  Fig.\ref{figbpG1}, it can be found that under the same observation inclination, an increase in magnetic charge $\mathcal{G}$ leads to a slight upward trend in both the inner shadow region of the accretion disk and the radius of the photon ring, as well as the corresponding observation intensity. The shape of the inner shadow will be altered and the asymmetry of the image will become more pronounced with an increase in observation inclination, while the morphology of the photon ring remains largely unaffected. When the observed inclination angle $\theta_o$ is a  large value, a distinct bright area resembling a crescent or an eye-brow emerges on the left side of the screen, where the direct and lensed images of the accretion disk can be clearly distinguished. In the corresponding observed flux images, see Fig.\ref{figbpG2}, it is show that an increase in magnetic charge leads to a moderate expansion of the transparent mirror region within the accretion disk, albeit without significant magnitudes. The observed flux of the lensed images gradually concentrates towards the lower half of the observation plane as the angle of observation increases, leaving only a small portion in the upper half. Furthermore, the  direct image also undergoes deformation in response to variations in the observation angle. The results of the effect of the observed inclination $\theta_o$ change on the black hole image show a certain agreement with the results shown in Fig. \ref{figbpA2}.


Subsequently, we direct our attention towards investigating the impact of variations in the dark matter parameter $\alpha$ on the black hole image, as illustrated in Fig. \ref{figbpP2}. The results indicate that an augmentation in the absolute value of the dark matter parameter $\alpha$ expands both the inner shadow region of the accretion disk and the radius of the photon ring, and this increase is considerably significant. Simultaneously, there is also an observed upward trend in intensity. When the observed inclination angle is altered, the shape of the inner shadow region undergoes a transition from circular to elliptical, gradually manifesting as an arch at $\theta_o=83^{\circ}$. It is worth noting that the variation of the observed inclination angle $\theta_o$ and parameter $\alpha$ does not have a significant influence on the morphology of the photon ring. Only when both parameter $\alpha$ and $\theta_o$ are significantly increased, a slight deformation in the shape of the photon ring can be observed. From the corresponding observed flux images, as shown in Fig. \ref{figbpP2}, it can be clearly seen that the width and radius of the transparent image of the accretion disk increase significantly when parameter $\alpha$ takes a larger value. The impact of changes in observation inclination on the measured flux of direct and lens images is still evident as a significant accumulation predominantly in the lower half of the image observation plane.
\begin{figure}[!t]
\centering 
\subfigure[$\alpha=-0.1$, $\theta_o = 0^{\circ}$]{\includegraphics[scale=0.375]{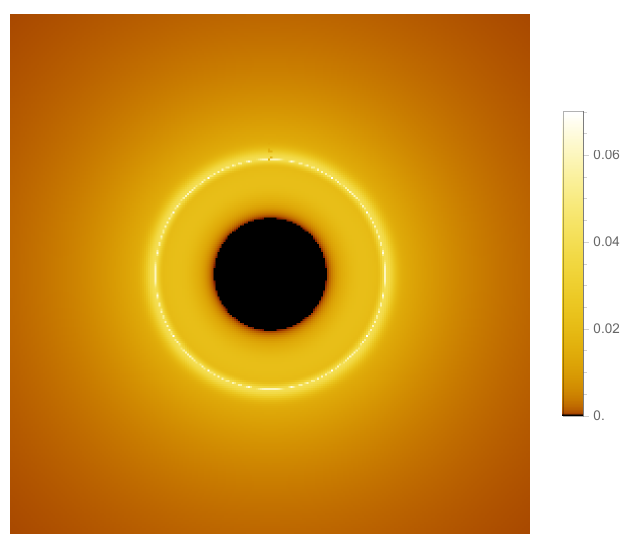}}
\subfigure[$\alpha=-0.3$, $\theta_o = 0^{\circ}$]{\includegraphics[scale=0.375]{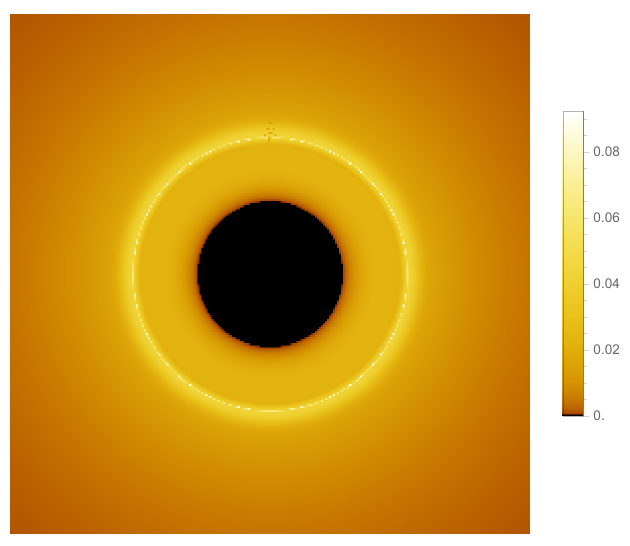}}
\subfigure[$\alpha=-0.5$, $\theta_o = 0^{\circ}$]{\includegraphics[scale=0.375]{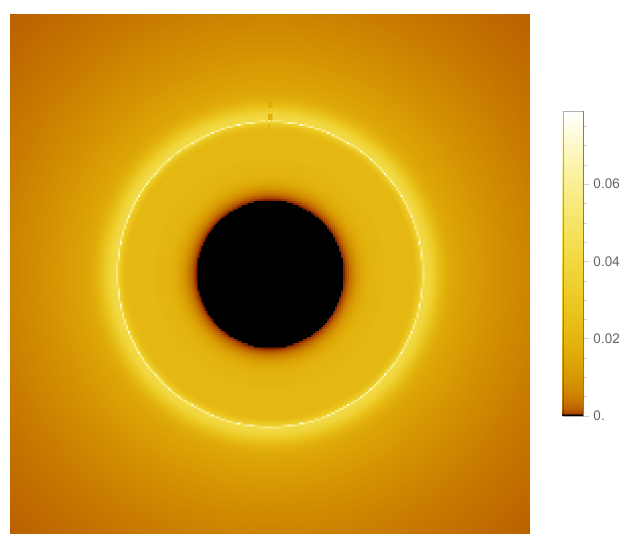}}
\subfigure[$\alpha=-0.9$, $\theta_o = 0^{\circ}$]{\includegraphics[scale=0.375]{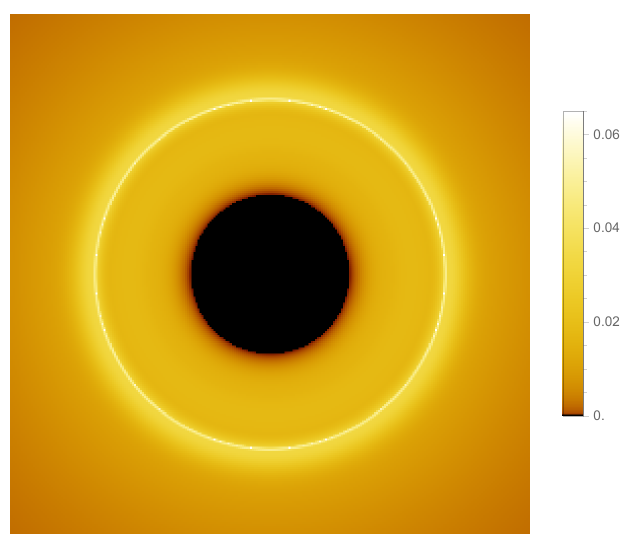}}
\subfigure[$\alpha=-0.1$, $\theta_o = 17^{\circ}$]{\includegraphics[scale=0.375]{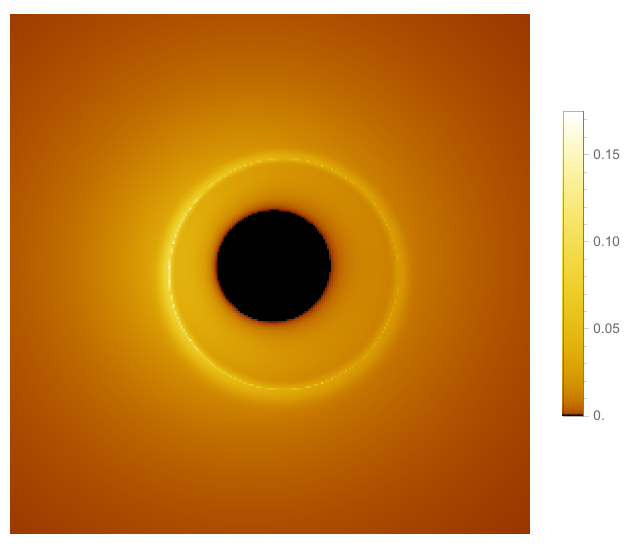}}
\subfigure[$\alpha=-0.3$, $\theta_o = 17^{\circ}$]{\includegraphics[scale=0.375]{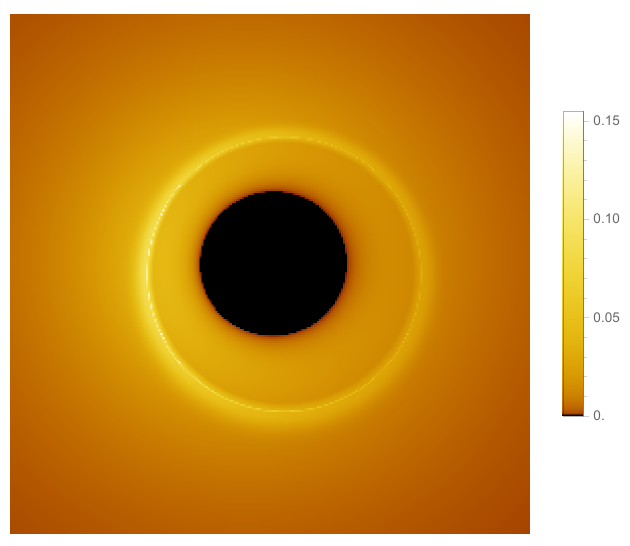}}
\subfigure[$\alpha=-0.5$, $\theta_o = 17^{\circ}$]{\includegraphics[scale=0.375]{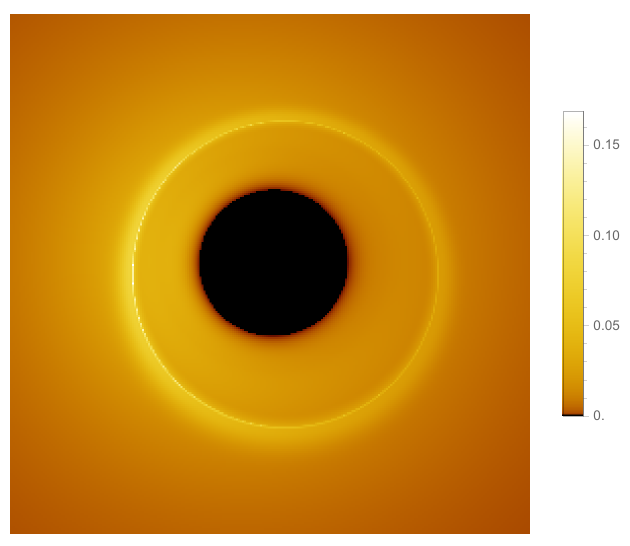}}
\subfigure[$\alpha=-0.9$, $\theta_o = 17^{\circ}$]{\includegraphics[scale=0.375]{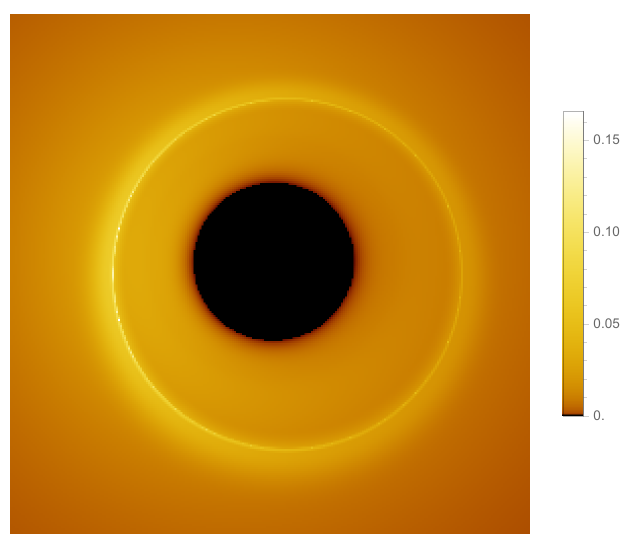}}
\subfigure[$\alpha=-0.1$, $\theta_o = 60^{\circ}$]{\includegraphics[scale=0.38]{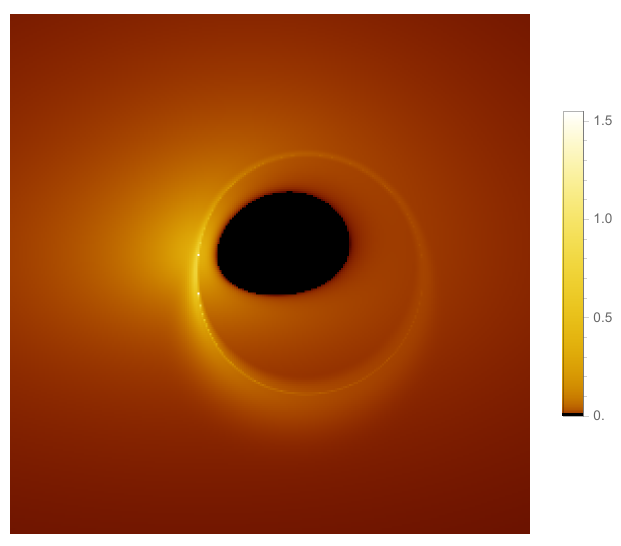}}
\subfigure[$\alpha=-0.3$, $\theta_o = 60^{\circ}$]{\includegraphics[scale=0.38]{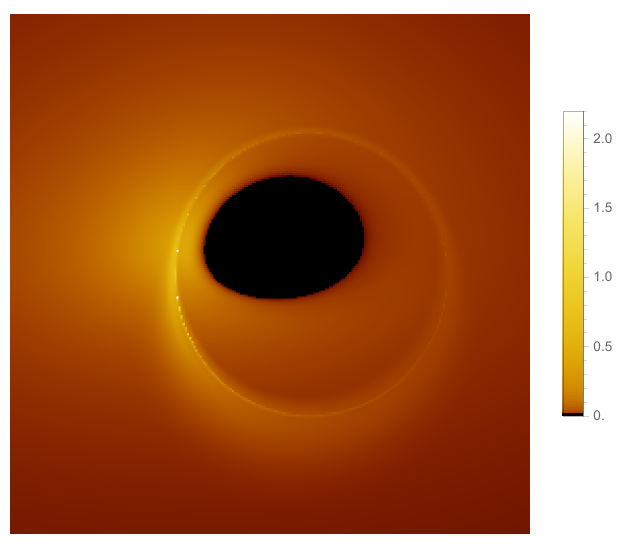}}
\subfigure[$\alpha=-0.5$, $\theta_o = 60^{\circ}$]{\includegraphics[scale=0.38]{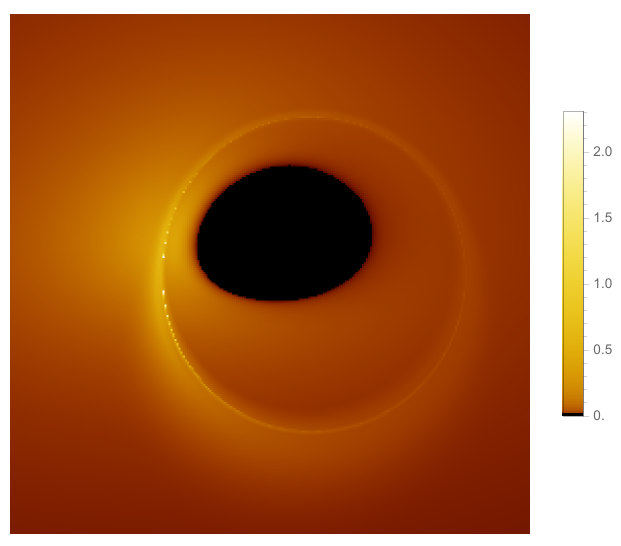}}
\subfigure[$\alpha=-0.9$, $\theta_o = 60^{\circ}$]{\includegraphics[scale=0.38]{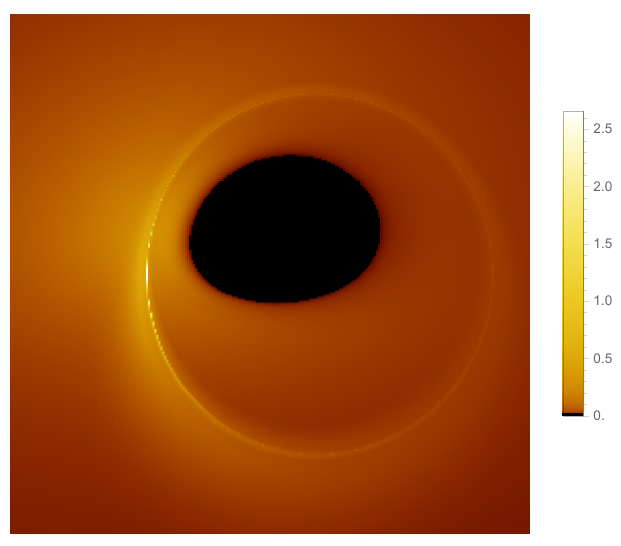}}
\subfigure[$\alpha=-0.1$, $\theta_o = 83^{\circ}$]{\includegraphics[scale=0.385]{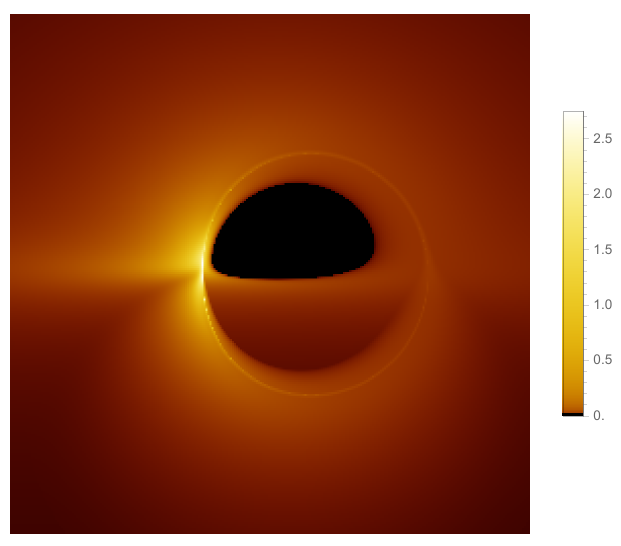}}
\subfigure[$\alpha=-0.3$, $\theta_o = 83^{\circ}$]{\includegraphics[scale=0.385]{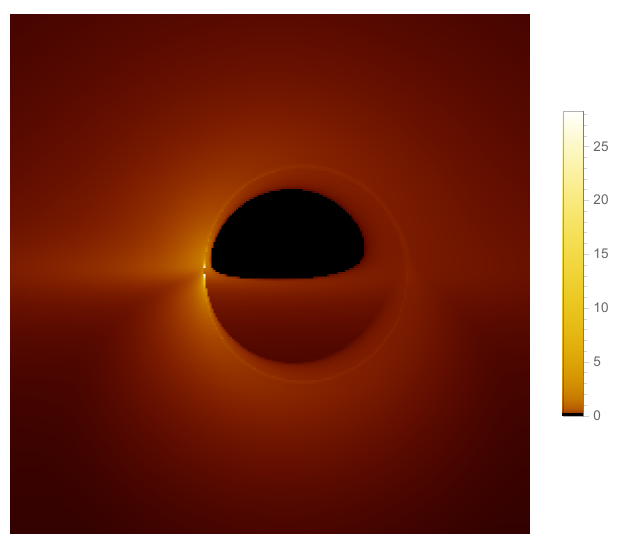}}
\subfigure[$\alpha=-0.5$, $\theta_o = 83^{\circ}$]{\includegraphics[scale=0.385]{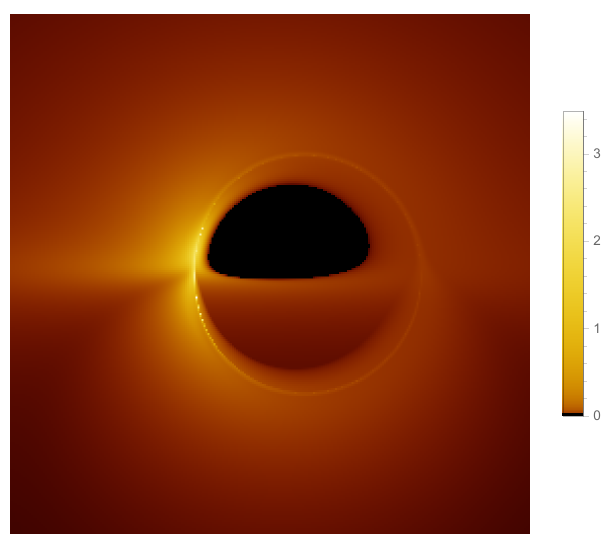}}
\subfigure[$\alpha=-0.9$, $\theta_o = 83^{\circ}$]{\includegraphics[scale=0.385]{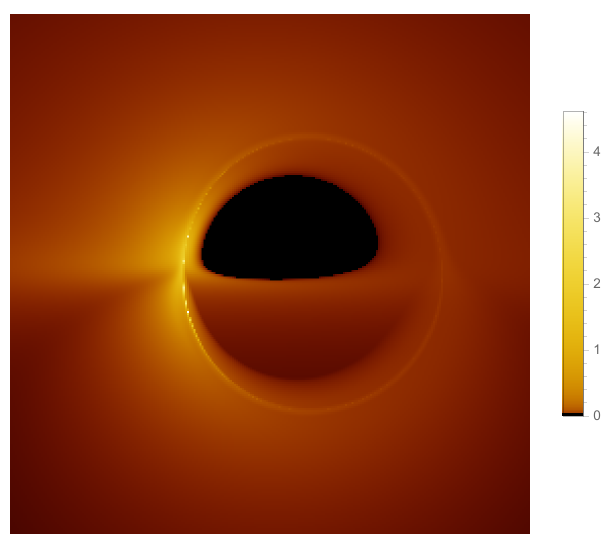}}
\caption{\label{figbpP1} In the thin accretion disk model, the images of a rotating Bardeen black holes surrounded by PFDM at 230 GHz across different parameter spaces. The observation angles in the first to fourth rows are $\theta_o = 0^{\circ}, 17^{\circ}, 60^{\circ}$ and $83^{\circ}$. And, the  dark matter parameter $\alpha$ in the first to fourth columns are $\alpha=-0.1, -0.3, -0.5$ and $-0.9$, respectively. In which, the other relevant parameters are fixed as $a=0.99$ and $\mathcal{G}=0.1$.}
\end{figure}
\begin{figure}[!t]
\centering 
\subfigure[$\alpha=-0.1$, $\theta_o = 0^{\circ}$]{\includegraphics[scale=0.38]{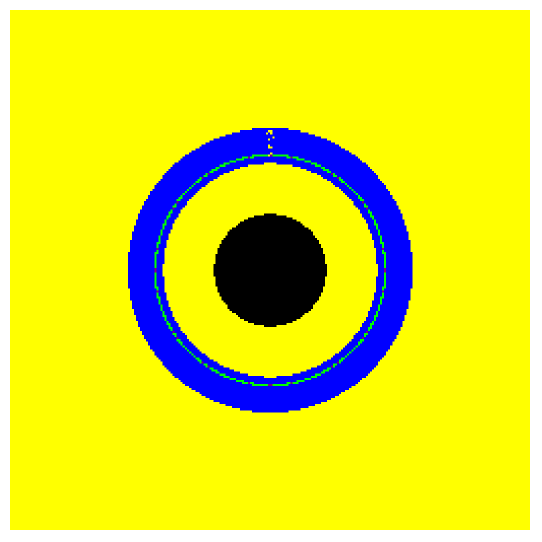}}
\subfigure[$\alpha=-0.3$, $\theta_o = 0^{\circ}$]{\includegraphics[scale=0.38]{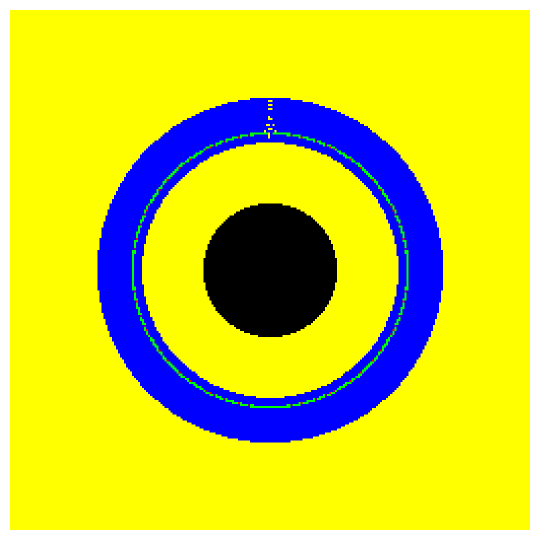}}
\subfigure[$\alpha=-0.5$, $\theta_o = 0^{\circ}$]{\includegraphics[scale=0.38]{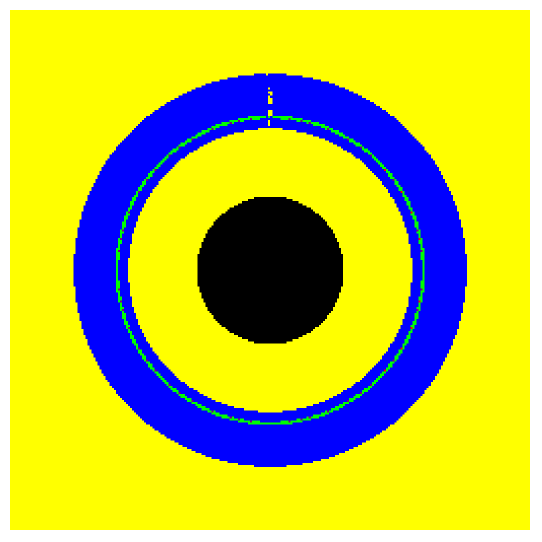}}
\subfigure[$\alpha=-0.9$, $\theta_o = 0^{\circ}$]{\includegraphics[scale=0.38]{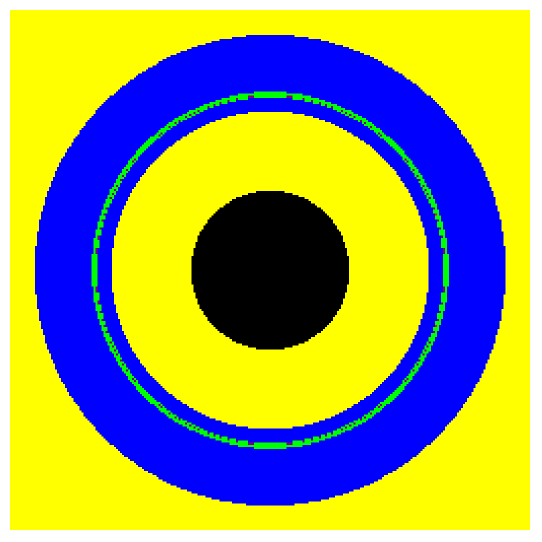}}
\subfigure[$\alpha=-0.1$, $\theta_o = 17^{\circ}$]{\includegraphics[scale=0.38]{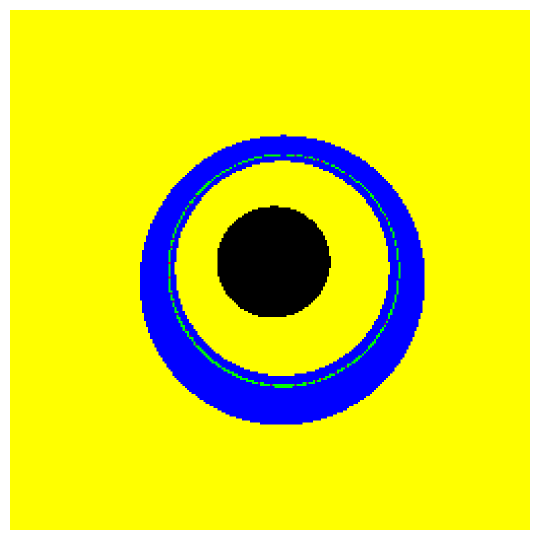}}
\subfigure[$\alpha=-0.3$, $\theta_o = 17^{\circ}$]{\includegraphics[scale=0.38]{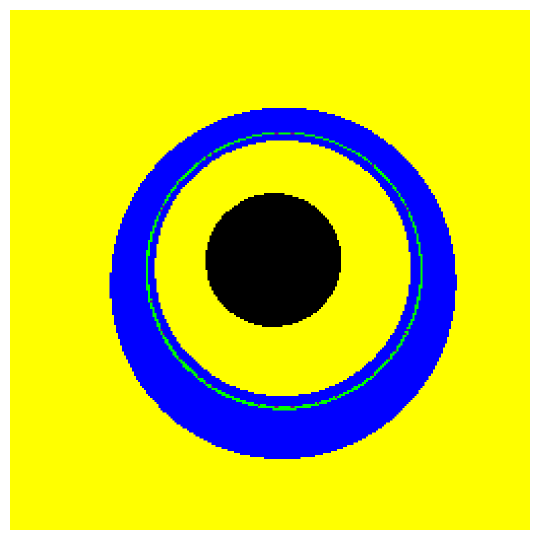}}
\subfigure[$\alpha=-0.5$, $\theta_o = 17^{\circ}$]{\includegraphics[scale=0.38]{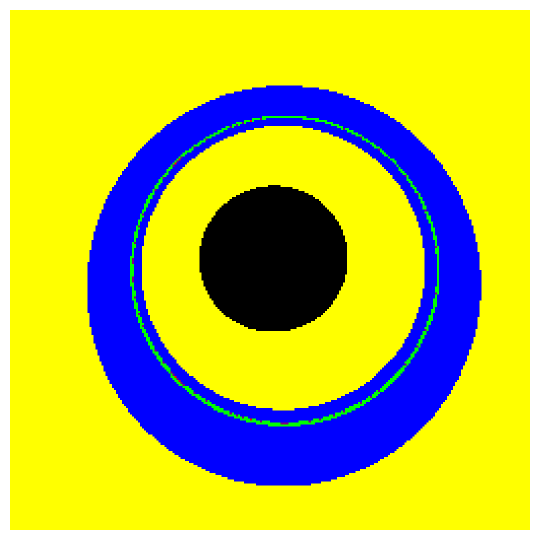}}
\subfigure[$\alpha=-0.9$, $\theta_o = 17^{\circ}$]{\includegraphics[scale=0.38]{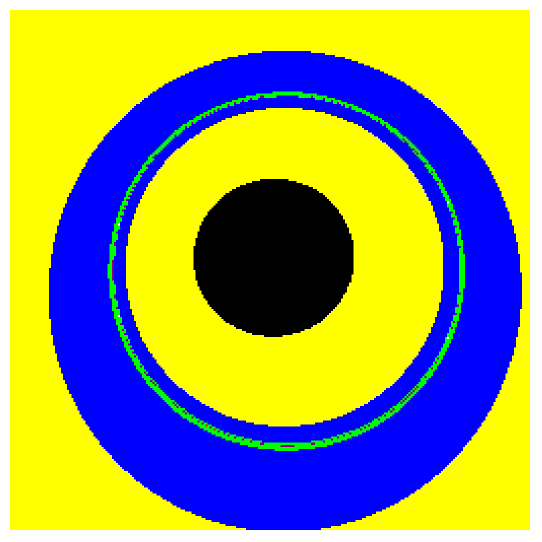}}
\subfigure[$\alpha=-0.1$, $\theta_o = 60^{\circ}$]{\includegraphics[scale=0.38]{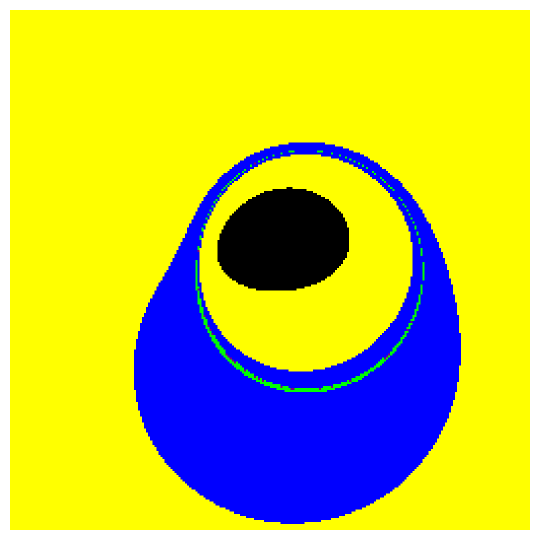}}
\subfigure[$\alpha=-0.3$, $\theta_o = 60^{\circ}$]{\includegraphics[scale=0.38]{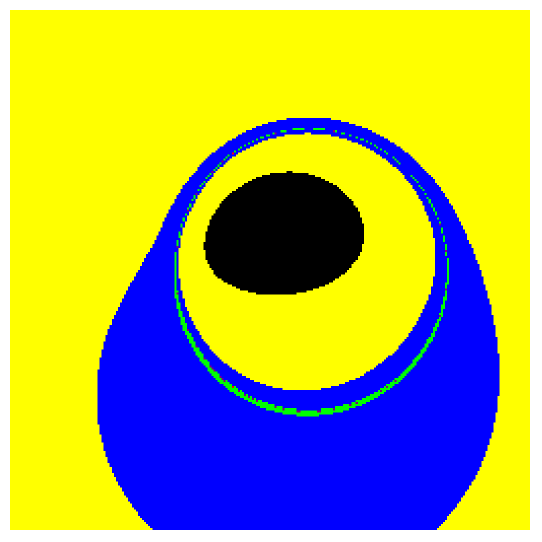}}
\subfigure[$\alpha=-0.5$, $\theta_o = 60^{\circ}$]{\includegraphics[scale=0.38]{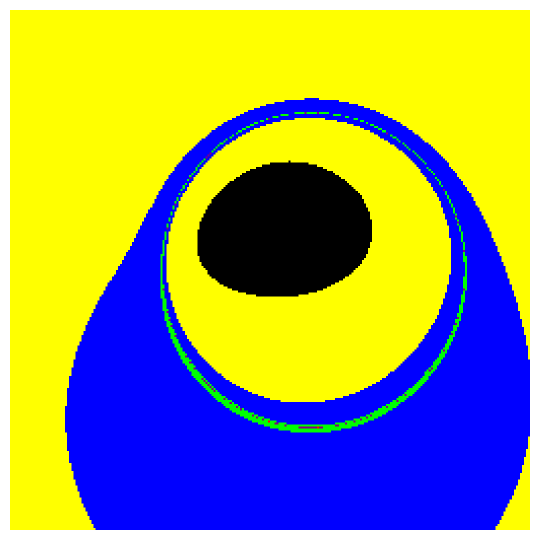}}
\subfigure[$\alpha=-0.9$, $\theta_o = 60^{\circ}$]{\includegraphics[scale=0.38]{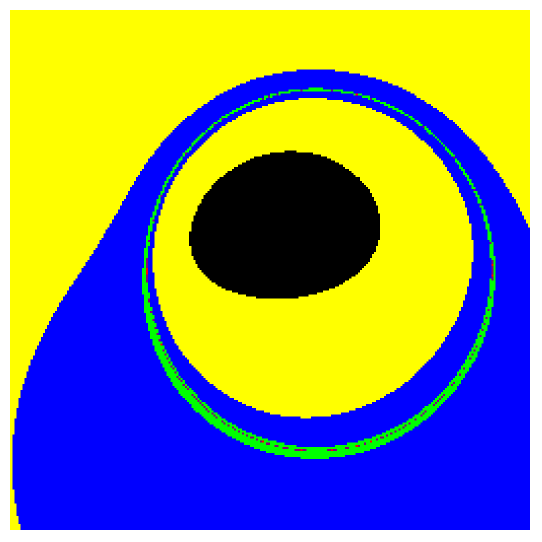}}
\subfigure[$\alpha=-0.1$, $\theta_o = 83^{\circ}$]{\includegraphics[scale=0.38]{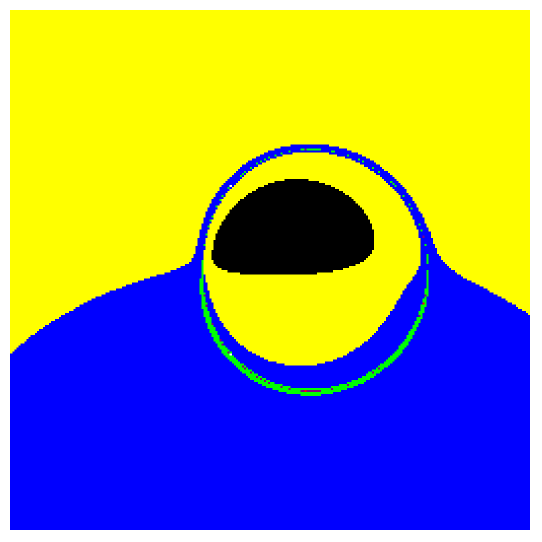}}
\subfigure[$\alpha=-0.3$, $\theta_o = 83^{\circ}$]{\includegraphics[scale=0.38]{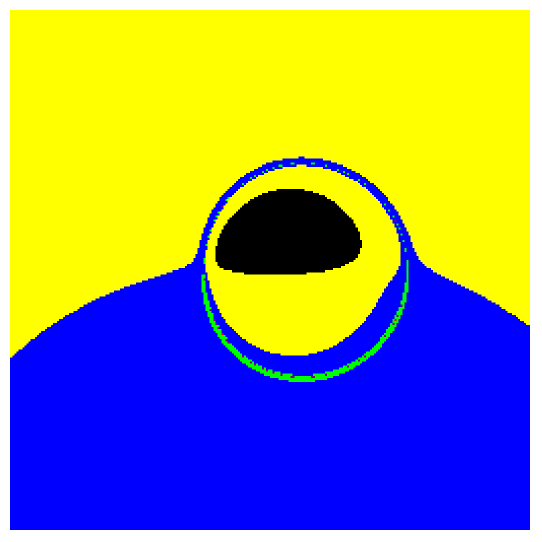}}
\subfigure[$\alpha=-0.5$, $\theta_o = 83^{\circ}$]{\includegraphics[scale=0.38]{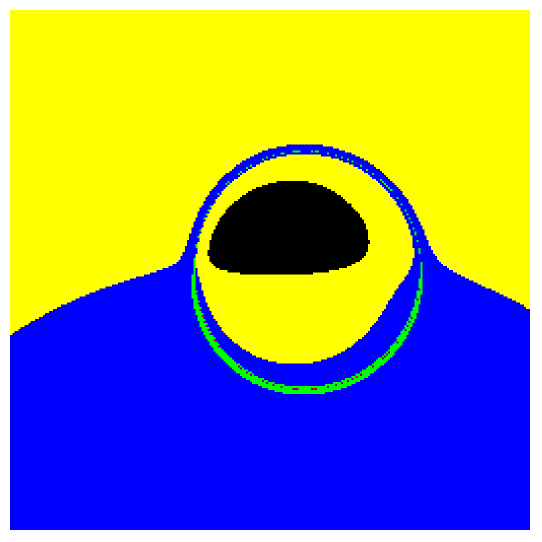}}
\subfigure[$\alpha=-0.9$, $\theta_o = 83^{\circ}$]{\includegraphics[scale=0.38]{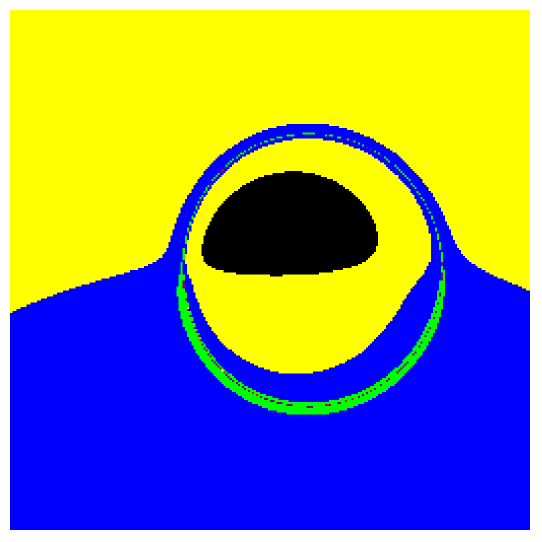}}
\caption{\label{figbpP2} The observed flux distribution of direct and lensed images of  the accretion disk. The colors yellow, blue and green correspond to the direct image, lensed image and photon ring, respectively. In addition, the absolute value of the dark matter parameter $a$  increases gradually in each column, while keeping the other relevant parameters fixed at $a = 0.99$ and $\mathcal{G}=0.1$, and each row represents a fixed observation angle. }
\end{figure}

\section{The distribution of redshift factors}
\begin{figure}[!t]
\centering 
\subfigure[$\alpha=-0.5$, $\mathcal{G}=0.1$, $a=0.99$,  $\theta_o = 17^{\circ}$]{\includegraphics[scale=0.38]{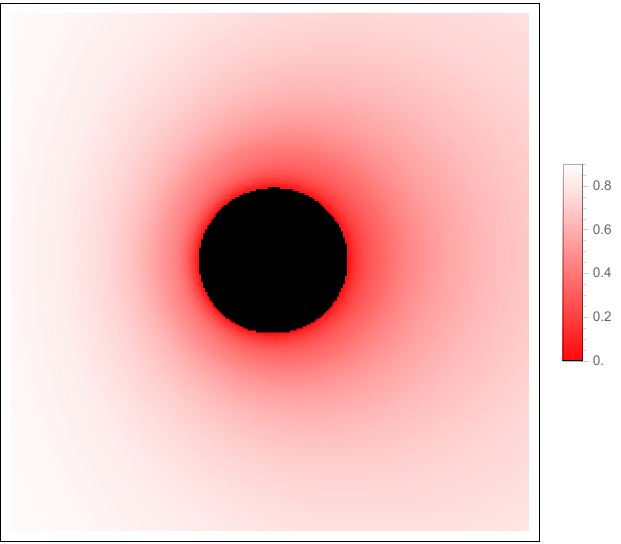}}
\subfigure[$\alpha=-0.5$, $\mathcal{G}=0.3$, $a=0.99$,  $\theta_o = 17^{\circ}$]{\includegraphics[scale=0.38]{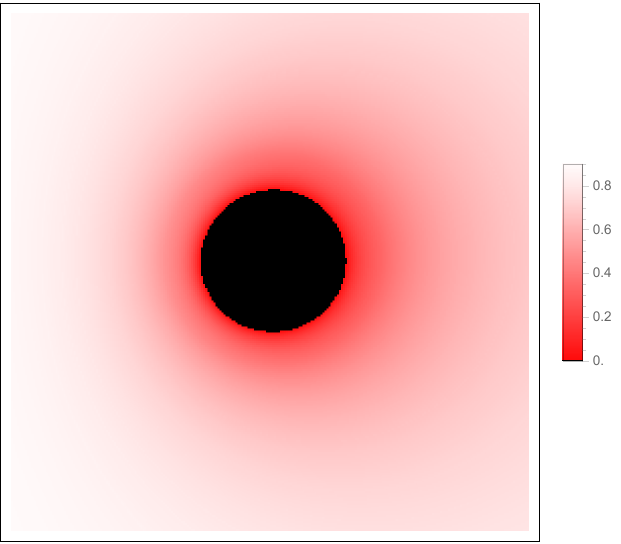}}
\subfigure[$\alpha=-0.5$, $\mathcal{G}=0.1$, $a=0.5$, $\theta_o = 17^{\circ}$]{\includegraphics[scale=0.38]{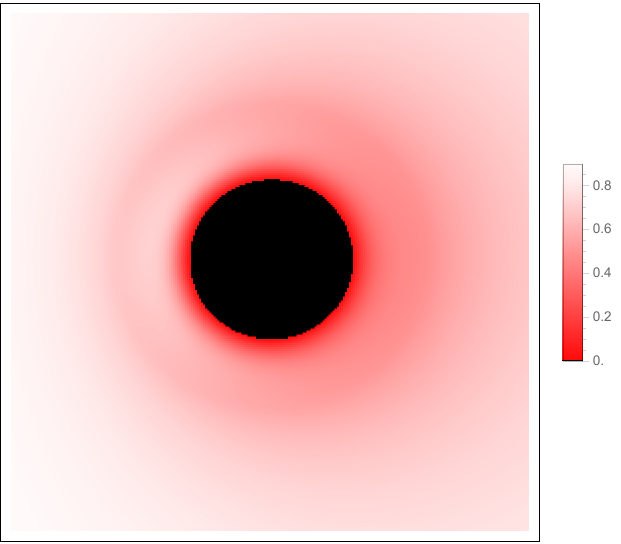}}
\subfigure[$\alpha=-0.9$, $\mathcal{G}=0.1$, $a=0.99$, $\theta_o = 17^{\circ}$]{\includegraphics[scale=0.38]{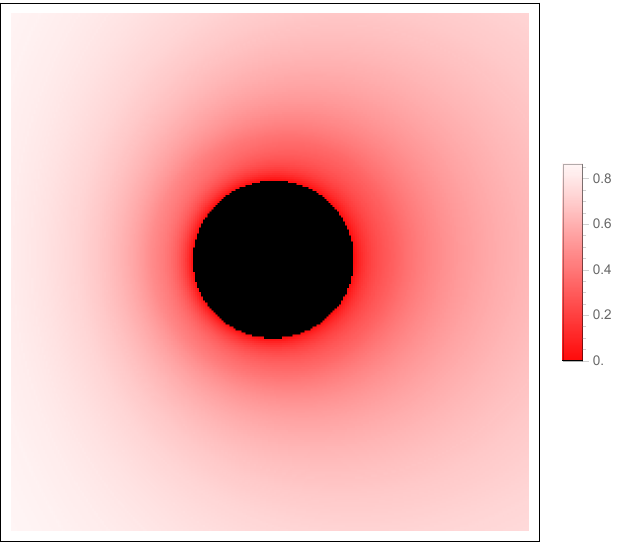}}
\subfigure[$\alpha=-0.5$, $\mathcal{G}=0.1$, $a=0.99$, $\theta_o = 60^{\circ}$]{\includegraphics[scale=0.38]{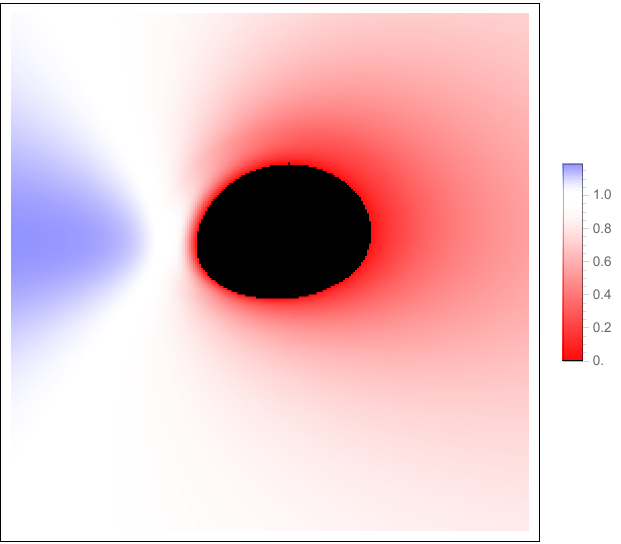}}
\subfigure[$\alpha=-0.5$, $\mathcal{G}=0.3$, $a=0.99$, $\theta_o = 60^{\circ}$]{\includegraphics[scale=0.38]{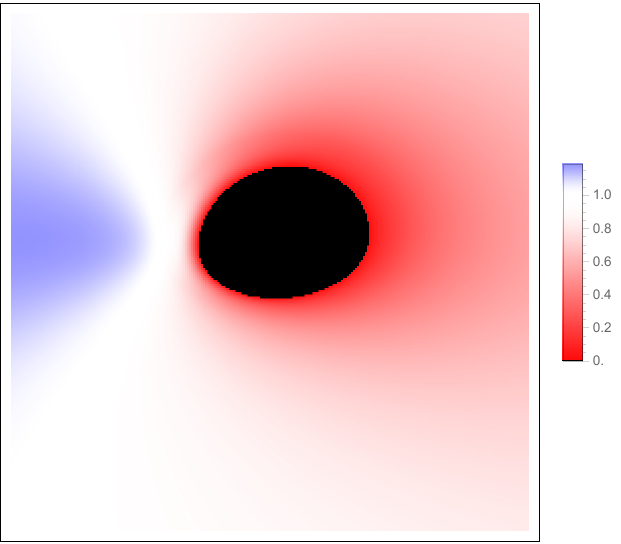}}
\subfigure[$\alpha=-0.5$, $\mathcal{G}=0.1$, $a=0.5$, $\theta_o = 60^{\circ}$]{\includegraphics[scale=0.38]{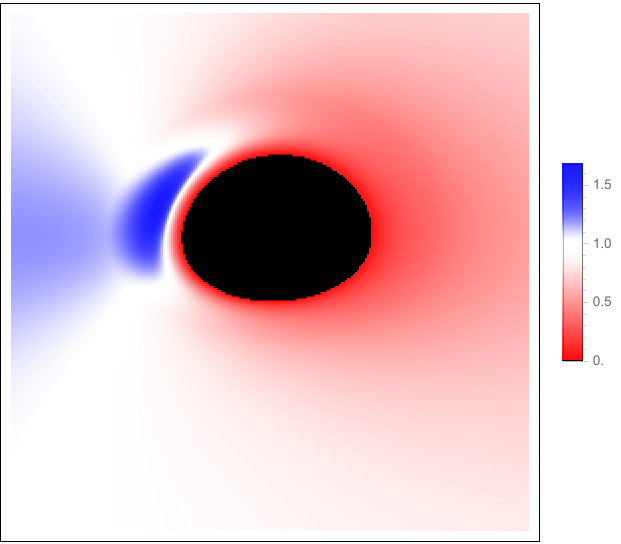}}
\subfigure[$\alpha=-0.9$, $\mathcal{G}=0.1$, $a=0.99$, $\theta_o = 60^{\circ}$]{\includegraphics[scale=0.38]{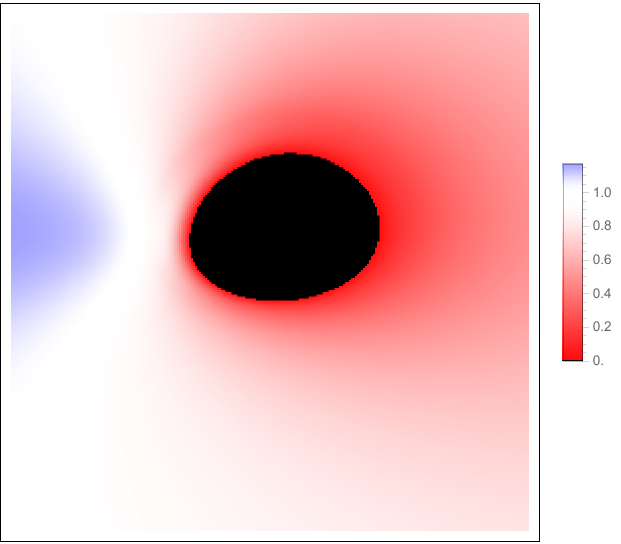}}
\subfigure[$\alpha=-0.5$, $\mathcal{G}=0.1$, $a=0.99$, $\theta_o = 83^{\circ}$]{\includegraphics[scale=0.38]{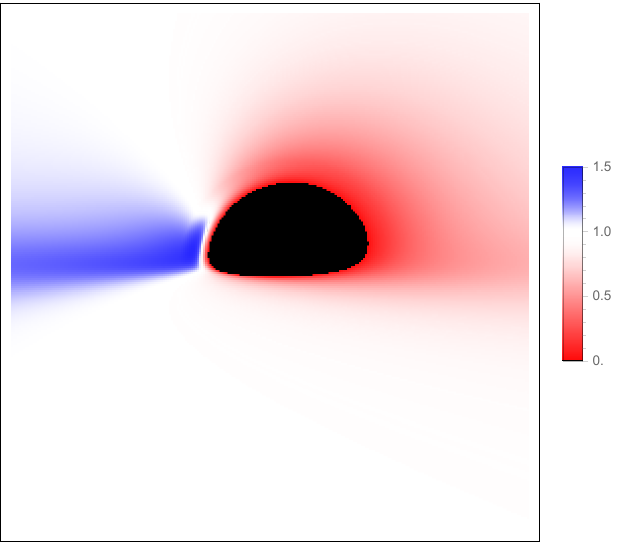}}
\subfigure[$\alpha=-0.5$, $\mathcal{G}=0.3$, $a=0.99$, $\theta_o = 83^{\circ}$]{\includegraphics[scale=0.38]{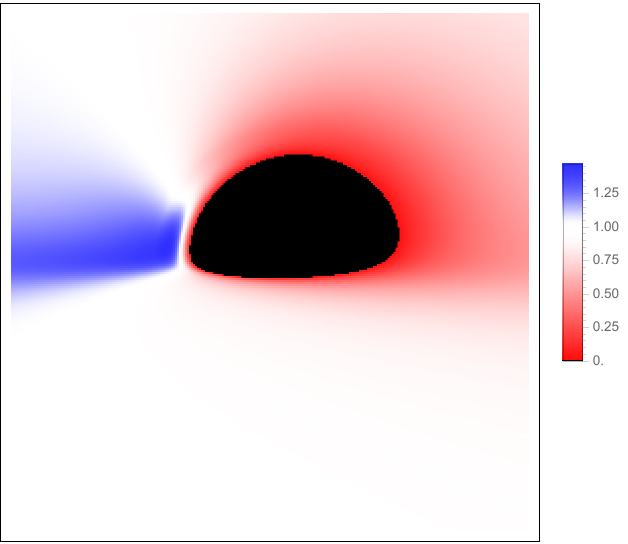}}
\subfigure[$\alpha=-0.5$, $\mathcal{G}=0.1$, $a=0.5$, $\theta_o = 83^{\circ}$]{\includegraphics[scale=0.38]{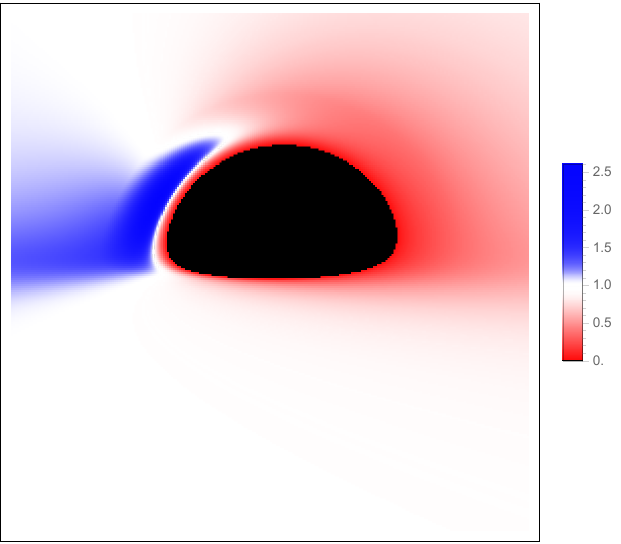}}
\subfigure[$\alpha=-0.9$, $\mathcal{G}=0.1$, $a=0.99$, $\theta_o = 83^{\circ}$]{\includegraphics[scale=0.38]{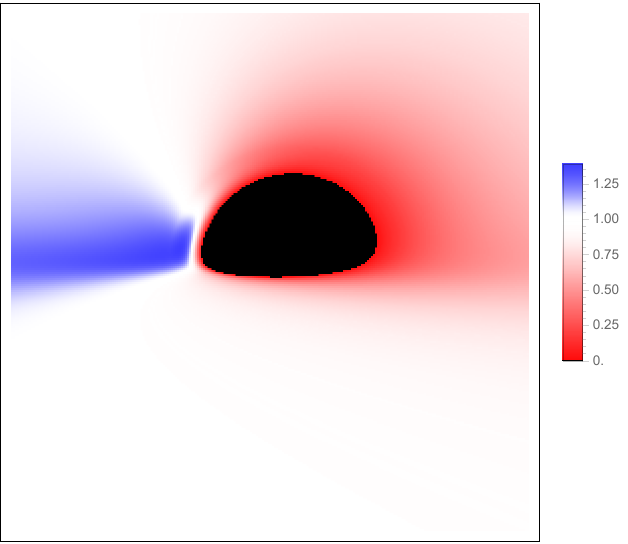}}
\caption{\label{figRDR1} The redshift factor distribution in direct images of accretion disks. The redshift factor is visually represented using a continuous and linear color spectrum, where the color red signifies redshift and the color blue signifies blueshift. The observation angles from the first to the third rows are $\theta_o = 17^{\circ}$, $\theta_o = 60^{\circ}$, and $\theta_o = 83^{\circ}$, respectively. The black regions are the inner shadows.}
\end{figure}

\begin{figure}[!t]
\centering 
\subfigure[$\alpha=-0.5$, $\mathcal{G}=0.1$, $a=0.99$,  $\theta_o = 17^{\circ}$]{\includegraphics[scale=0.37]{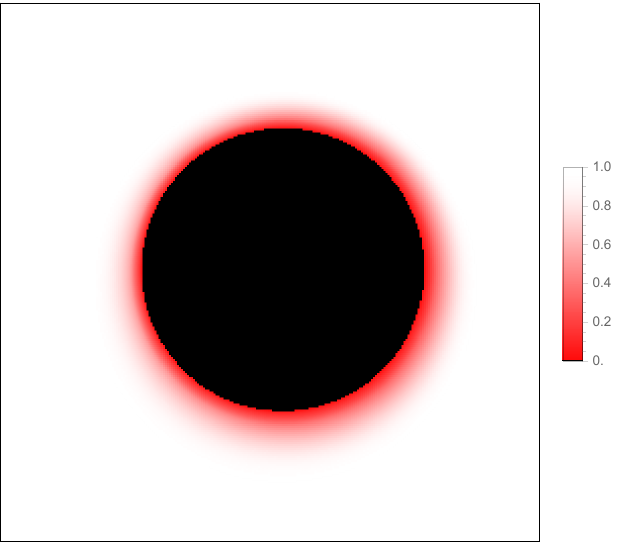}}
\subfigure[$\alpha=-0.5$, $\mathcal{G}=0.3$, $a=0.99$,  $\theta_o = 17^{\circ}$]{\includegraphics[scale=0.37]{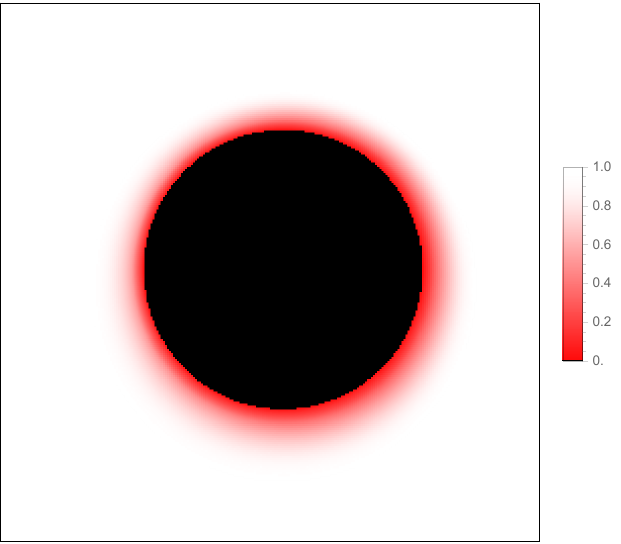}}
\subfigure[$\alpha=-0.5$, $\mathcal{G}=0.1$, $a=0.5$, $\theta_o = 17^{\circ}$]{\includegraphics[scale=0.37]{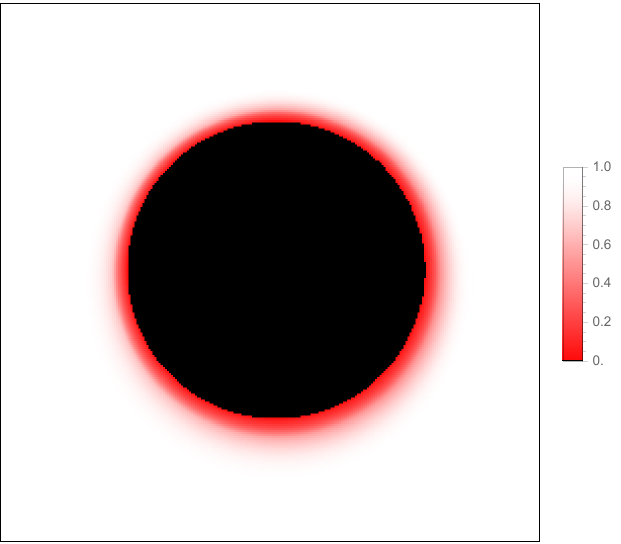}}
\subfigure[$\alpha=-0.9$, $\mathcal{G}=0.1$, $a=0.99$, $\theta_o = 17^{\circ}$]{\includegraphics[scale=0.37]{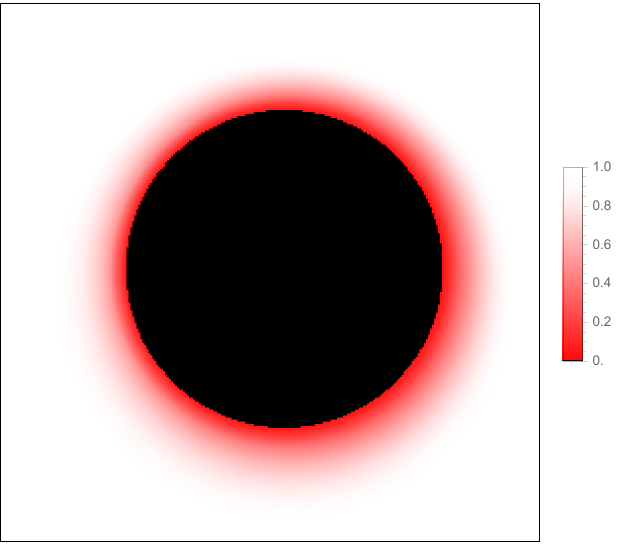}}
\subfigure[$\alpha=-0.5$, $\mathcal{G}=0.1$, $a=0.99$, $\theta_o = 60^{\circ}$]{\includegraphics[scale=0.37]{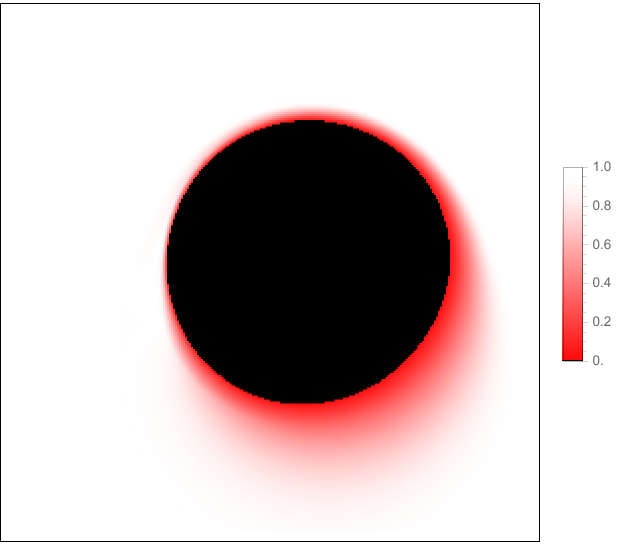}}
\subfigure[$\alpha=-0.5$, $\mathcal{G}=0.3$, $a=0.99$, $\theta_o = 60^{\circ}$]{\includegraphics[scale=0.37]{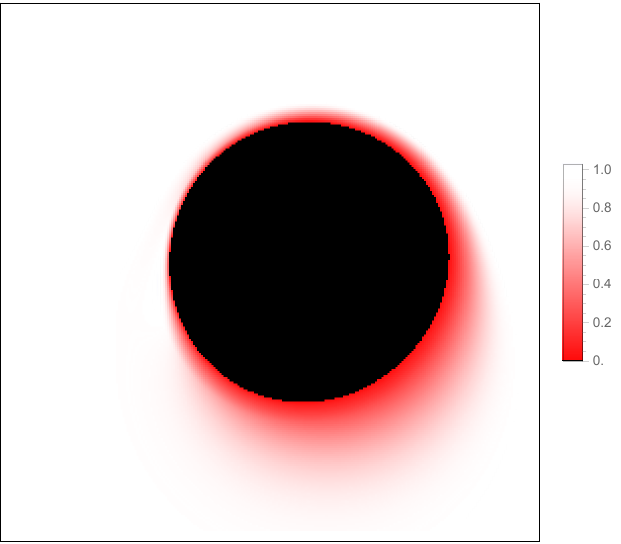}}
\subfigure[$\alpha=-0.5$, $\mathcal{G}=0.1$, $a=0.5$, $\theta_o = 60^{\circ}$]{\includegraphics[scale=0.37]{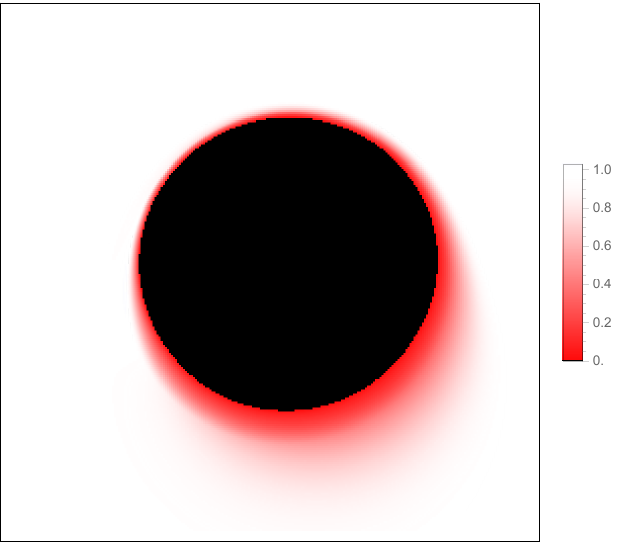}}
\subfigure[$\alpha=-0.9$, $\mathcal{G}=0.1$, $a=0.99$, $\theta_o = 60^{\circ}$]{\includegraphics[scale=0.37]{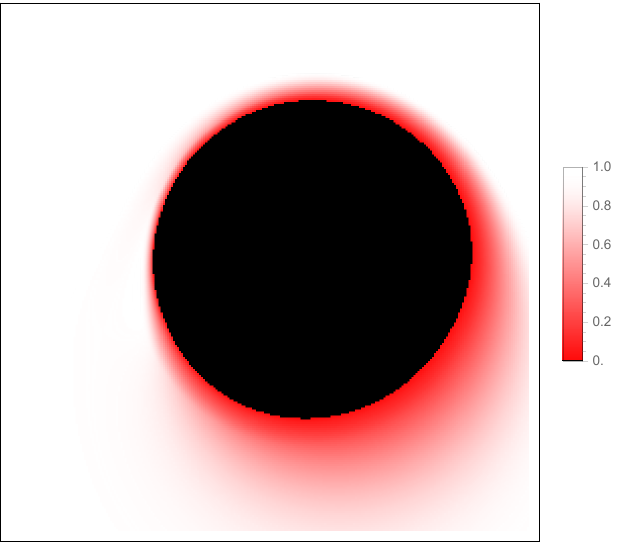}}
\subfigure[$\alpha=-0.5$, $\mathcal{G}=0.1$, $a=0.99$, $\theta_o = 83^{\circ}$]{\includegraphics[scale=0.37]{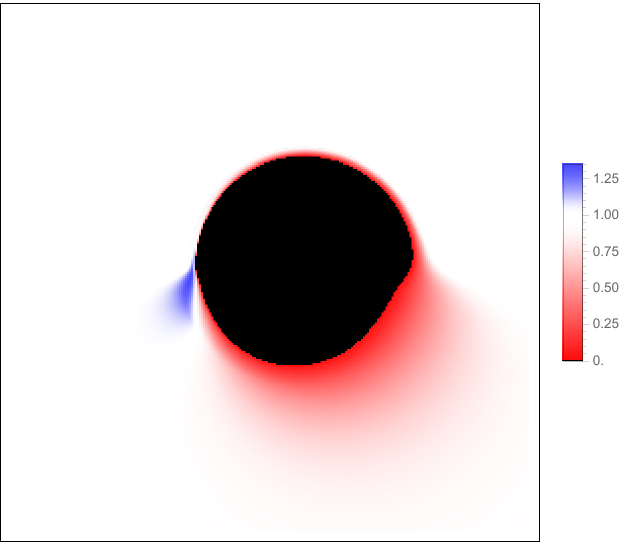}}
\subfigure[$\alpha=-0.5$, $\mathcal{G}=0.3$, $a=0.99$, $\theta_o = 83^{\circ}$]{\includegraphics[scale=0.37]{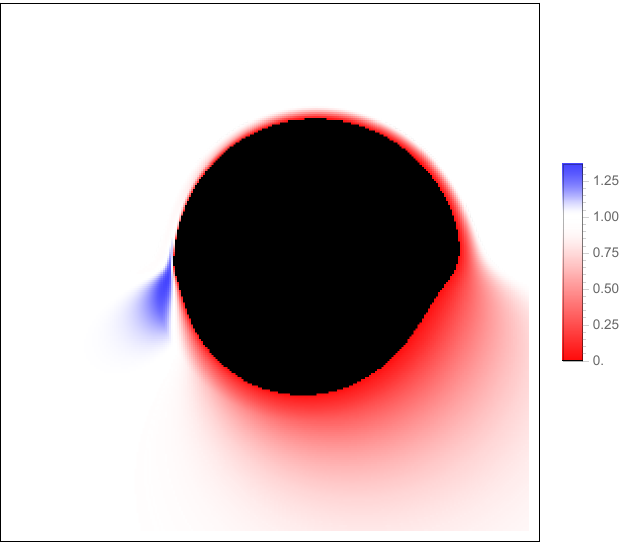}}
\subfigure[$\alpha=-0.5$, $\mathcal{G}=0.1$, $a=0.5$, $\theta_o = 83^{\circ}$]{\includegraphics[scale=0.37]{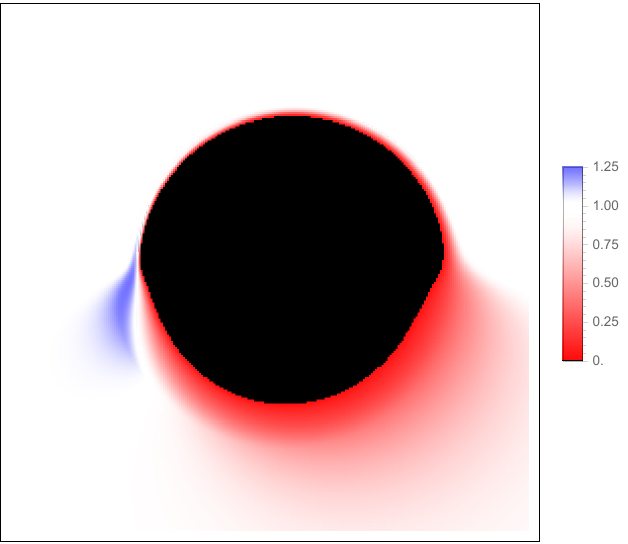}}
\subfigure[$\alpha=-0.9$, $\mathcal{G}=0.1$, $a=0.99$, $\theta_o = 83^{\circ}$]{\includegraphics[scale=0.37]{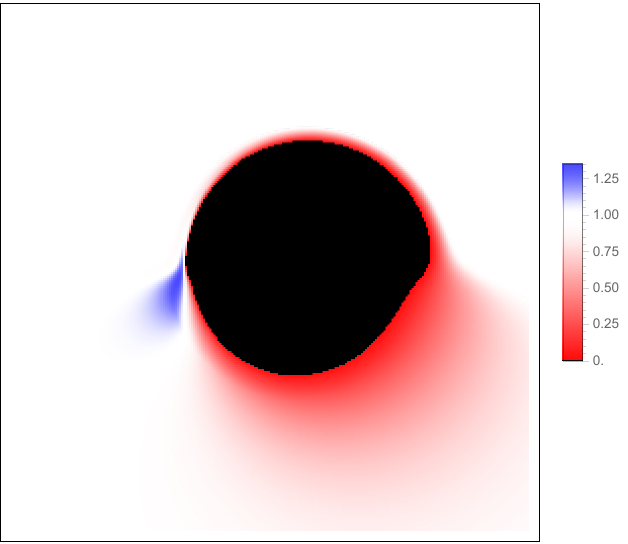}}
\caption{\label{figRDR2} The redshift factor distribution in lensed images of accretion disks. The redshift factor is visually represented using a continuous and linear color spectrum, where the color red signifies redshift and the color blue signifies blueshift. The observation angles from the first to the third rows are $\theta_o = 17^{\circ}$, $\theta_o = 60^{\circ}$, and $\theta_o = 83^{\circ}$, respectively. The boundaries of the black regions represent the lensed images of event horizon $r_h$.}
\end{figure}

The Doppler effect is introduced by the relative motion between the accreting particles and the observer, which can have a significant impact on the characteristics of the black hole image. Naturally, one crucial aspect that necessitates meticulous attention in the imaging process of a black hole is the precise determination of the redshift associated with the captured light. In Fig.\ref{figRDR1} and Fig.\ref{figRDR2}, the redshifts of direct and lensed images are shown when the relevant parameters are taken to different values.

The images in Fig. \ref{figRDR1} and \ref{figRDR2} are arranged in three rows, where the first row corresponds to an observed inclination of $\theta_o$ with $\theta_o = 17^{\circ}$, the second row corresponds to $\theta_o = 60^{\circ}$, and the third row corresponds to $\theta_o = 83^{\circ}$.The redshift factor is visually represented by a continuous linear color map, with the color red indicating redshift and the color blue indicating blueshift. The central black region of each observing plane represents the inner shadow cast by the accretion disk, which is bounded by the projection of the event horizon of the black hole. The results indicate that the size of the inner shadow region is correlated with the values of  relevant parameters, while its shape is influenced by the observation inclination.

At smaller observation inclinations, such as $\theta_o = 17^{\circ}$, the outer boundary of the inner shadow always exhibits a prominent "red ring" due to the emission of light by particles within the falling region, resulting in a significant redshift. Meanwhile, both the direct and lensed images of the accretion disk predominantly exhibits redshift features, while the discernibility of blueshift features is considerably diminished. The emergence of this phenomenon is attributed to the component of the accretion disk's motion projected along the line of sight is insignificant at smaller observation angles, thereby gravitational redshift to be the prevailing effect. In particular, the blueshift effect is not observed in lensed images, even when the observed inclination is $\theta_o = 60^{\circ}$. When the observed inclination is a large value $\theta_o = 83^{\circ}$, the distribution of the blue shift and red shift regions is primarily determined by the rotation direction of the accretion disk, with the former located on the left side and the latter on the right side of the image. Interestingly, the phenomena of redshift and blueshift will be suppressed with the increase of the absolute values of rotation parameter $a$, magnetic charge $\mathcal{G}$, and dark matter parameter $\alpha$, while they will be enhanced with an increase in observation angle.  The impact of observation inclination on the distribution of redshift factor is highly intuitive. With a larger observation inclination, the line of sight approaches closer to the disk plane, resulting in an increased velocity component of accretion material along the line of sight and consequently enhancing the Doppler effect.

\section{Conclusions and discussions}
\label{sec:conclusion}
The groundbreaking accomplishments of the EHT have paved new avenues for harnessing black hole imagery in the exploration of high-energy physics and gravitational theories, igniting widespread interest in numerical simulations of black hole images within the scientific community. In the imaging process of black holes, the accretion of matter surrounded by high-energy radiation plays a pivotal role.  After considering two distinct background light sources, namely the celestial light source and the thin accretion disk, we thoroughly examine the observed properties of the rotating Bardeen black holes surrounded by PFDM and elucidate how changes in relevant parameters impact the characteristics of its image.

As the only background light source, the celestial sphere was  employed to illuminate the black hole, while employing the backward ray-tracing technique enabled us to ascertain the precise shape of the shadow projected onto the observer's screen. The results indicate that the observation plane of ZAMO consistently exhibits a central region characterized by darkness, accompanied by a ring-like structure. The dark region corresponds to the shadow cast by the black hole, while the white ring structure represents the Einstein ring. During the process of increasing the observation angle from $\theta_o = 0^{\circ}$ (the pole of the black hole) to $\theta_o = 90^{\circ}$ (the equatorial plane of the black hole), it is evident that the silhouette of the black hole shadow gradually transitions from an axisymmetric closed circle to a D-shaped configuration, and its size also increases slightly. We also investigate the correlation between black hole shadow characteristics and other spacetime parameters, namely magnetic charge $\mathcal{G}$, rotation parameter $a$, and dark matter parameter $\alpha$. The observation clearly indicates that, for a constant magnetic charge $\mathcal{G}$, an increase in the absolute value of the dark matter parameter $\alpha$ leads to an enlargement in the size of the shadow. As the parameter $a$ increases, there is an enhancement in the degree of deformation of the shadow shape, and the image of the background light source surrounding the shadow also undergoes distortion due to the drag effect. For a constant dark matter parameter $\alpha$, although the size of the shadow remains relatively unchanged, there is a discernible increase in the level of distortion induced by magnetic charge $\mathcal{G}$ as the rotation parameter $a$ escalates.

On the other hand, we take into account the presence of optically and geometrically thin accretion disks as the sole source of background illumination for the black hole. The accretion disk is situated in the equatorial plane of the black hole, and particle motion within the disk can be classified into two categories based on  the ISCO, that is, the particles located outside the ISCO will exhibit stable circular orbits, whereas particles inside the ISCO will undergo the critical plunging orbits.  The radiation emitted by the accretion disk is simulated based on the methodology outlined in reference \cite{Hou:2022eev}. Within the ZAMO framework, we compute the intensity distribution on the observer screen by meticulously tracing a complete light path between the screen and the accretion disk. In the image of a rotating BD black hole surrounded by PFDM, our primary focus lies in examining the variations observed in the  direct image, lens image, critical curve and inner shadows  under different relevant parameters. When the angle of observation increases, the observed flux of the direct and lensed images of the accretion disk starts to converge towards the lower region of the observation plane, while simultaneously transforming the circular shape of the inner shadow into an ellipsoid or arch-like form.  For a  larger observation inclinations, such as $\theta_o = 60^{\circ}$ and $\theta_o = 83^{\circ}$, the left side of the image will exhibit a crescent or eyebrow-shaped bright area, which can be attributed to the amplification of the angle causing an increasingly pronounced Doppler effect. Interestingly, when the observation inclination is small $\theta_o = 17^{\circ}$, due to the distribution of observation intensity is not sharp enough, the separation degree of direct and lens images is not obvious enough, that is, the direct image and the transparent image of the accretion disk cannot be directly distinguished. It should be emphasized that the change of the observation inclination does not affect the position of the critical curve, but its profile will be slightly deformed at a large observation inclination. In addition, one can find that an increase in the absolute value of the dark matter parameter $\alpha$ leads to a corresponding augmentation in both the critical curve  and  inner shadow region, thereby significantly expanding even the lensed image region. Although the increase in magnetic charge yields a similar effect to parameter A, its impact is relatively negligible.

Furthermore,  we conducted an investigation into the redshift of both the direct and lensed images originating from the accretion disk. It can be observed that  the redshift or blueshift effect of both direct and lens images is amplified with the increase in observation inclination. The presence of a smaller observation inclinations, such as $\theta_o=17^{\circ}$, leads to the formation of a prominent red ring structure surrounding the inner shadow, which is attributed to the emission of light by particles within the critical plunging orbits. Consequently, there is a significant redshift observed. Meanwhile, the negligible projection of the accretion disk motion along the line of sight simultaneously renders the blue shift feature almost imperceptible renders the blue shift feature almost imperceptible. In particular, in the lens image of the accretion disk, even if the observed inclination reaches $\theta_o = 60^{\circ}$, only the redshift effect remains observable. Additionally, changes in relevant parameters significantly influence the redshift or blueshift of the image. It is demonstrated that augmenting the absolute value of the rotation parameter $a$, magnetic charge $\mathcal{G}$, and dark matter parameter $\alpha$ will hinder both redshift and blueshift phenomena.

\section*{Acknowledgements}{This work is supported by the National Natural Science Foundation of China (Grant Nos. 11875095 and 11903025), and by the starting fund of China West Normal University (Grant No.18Q062), and by the Sichuan Youth Science and Technology Innovation Research Team (21CXTD0038), and by the Chongqing science and Technology Bureau (cstc2022ycjh-bgzxm0161), and by the Natural Science Foundation of SiChuan Province(2022NSFSC1833), and by the Natural Science Foundation of Chongqing (CSTB2023NSCQ-MSX0594), and by the Sichuan Science and Technology Program (No. 2023ZYD0023).


\end{document}